# Software Design Document, Testing, Deployment

# And Configuration Management,

# And User Manual of the UUIS

# -- A Team 4 COMP5541-W10 Project Approach

**Computer Science & Software Engineering**

## Team Members

| | |
|---|---|
| Requirements Analyst | Kanj Sobh |
| System Architect | Deyvisson Oliveira |
| Development Lead | Bing Liu |
| UI Specialist | Max Mayantz |
| Quality Assurance Specialist | Yu Ming Zhang |
| Database Administrator | Ali Alhazmi |
| System Administrator | Robin de Bled |
| Project Manager | Abdulrahman Al-Sharawi |

# Table of Contents

# 1 Introduction

This introduction provides an overview of the *System Architecture Document* for Unified University Inventory System. It includes the purpose, scope, target audience, design approach, main component design and high level system design considerations of the system.

## Document Scope and Purpose

This document provides a description of the technical design for Unified University Inventory System – Web Portal. This document's primary purpose is to describe the technical vision for how business requirements will be realized. This document provides an architectural overview of the system to depict different aspects of the system. This document also functions as a foundational reference point for developers.

Please note that this is a baseline document and may be updated as development progresses.

## Target Audience

This document is targeted (but not limited) to technical stakeholders:

- Development Team
- IT Management
- Support Staff

It is assumed that the reader has a technical background in software design and development.

## Acronyms/Abbreviations

| Acronym | Meaning |
| --- | --- |
| **UUIS** | Unified University Inventory System |
| **IUFA** | Imaginary University of Arctica |

## Reference Documents

- System requirement document of UUIS
- Development Standards and Guidelines

## System Environment

- Development: Netbeans 6.8 + Grails 1.2.1
- Unit Test: Junit
- Diagrams: Visio 2007 / ConceptDraw Pro (Mac)
- Database Management: MySQL Workench:
- Database: MySQL 5
- Server: Windows Server 2008
- Server: Apache Tomcat 6
- Revision control: Sourceforge Subversion - https://comp5541-team4.svn.sourceforge.net/svnroot/comp5541-team4
- Discussion: Google groupss

## Design Approach

The design approach used here is based on the following:

## Data Flow Design

The data flow of the UUIS is Internet-based. Hibernate technologies will be utilized to retrieve and cache data from MySQL database to be displayed by the Web portal user interface. Hibernate would also allow updating the data where applicable.

## Architecture Design

The Customer Support System application will follow a Four Layer Architecture so that the objects in the system as a whole can be organized to best separate concerns and prepare for distribution and reuse. A principal advantage to this design is the relative stability of the components as seen by the applications developer. Implementations may change considerably to enhance the performance or in response to changes in the architecture. These changes are less likely to cause major impact to the applications' programs.

## UI Design

Wire Frames are used for UI design. Wire frames are an effective tool for collecting and presenting functionality, navigation, and content of an application or web site. Annotations or notes attached to elements or widgets on the wire frame help to communicate specific functions.

Some Screen Shoots.

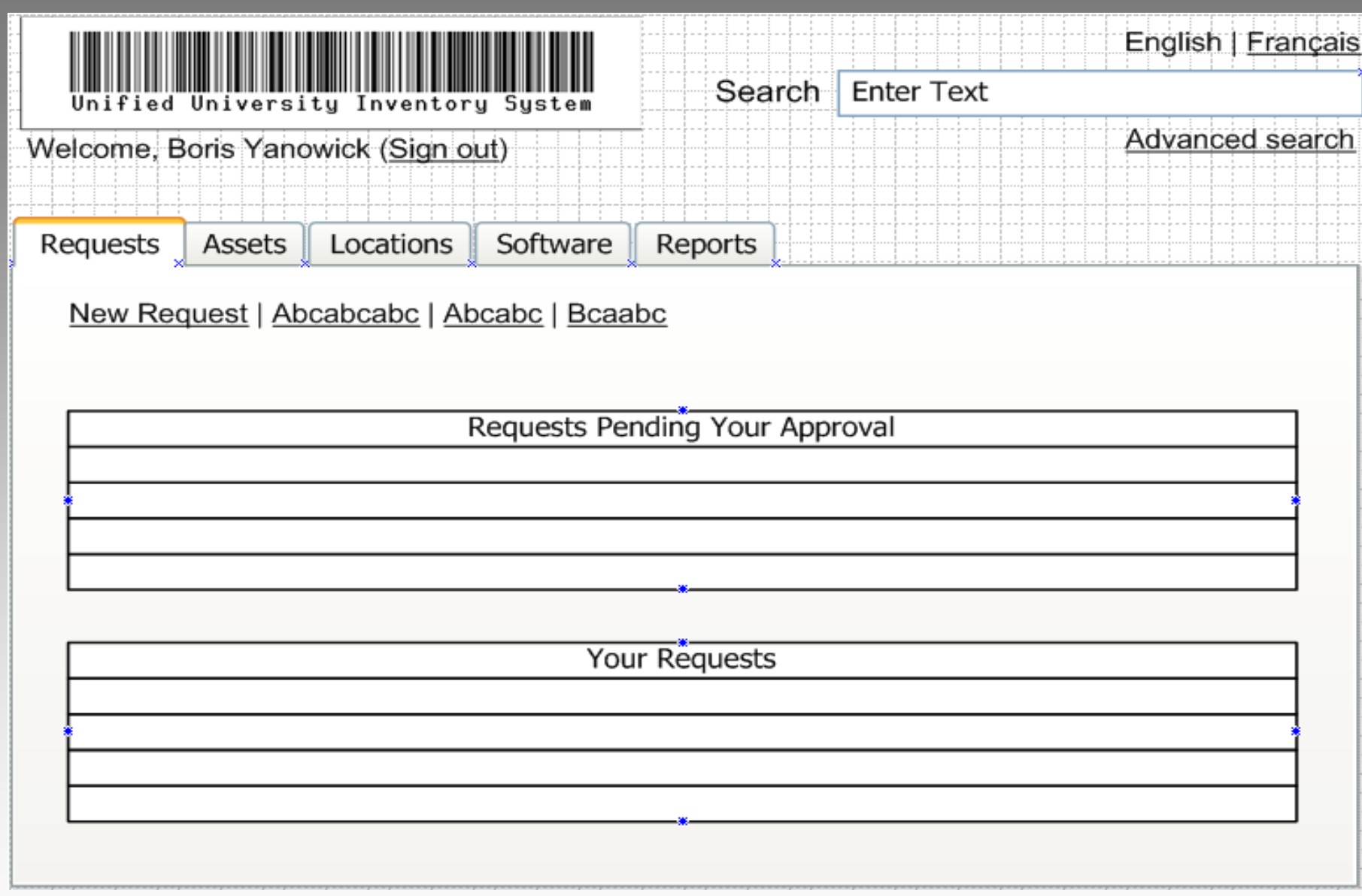

Screen Mockup – Request *Figure 1.1*

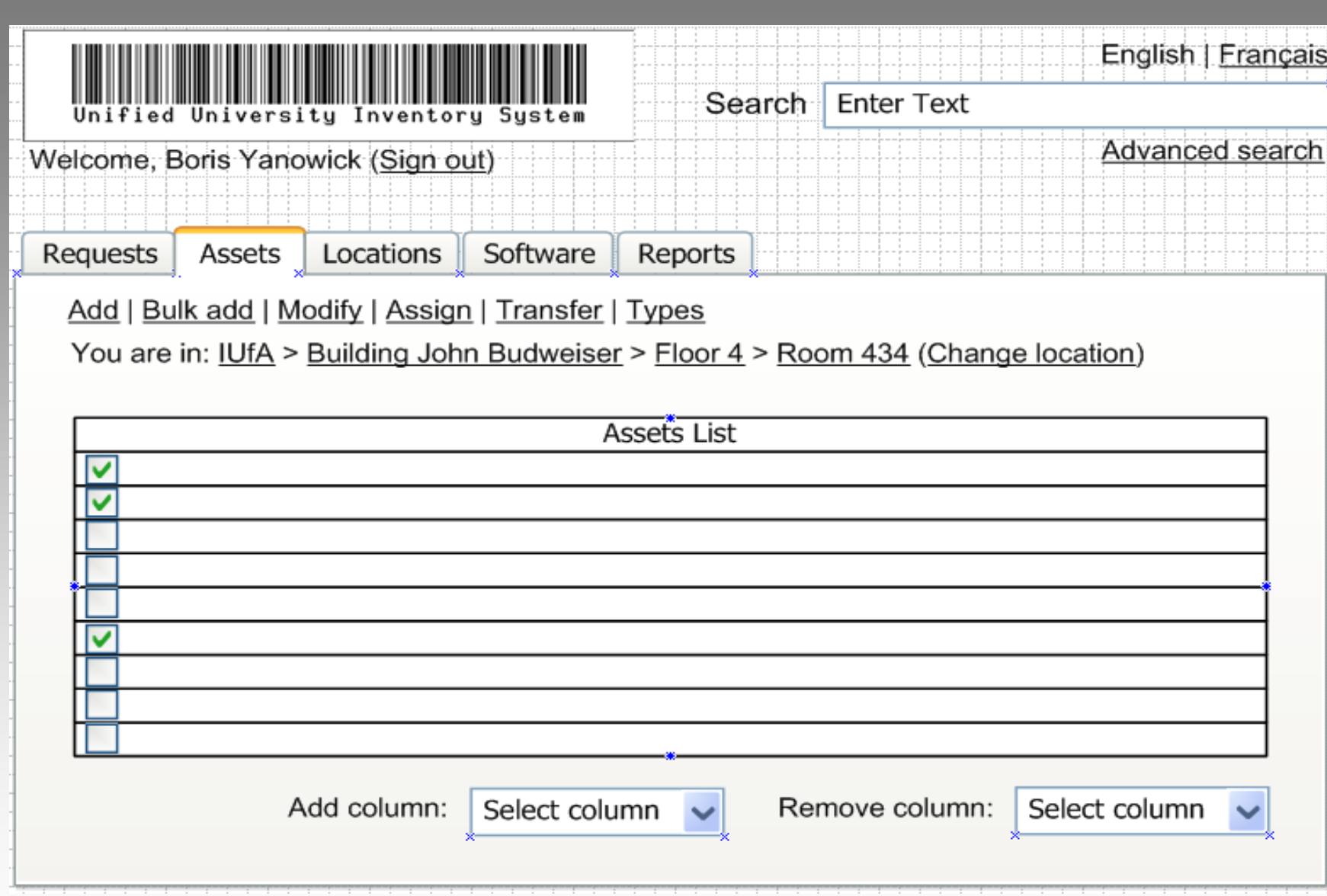

Screen Mockup – Assets *Figure 1.2*

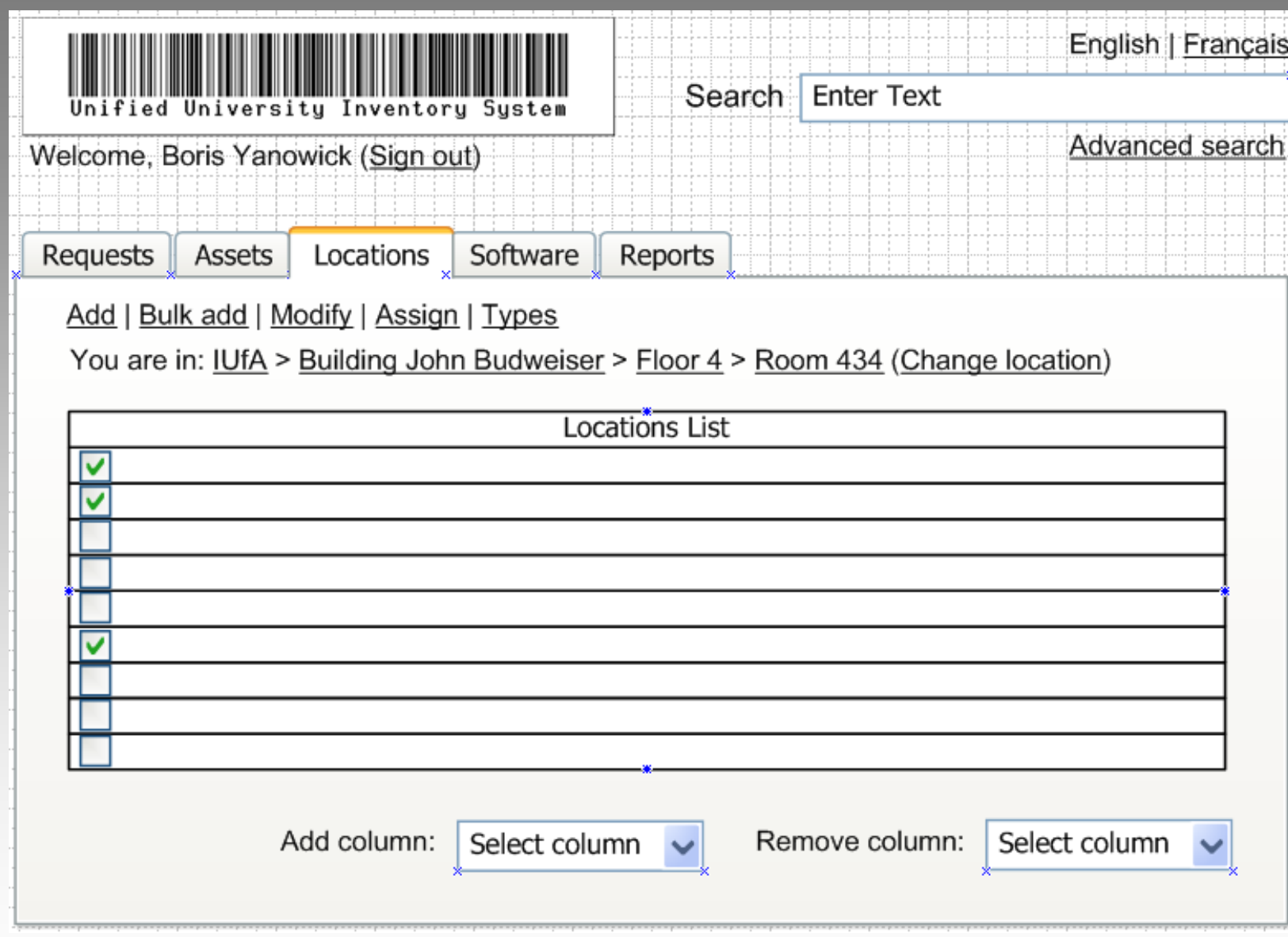

Screen Mockup – Location *Figure 1.3*

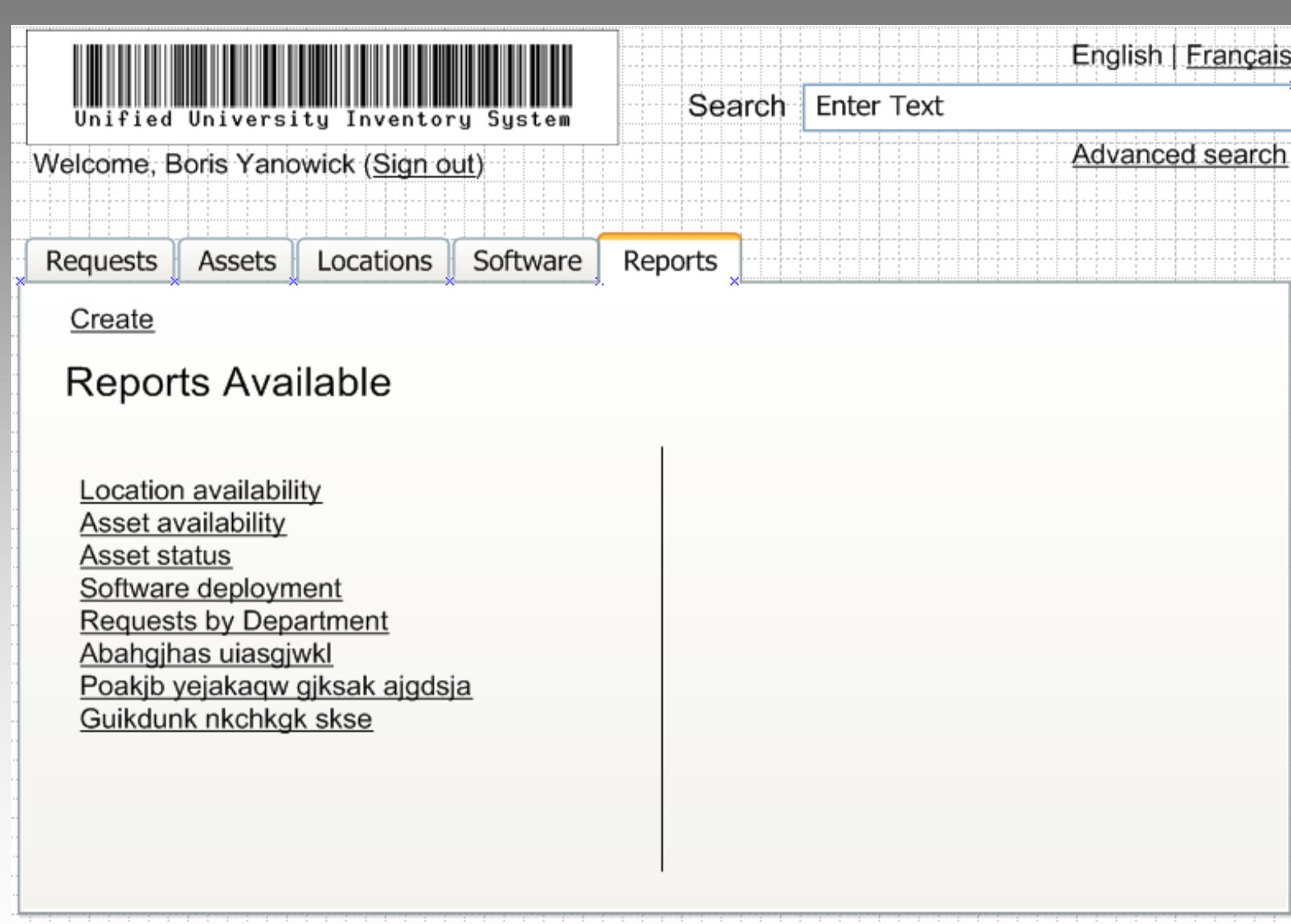

Screen Mockup – Report *Figure 1.4*

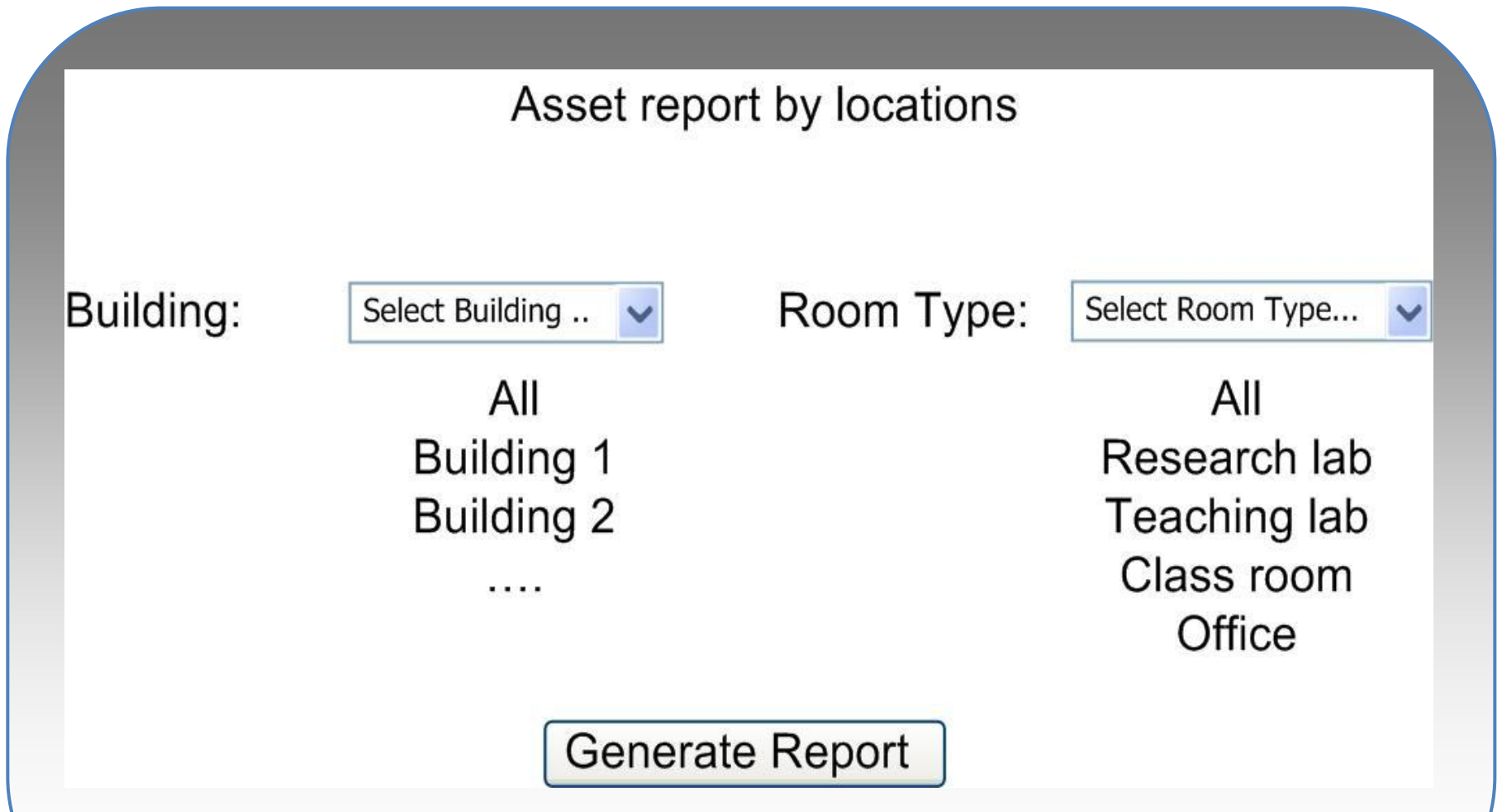

Screen Mockup – Asset report *Figure 1.5*

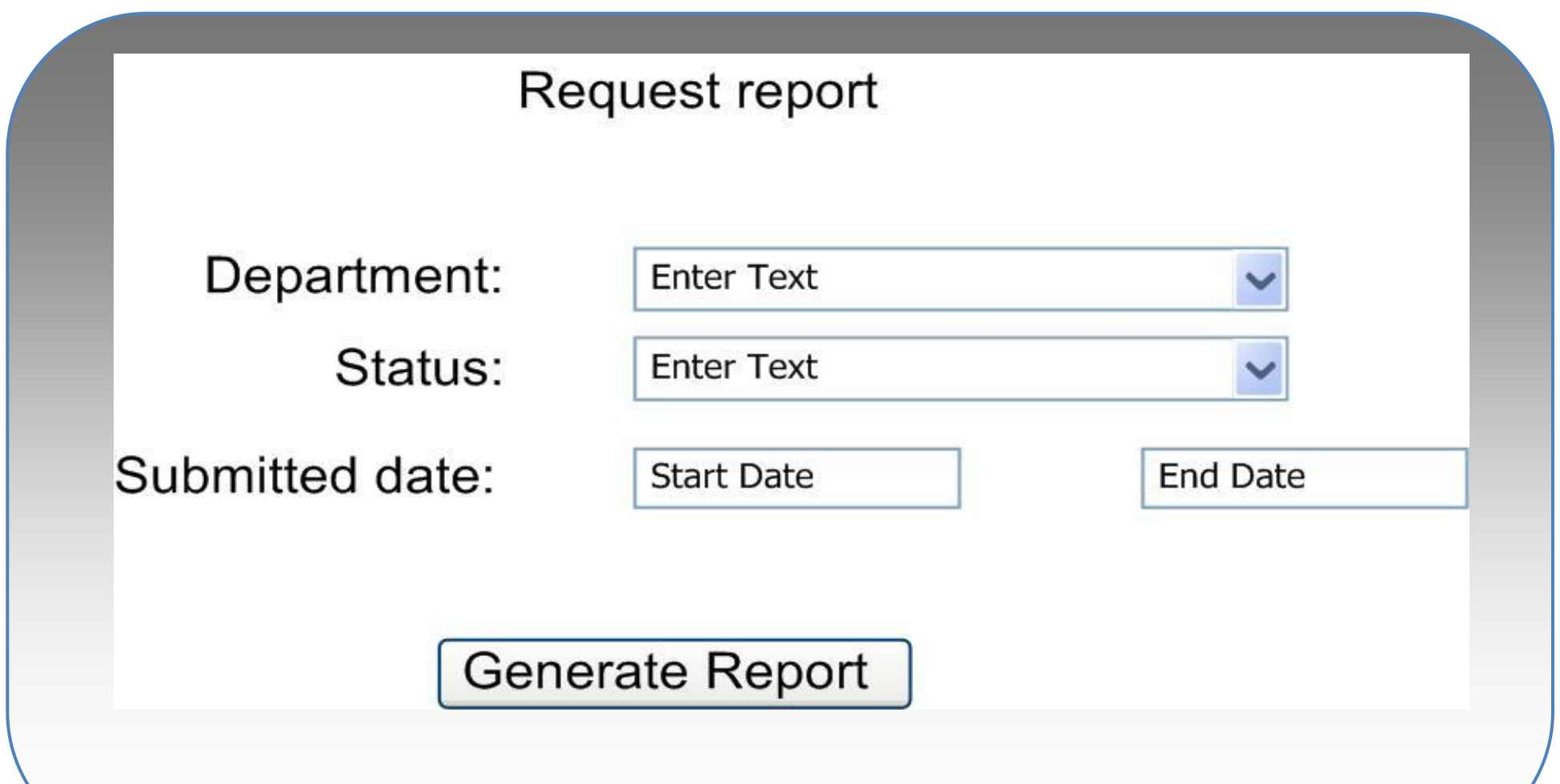

Screen Mockup – Request report *Figure 1.6*

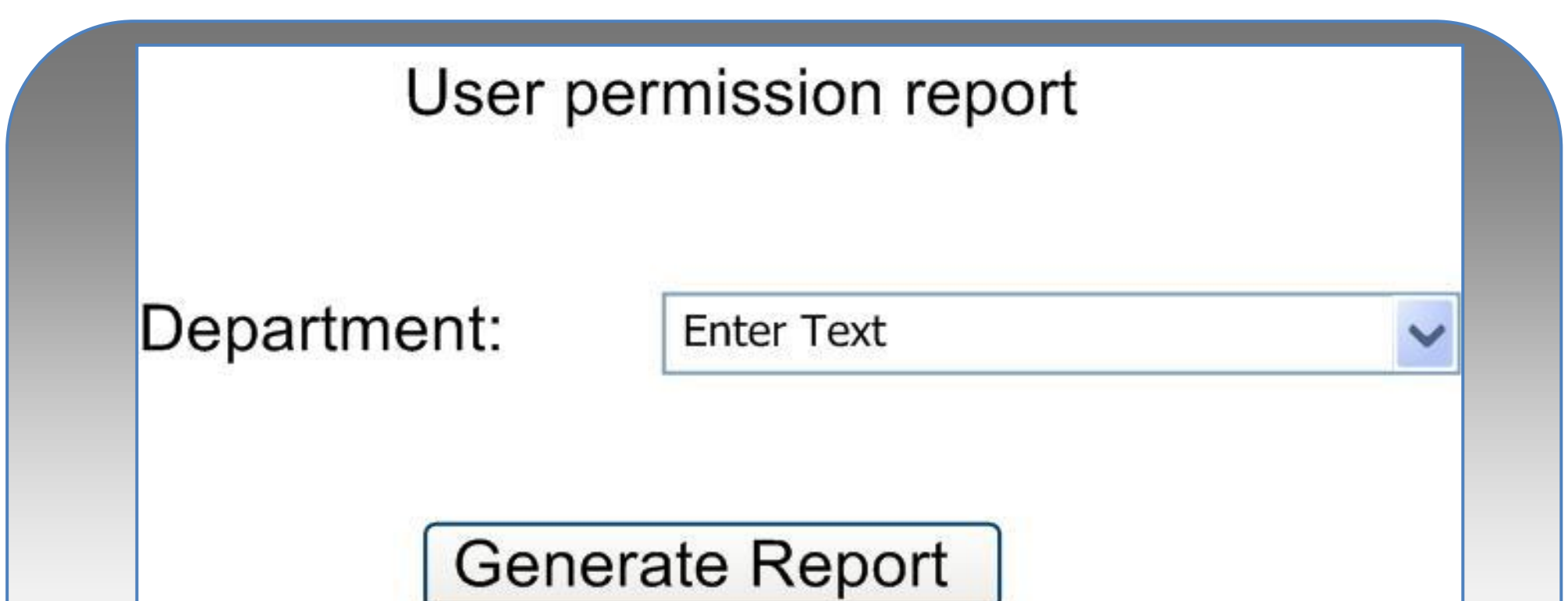

Screen Mockup – User permission report *Figure 1.7*

## Design Patterns

This application is designed as an object-oriented system for an Internet-based architecture using four-layer architecture by factoring application classes into the following layers:

**The Presentation layer**. This is the layer where the physical window and widget objects live. It will also contain Controller classes as in classical MVC. Any new user interface widgets developed for this application are put in this layer. In most cases this layer is completely developed within Grails.

**The Domain Mode**. Most objects identified in the OO analysis and design will reside. To a great extent, the objects in this layer can be application-independent. Generic objects may be used in this application to reap the benefits of Object Oriented programming.

**The Hibernate layer**. The domain model has been mapped using Hibernate. This will map Grails domain classes onto a wider range of legacy systems and will be more flexible in the creation of your database schema.

**The Data layer**. The data is managed by MySQL.

## UUIS High level View

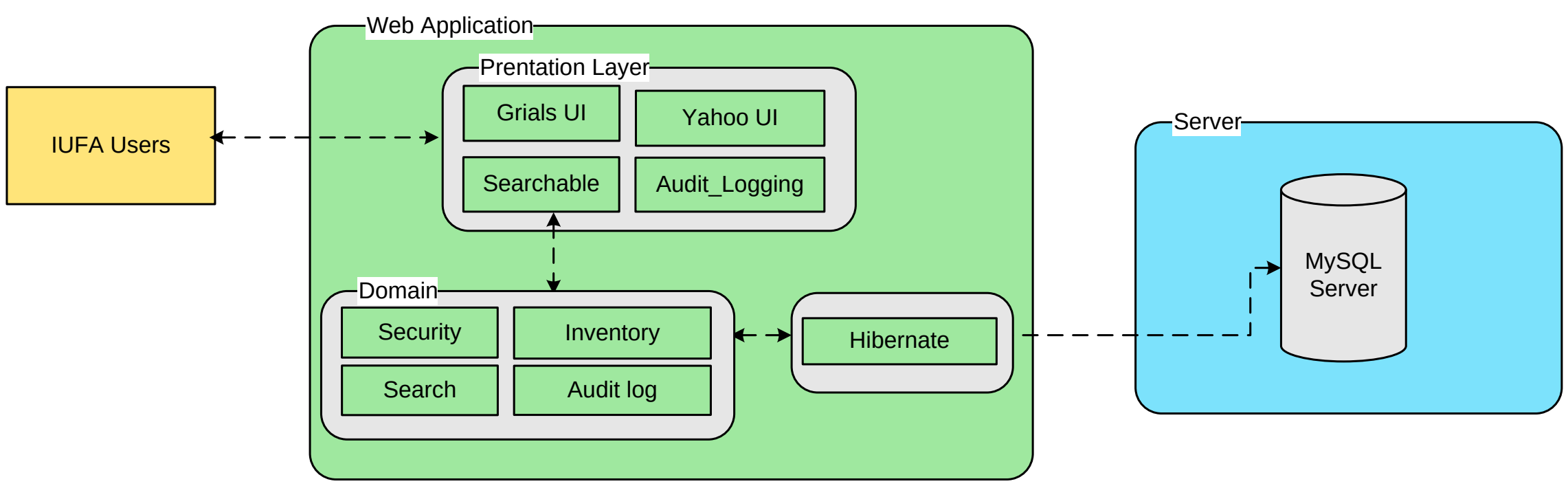

The high level view of UUIS *Figure 1.8*

Grails UI plug in: to present the user interface

Yahoo UI: A set of utilities and controls, written with JavaScript and CSS, for building richly interactive web applications using techniques such as DOM scripting, DHTML and AJAX

Searchable plug in: to implement the search component.

DynamicJasper plug in: to generate the reports.

Audit_Logging plug in: easy to implement the audit log module.

# 2 System Design Considerations

## Web-site directories

The following web-site directories will be used to organize the JSP Pages used in the web-site:

User site Directories:

- Root Directory - All web pages related to user functions.
- Images Directory – Contains the images used by the system.
- Reports: Contains the local reports that would be used to export to CSV format.

## Exception Handling

A good way to track exceptions is to implement an exception handler at the application level. This will allow consolidation of the logging and notification parts of the exception handling in one convenient place.

The global exception handler can handle both specific exceptions that UUIS trap in the code as well as generic un-handled exceptions. After global exception handler has done its work, it would redirect the users of the website to a friendly error page that tells them that something has gone wrong and instructions on what to do.

All unhandled exceptions would be logged in the Windows Application Event Log or log files with some detail on the nature of the failure.

# 3 Modules

## Search

### General Search

The general search included that retrieve records from the database according to user-specified search criteria. Further, the search may encompass other information collections like on-screen data.

### Advanced Search

The advanced search may need to be added to the system in order to remain as flexible as possible, the search module is designed in such a way as to isolate the specifics of search implementation from the application. The search module has a limited API exposed, and depending on its configuration it can implement the required search class without further programmer intervention. It should be possible to add other classes of search, relating to other types of data, with relatively little effort.

## Inventory

### Assets

The assets included the management of asserts (et. Computers, tables, office supplies, software …). The granted user can add, edit and modify the assets within his scope. When create a new asset, a unit bar code will be generated. The granted user can also add, delete or modify an asset type. For each asset, there should have a barcode property (optional). To generate the barcode from the asset ID, we can employ the barcode generator at http://www.barcodesinc.com/generator/index.php

### Locations

The location included the management of location resources (et. Buildings, offices, teaching lab, research lab …). The granted user can add, edit and modify the locations within his scope. And modify the location type, capacity, description and property of. The granted user can also add, delete and modify the location type.

### Requests

The request included the management of requests. The general user can make, submit and view requests. The granted administrators can approve, reject and view requests within his scope.

### Reports

This module generates reports on assets, locations and user permission, and export to pdf to easy use. There will be three reports (also refer the UI mockup):

1) Asset report: by selecting the building and room type, generate the asset report by calculated the numbers of computer, locker, and tables … in that building.
2) Request report: filtering by the department, request status and request data range, generate the request report and export to pdf.
3) User permission report: by selecting the department, generate the user permission report to illustrate all the user permission.
4) User report: by selecting the building and room type, generate the user report by calculated the numbers of students, professors, and stuffs in that building.

### BulkLoad

Batch import the assets from CVS and also able to batch modify the assets from CVS.

## Security

### University Stucture

This module can manage the university, faculties and departments hierarchy structure. Change the relationship of university structure.

### Users & Permissions

This module can assign, remove and editing the user permission. There are four levels of user from the access permission respective: 1) Level 1-the student and general user, who are only able to make request and view their own requests; 2) Level 2-department administrators who are able to manage the requests, view audit log and search assets in their own department; 3) Level 3- faculty administrators who are able to manage the requests, view audit log and search assets in their own faculty; 4) Level 4- IT stuffs, inventory stuffs and university administrators who are able to manage the requests, view audit log and search assets in the whole university.

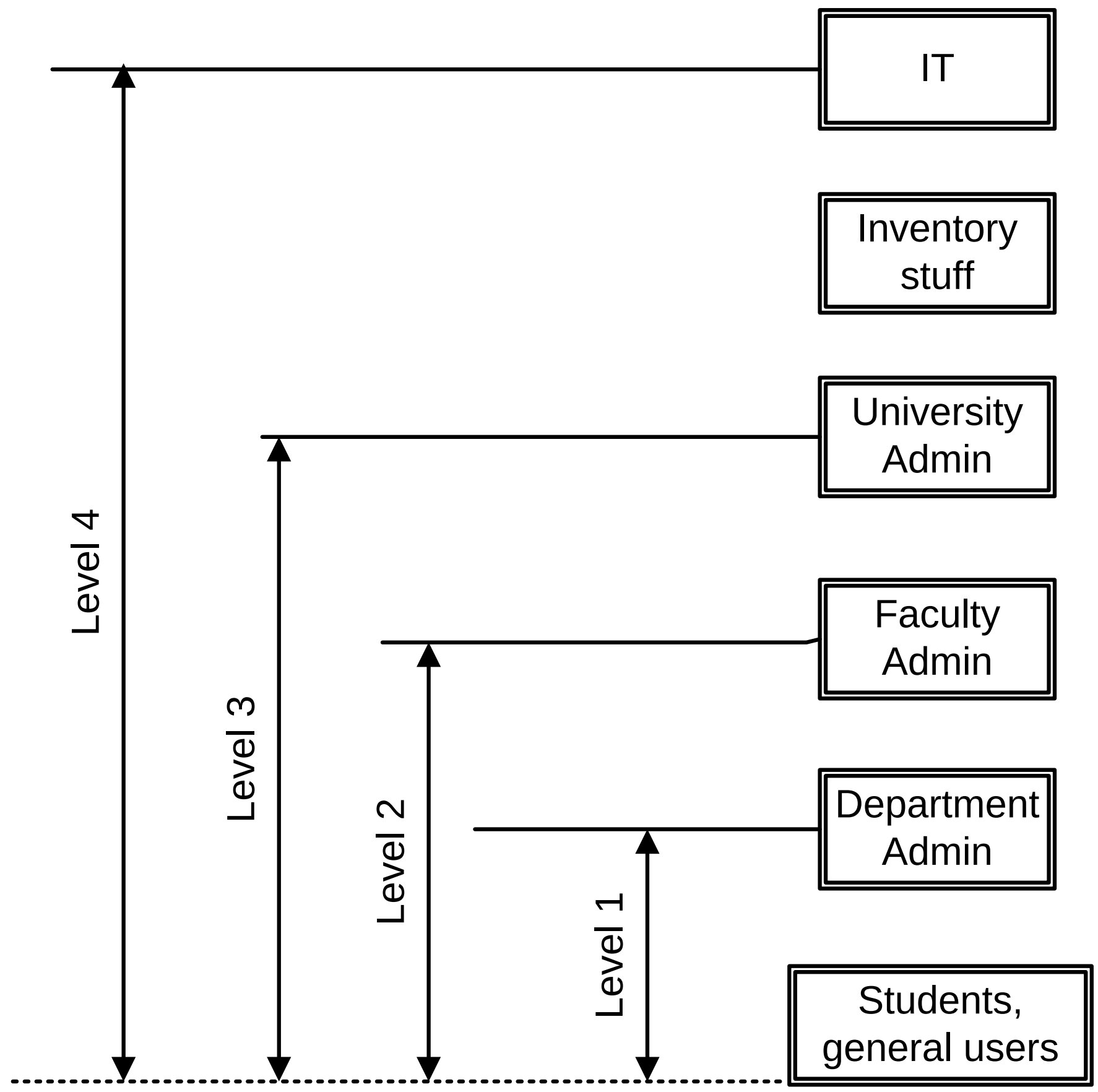

User Hierarchy Level *Figure 3.1*

## Audit log

### **Auditing**

This module can list all the user activities in the web application.

# 4 Activity Diagram

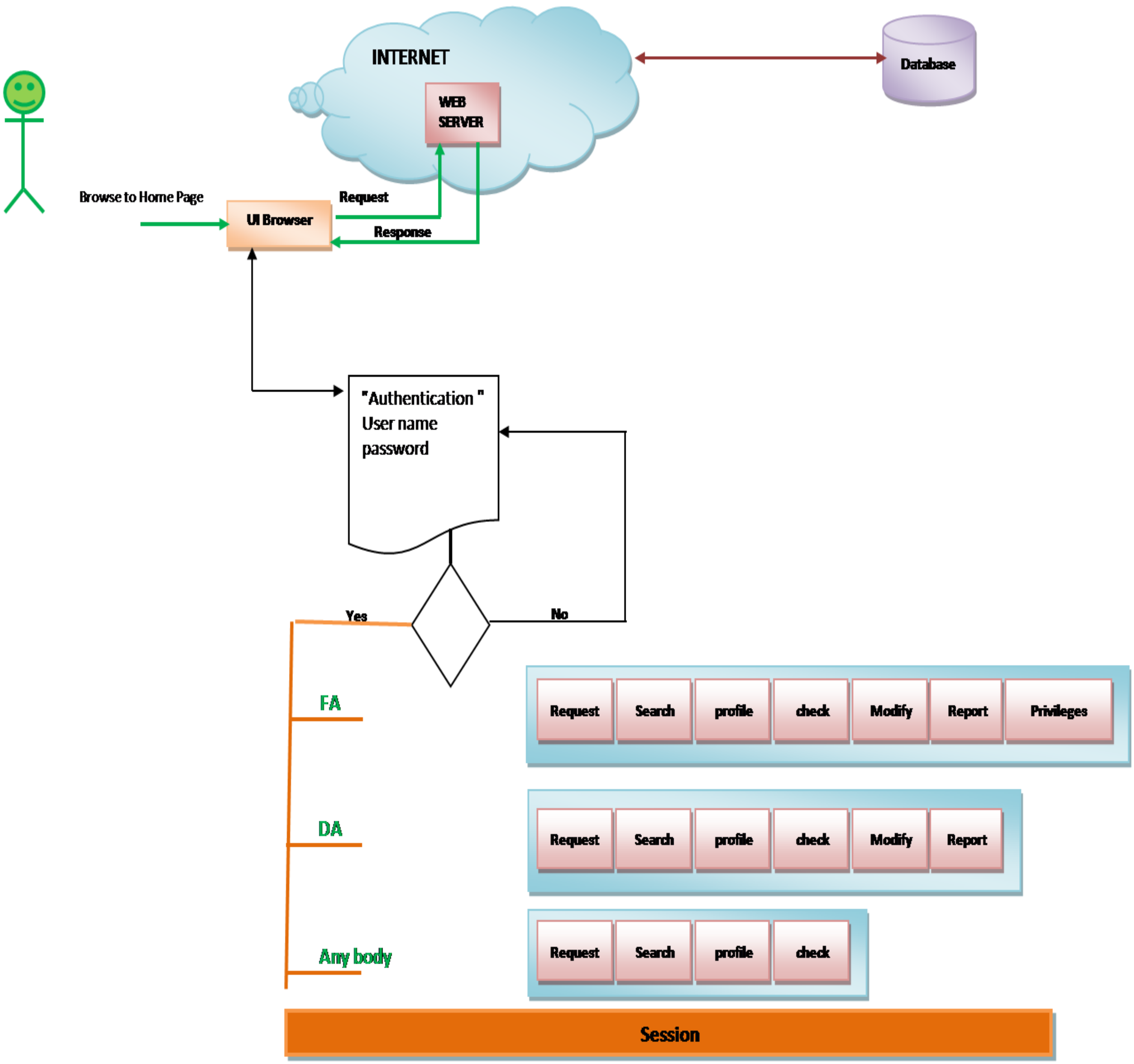

Activity Diagram of UUIS *Figure 4.1*

# 5 Entity Diagram

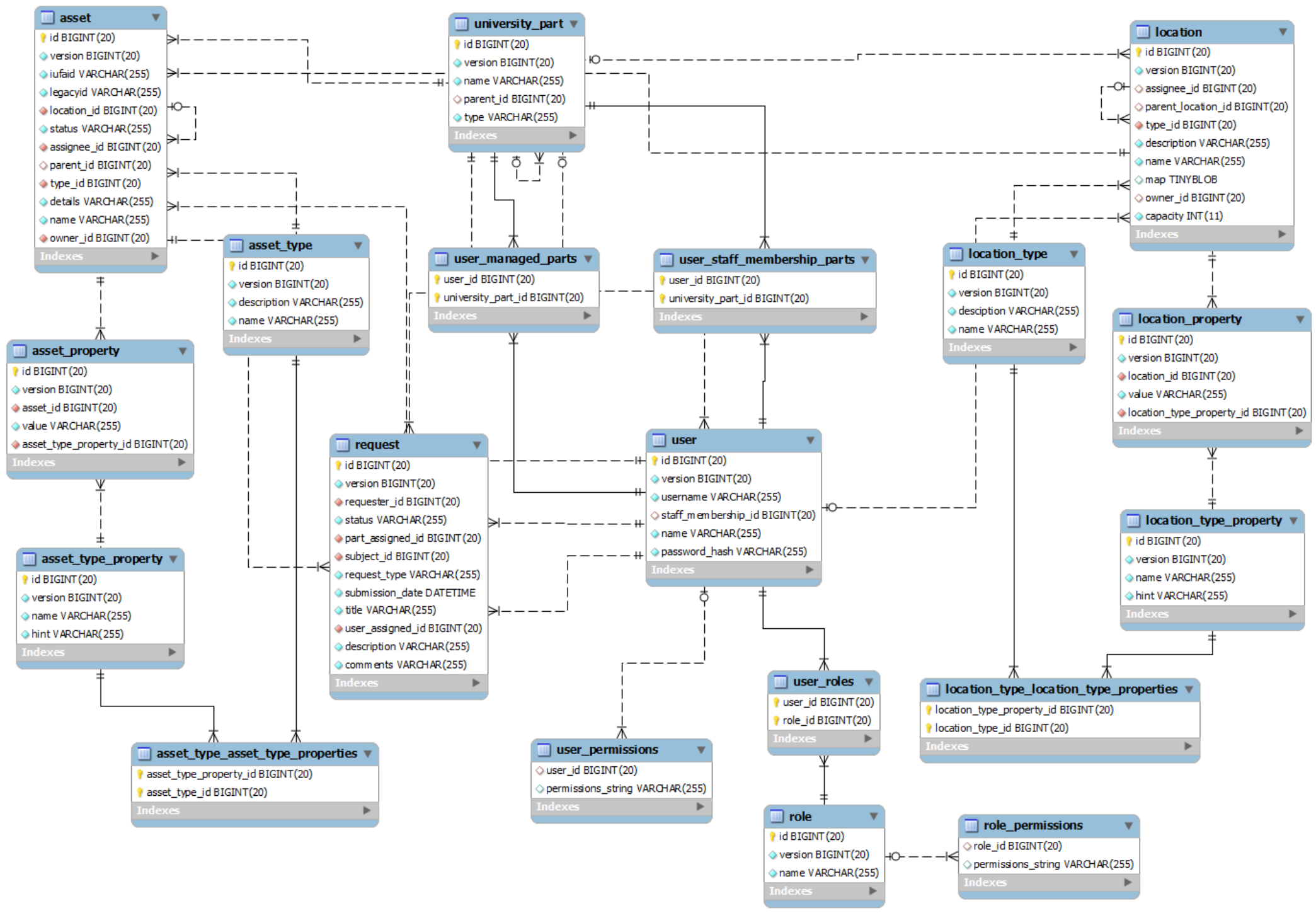

Entity Diagram of UUIS *Figure 5.1*

# 6 Class Diagram

Grails controllers, domain classes and services have dynamic methods and properties that won't be depicted on the previous class diagrams in order to keep them clear. For example, controller classes provide methods such as render and redirect to deal with the application flow, and domain classes provide methods such as find, get, save and delete to deal with persistence. [11]

## Domain Classes

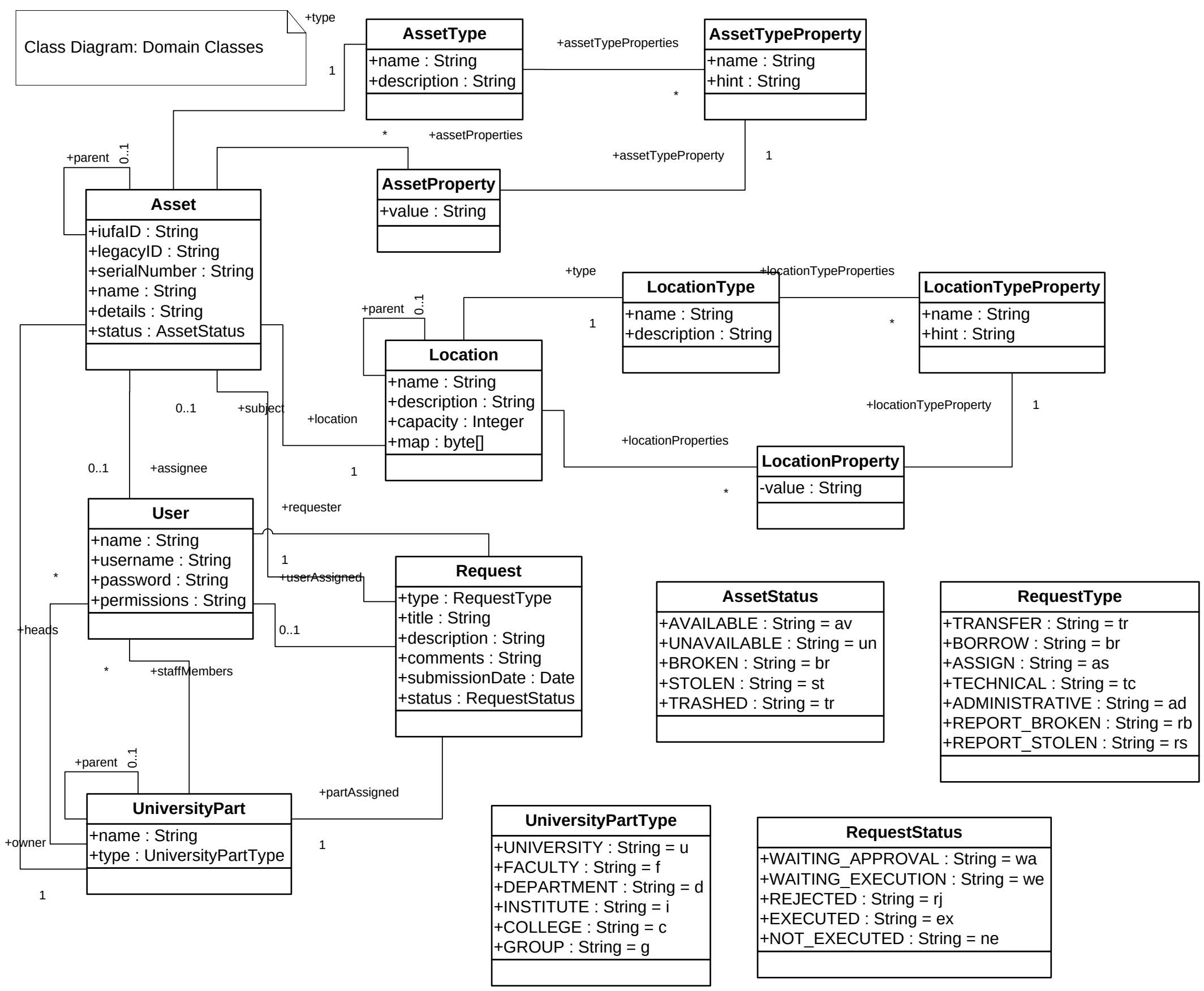

Domain Classes *Figure 6.1*

# Controllers

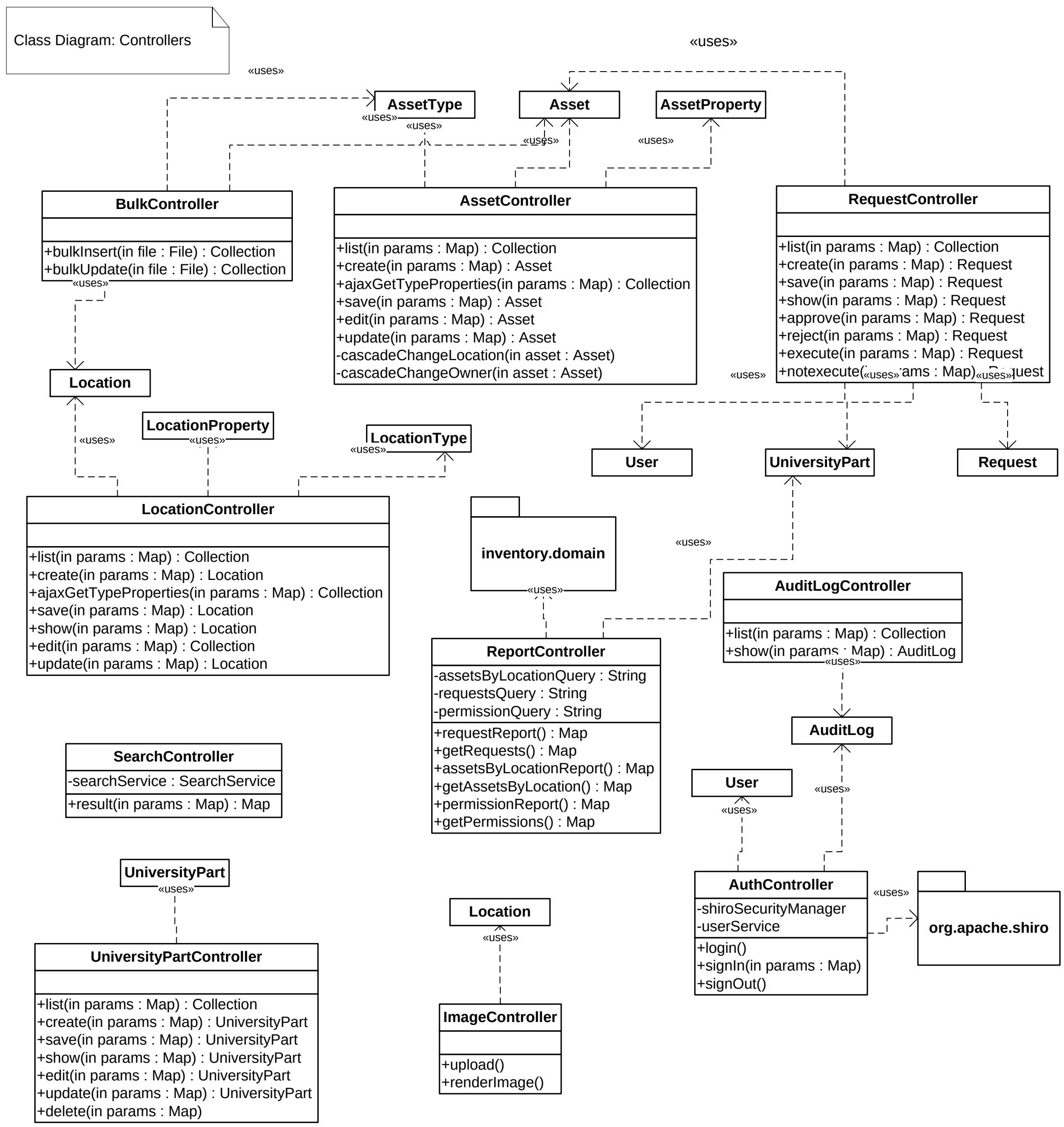

Controllers *Figure 6.2*

The controllers actions shown as methods on the previous diagram actually need to be implemented as closures. [12]

# Services

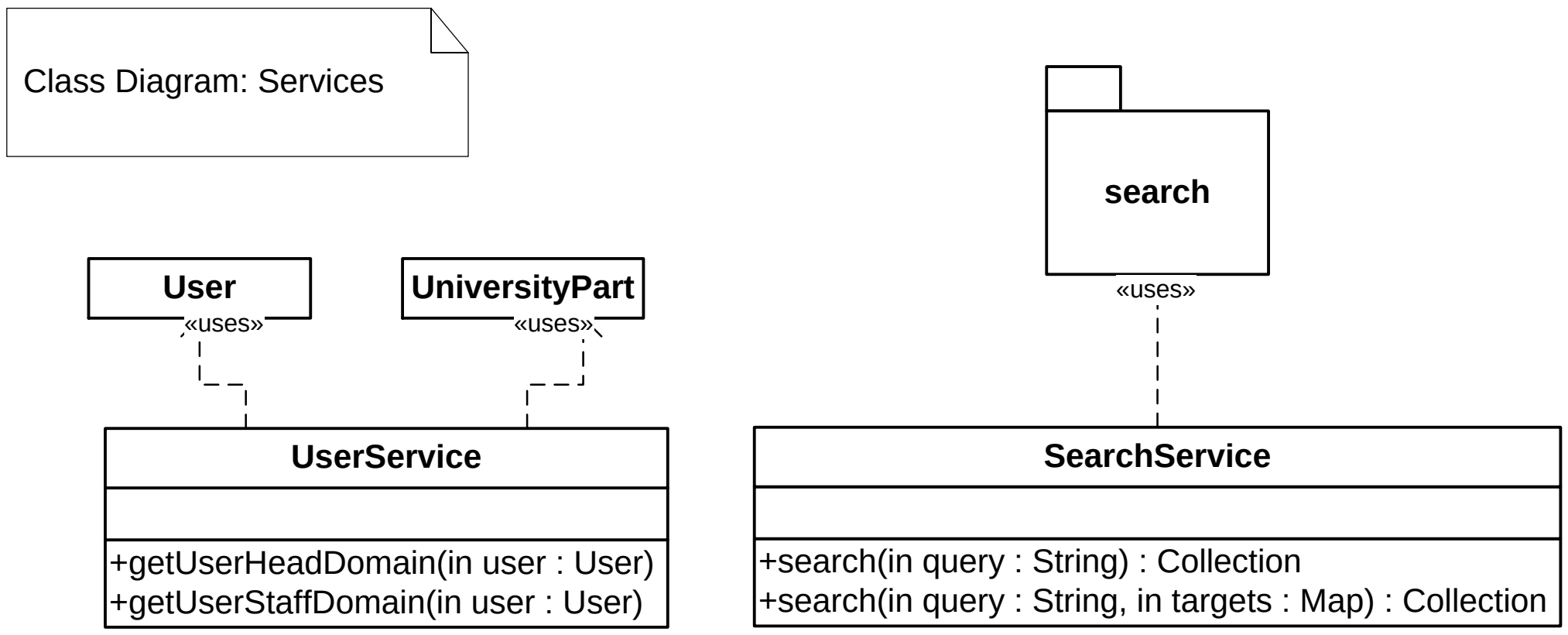

Services *Figure 6.3*

# 7 Sequence Diagrams

## Login

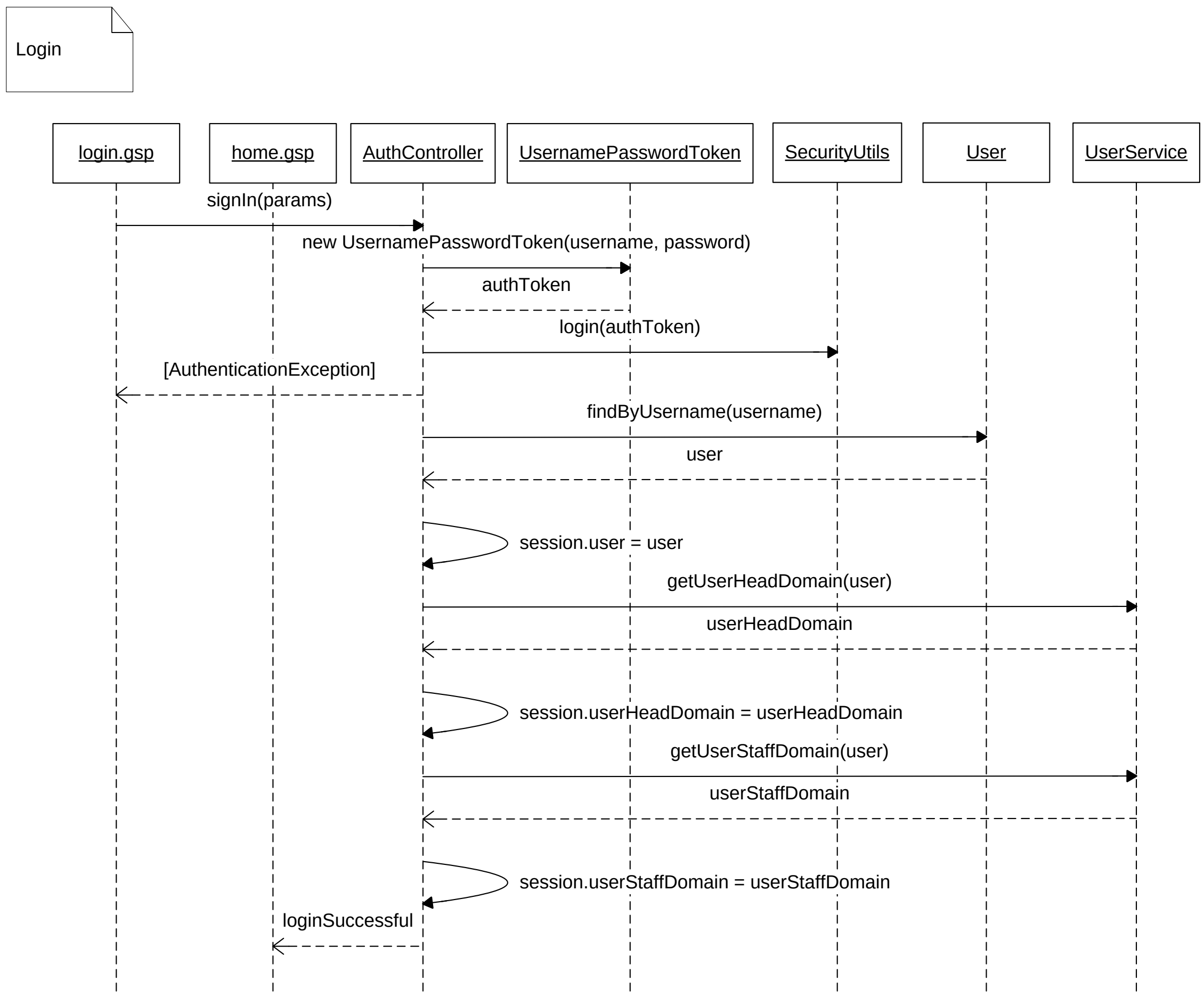

Sequence Diagram: Login *Figure 7.1*

# Logout

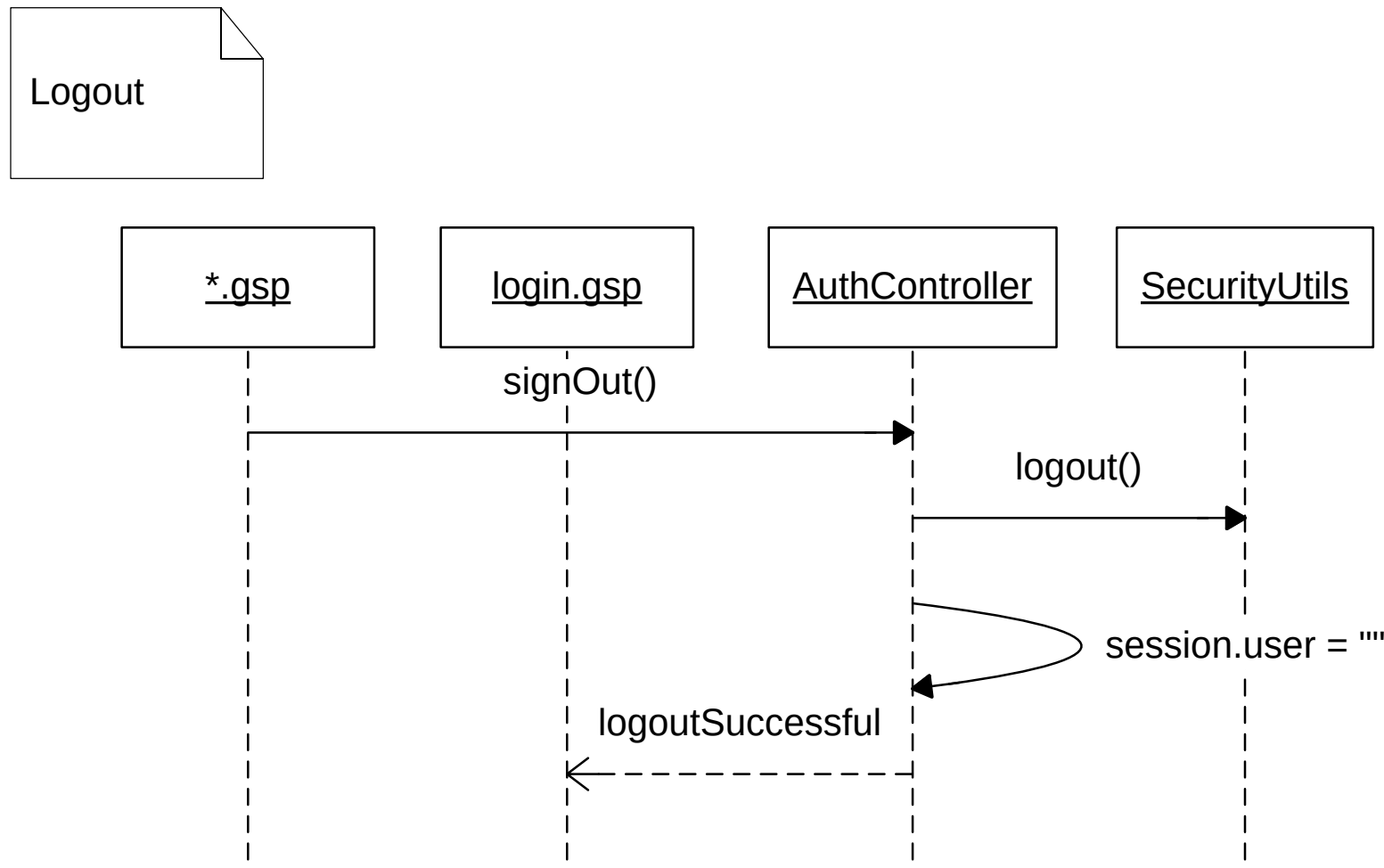

Sequence Diagram: Logout *Figure 7.2*

# List Asset

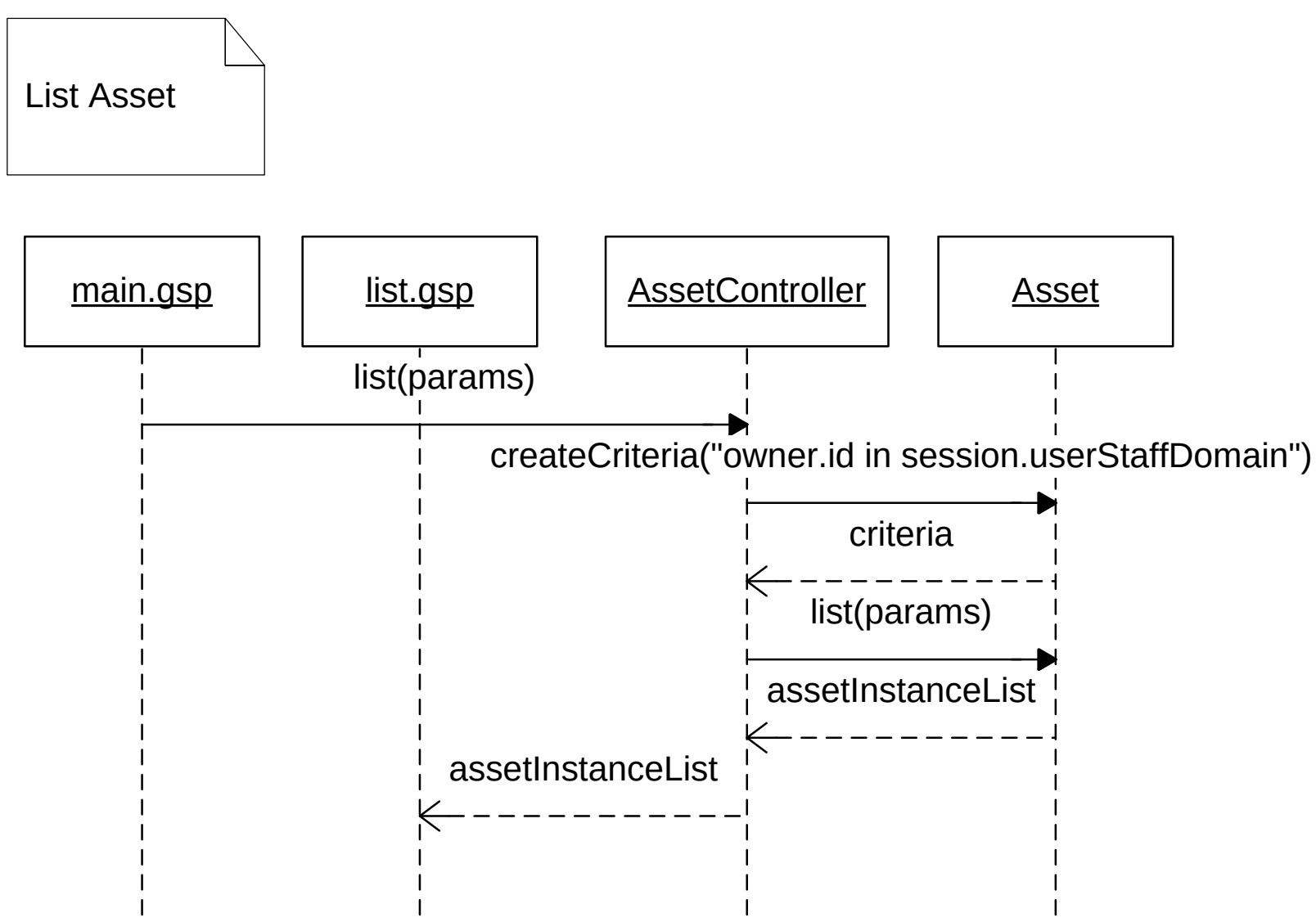

Sequence Diagram: List Asset *Figure 7.3*

# Create Asset

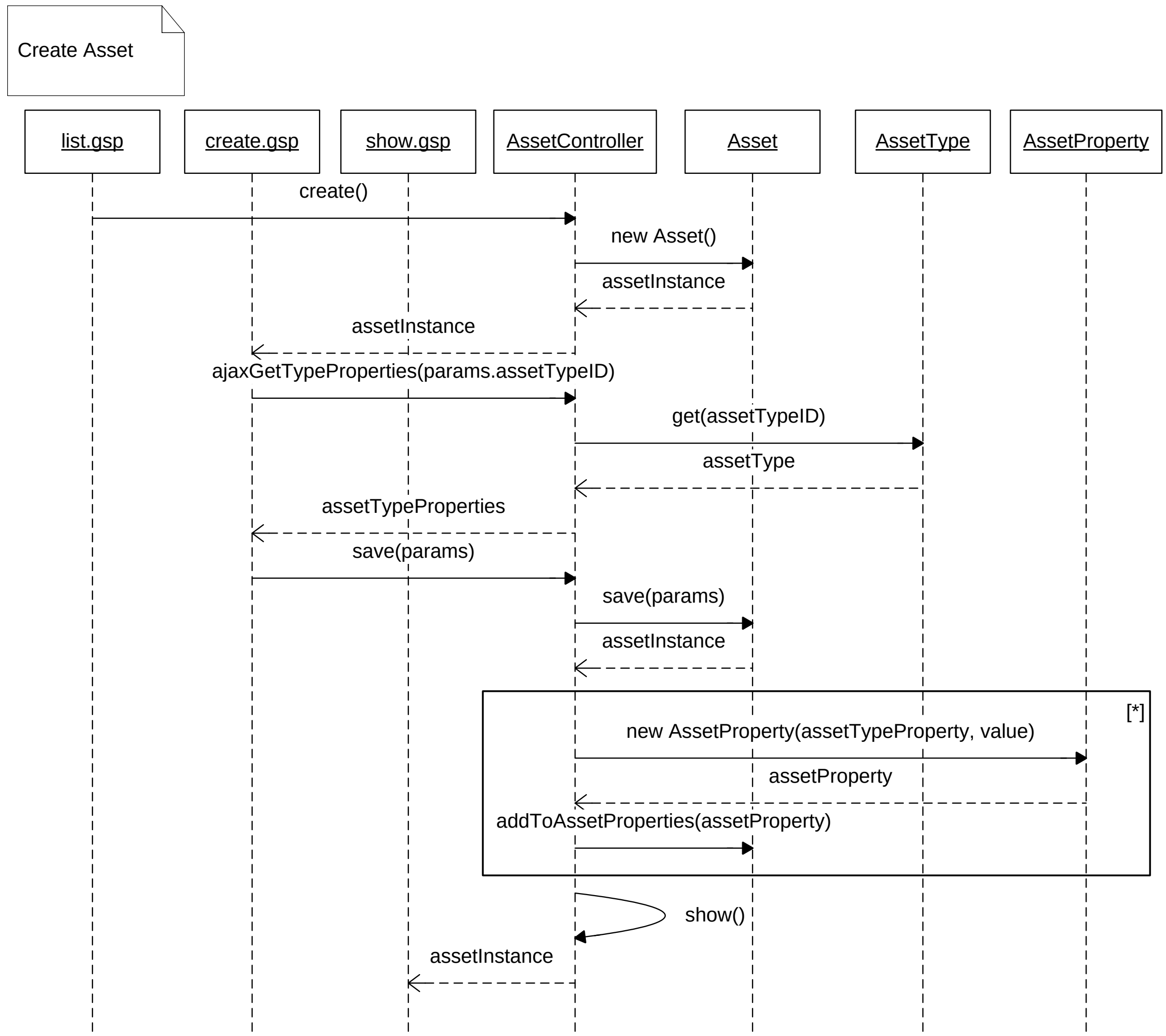

Sequence Diagram: Create Asset *Figure 7.4*

# Edit Asset

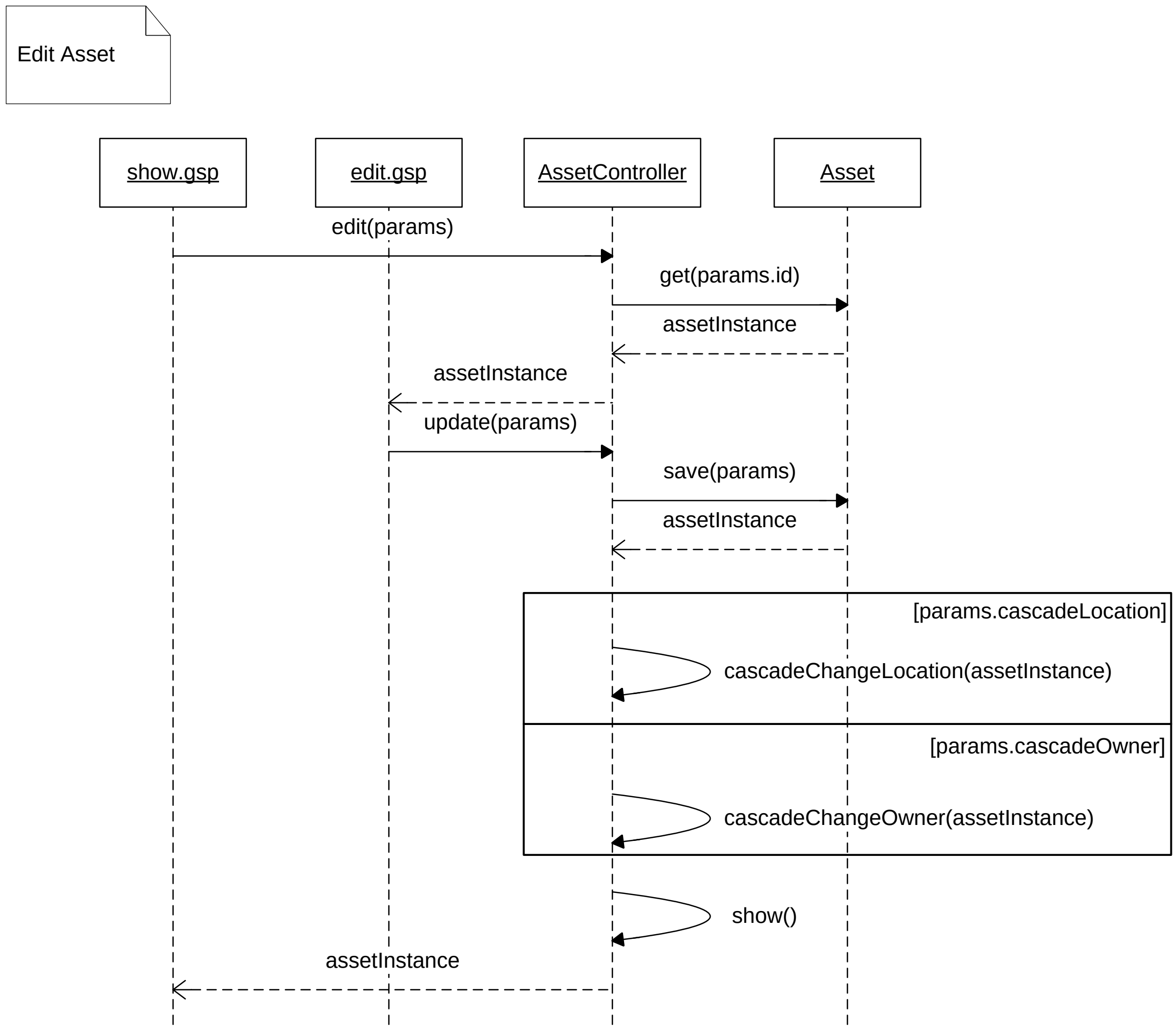

Sequence Diagram: Edit Asset *Figure 7.5*

# Show Asset

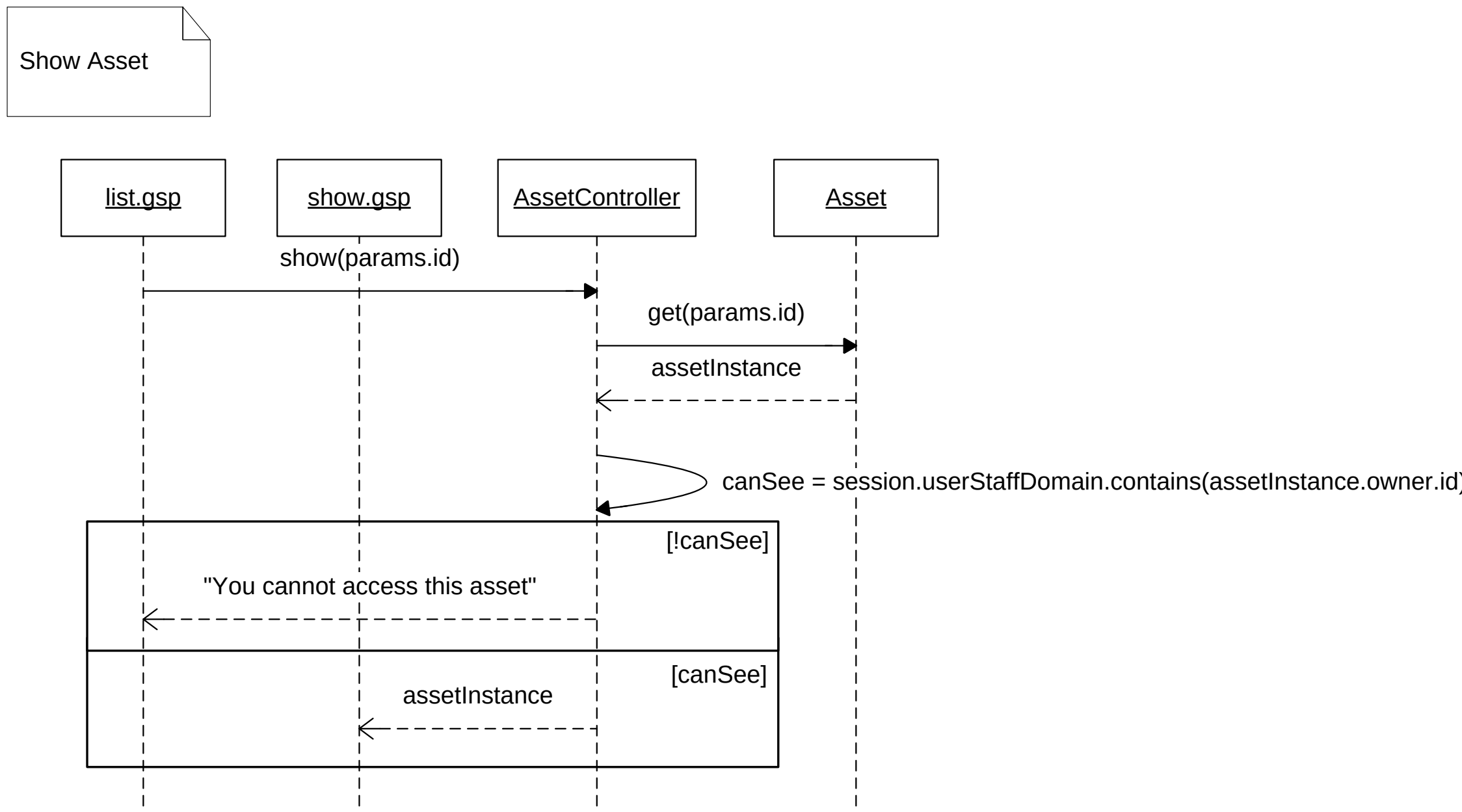

Sequence Diagram: Show Asset *Figure 7.6*

# List Location

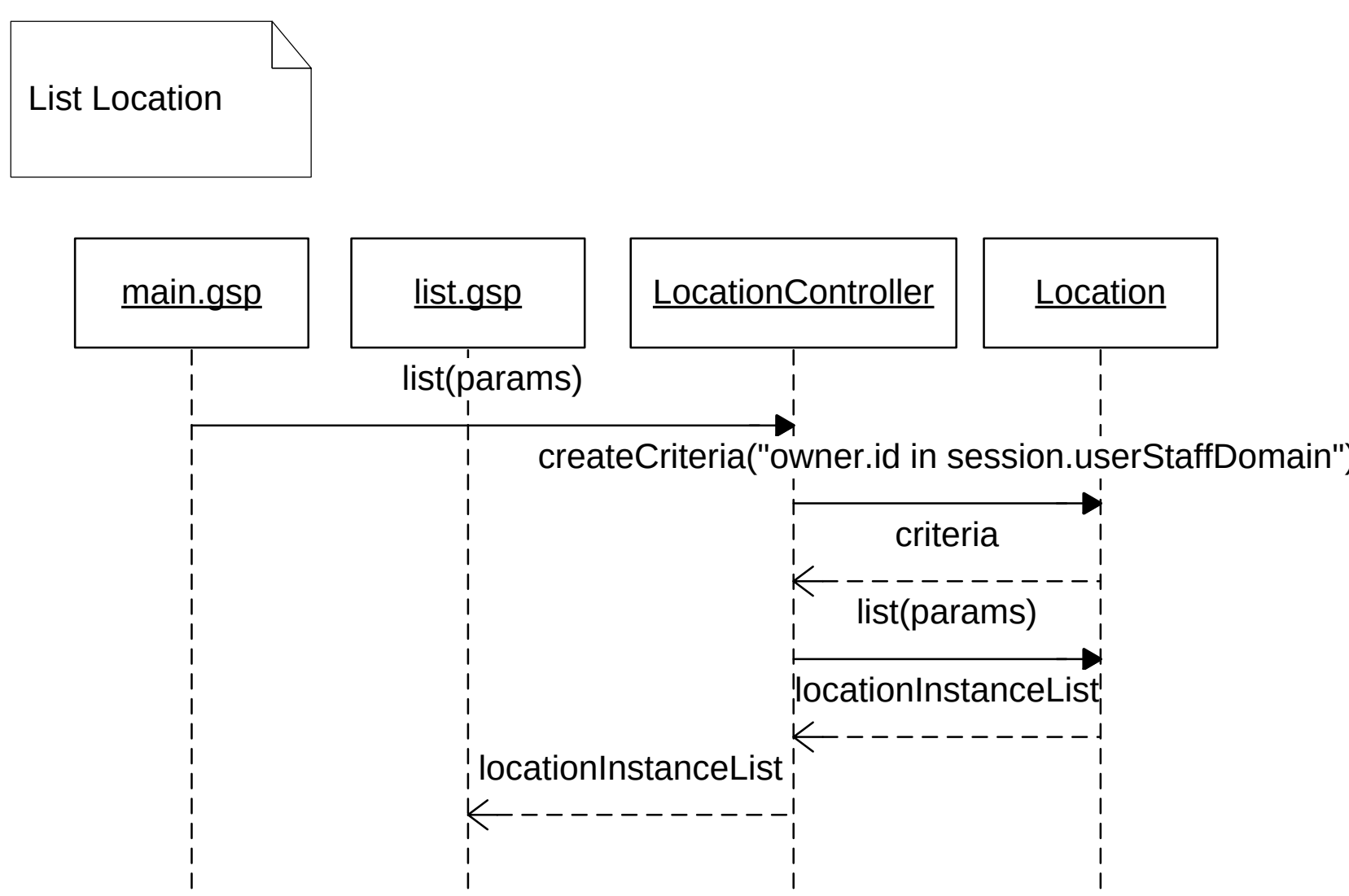

Sequence Diagram: List Location *Figure 7.7*

# Create Location

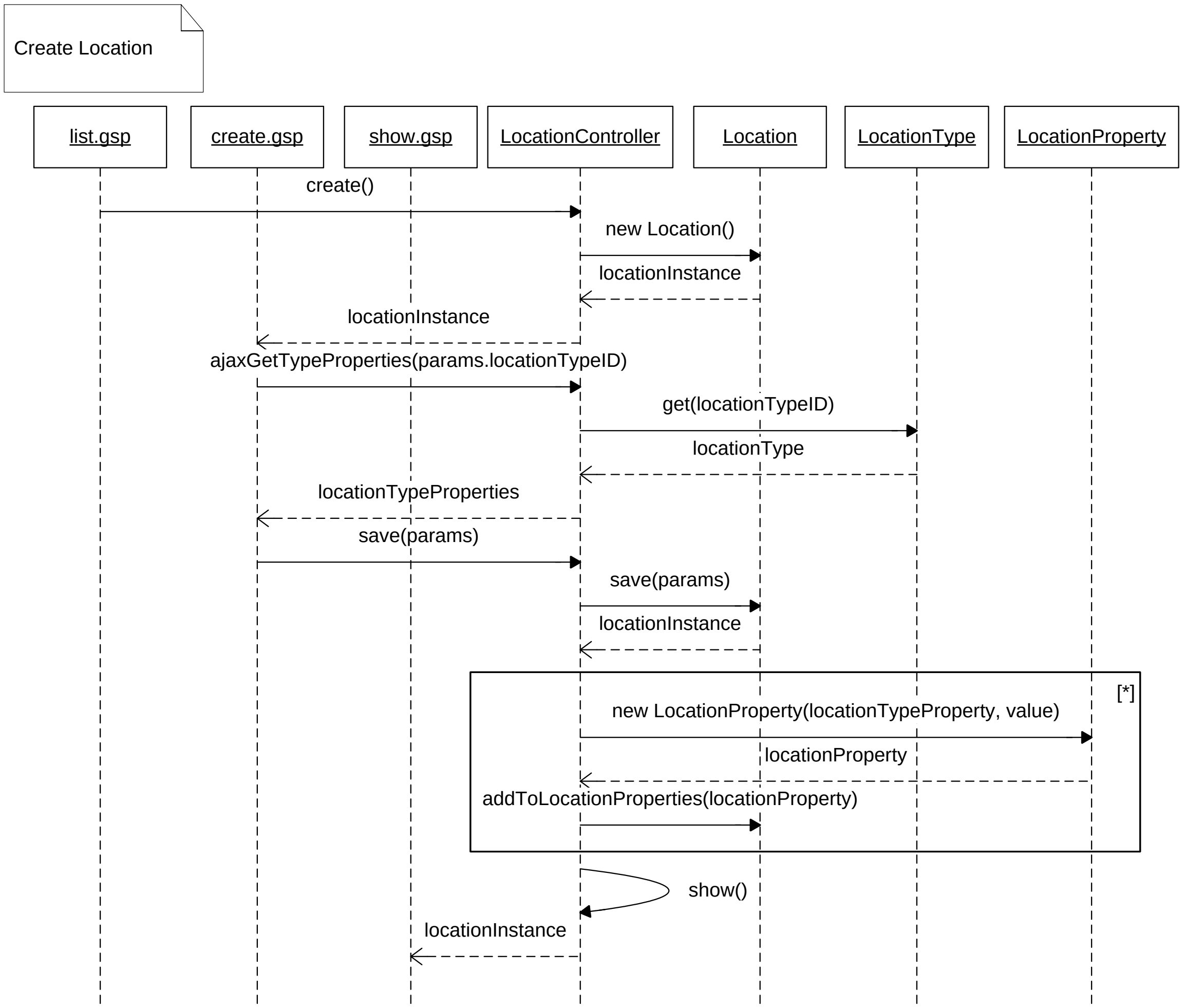

Sequence Diagram: Create Location *Figure 7.8*

# Edit Location

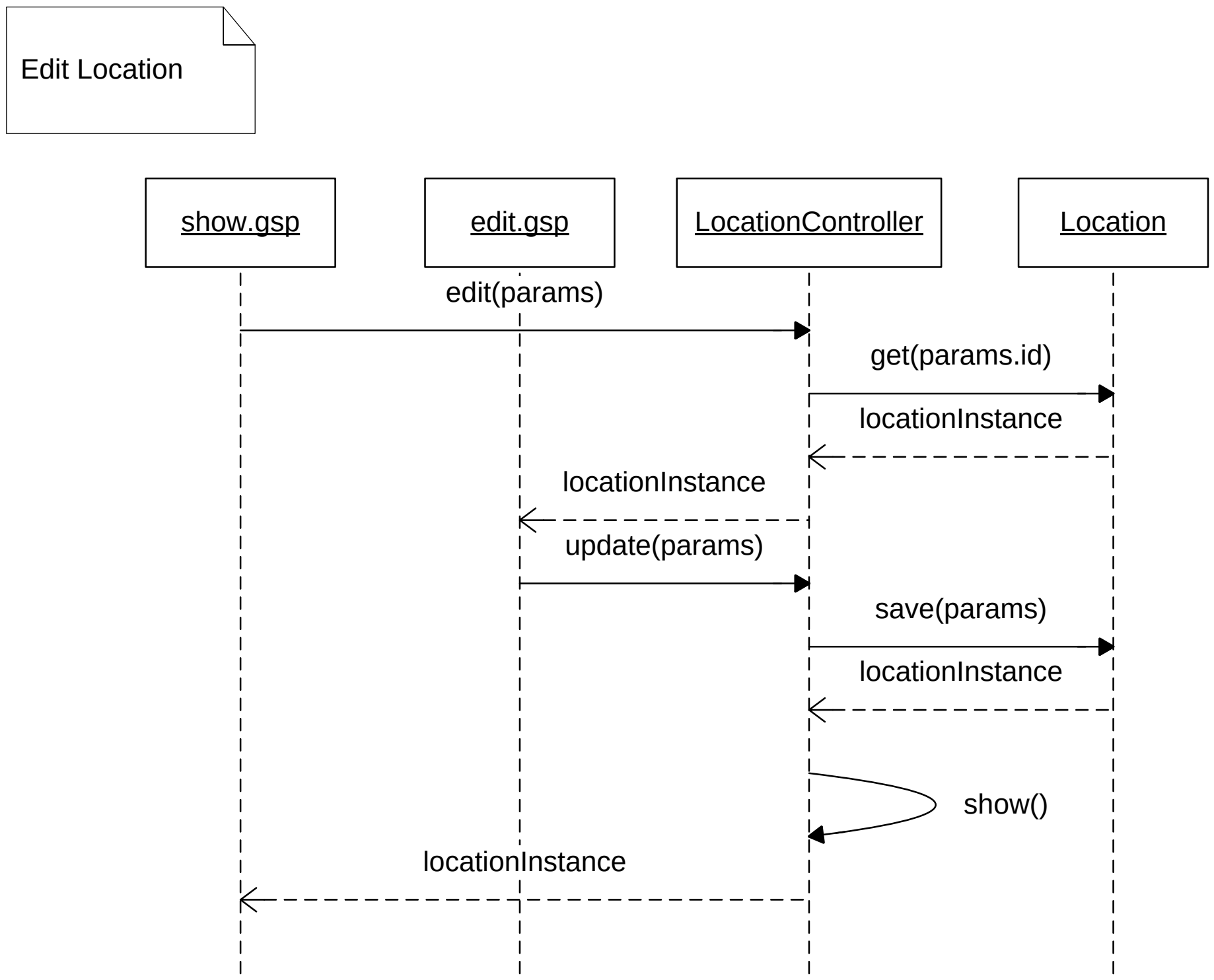

Sequence Diagram: Edit Location *Figure 7.9*

# Show Location

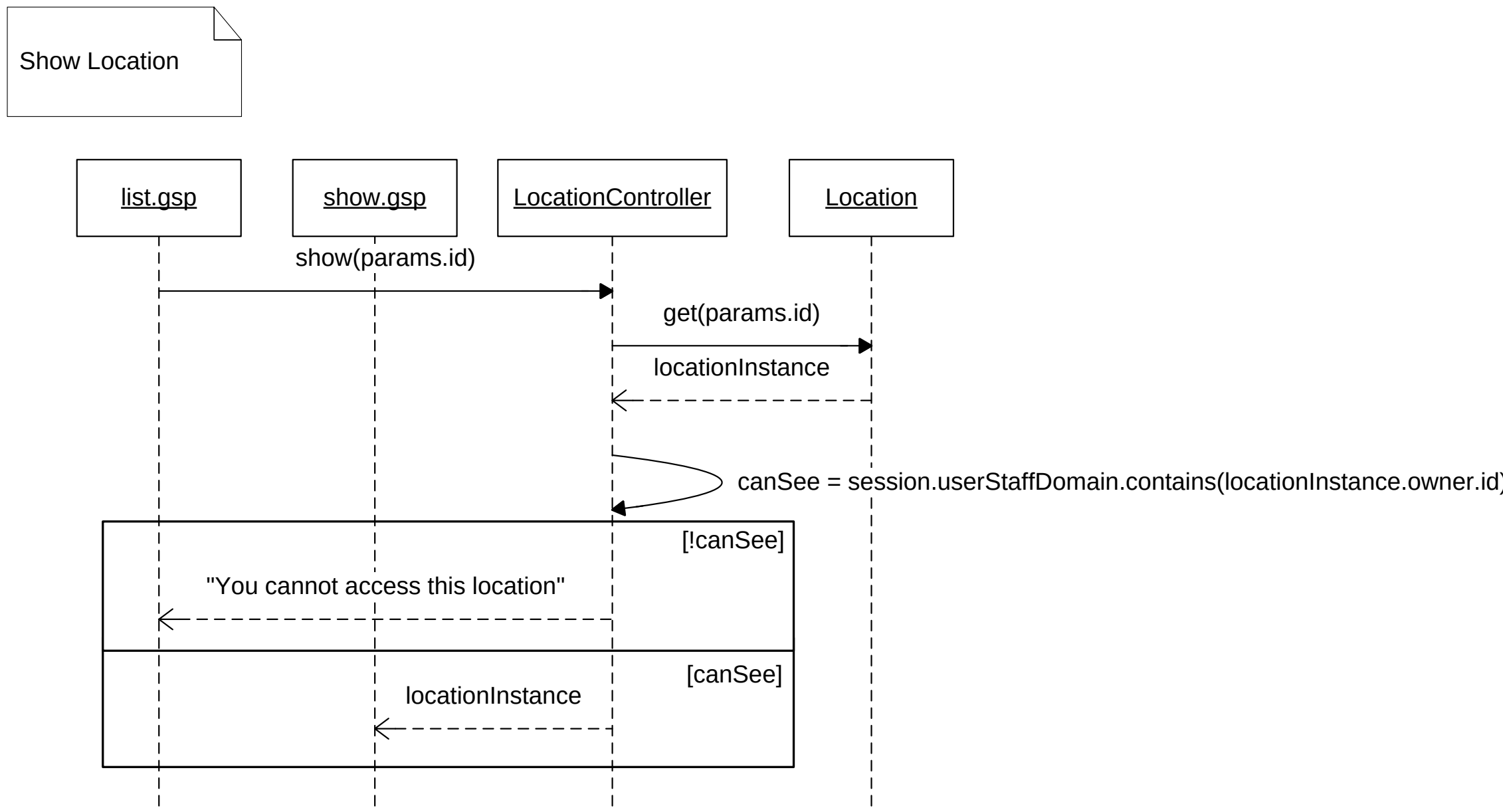

Sequence Diagram: Show Location *Figure 7.10*

# List Request

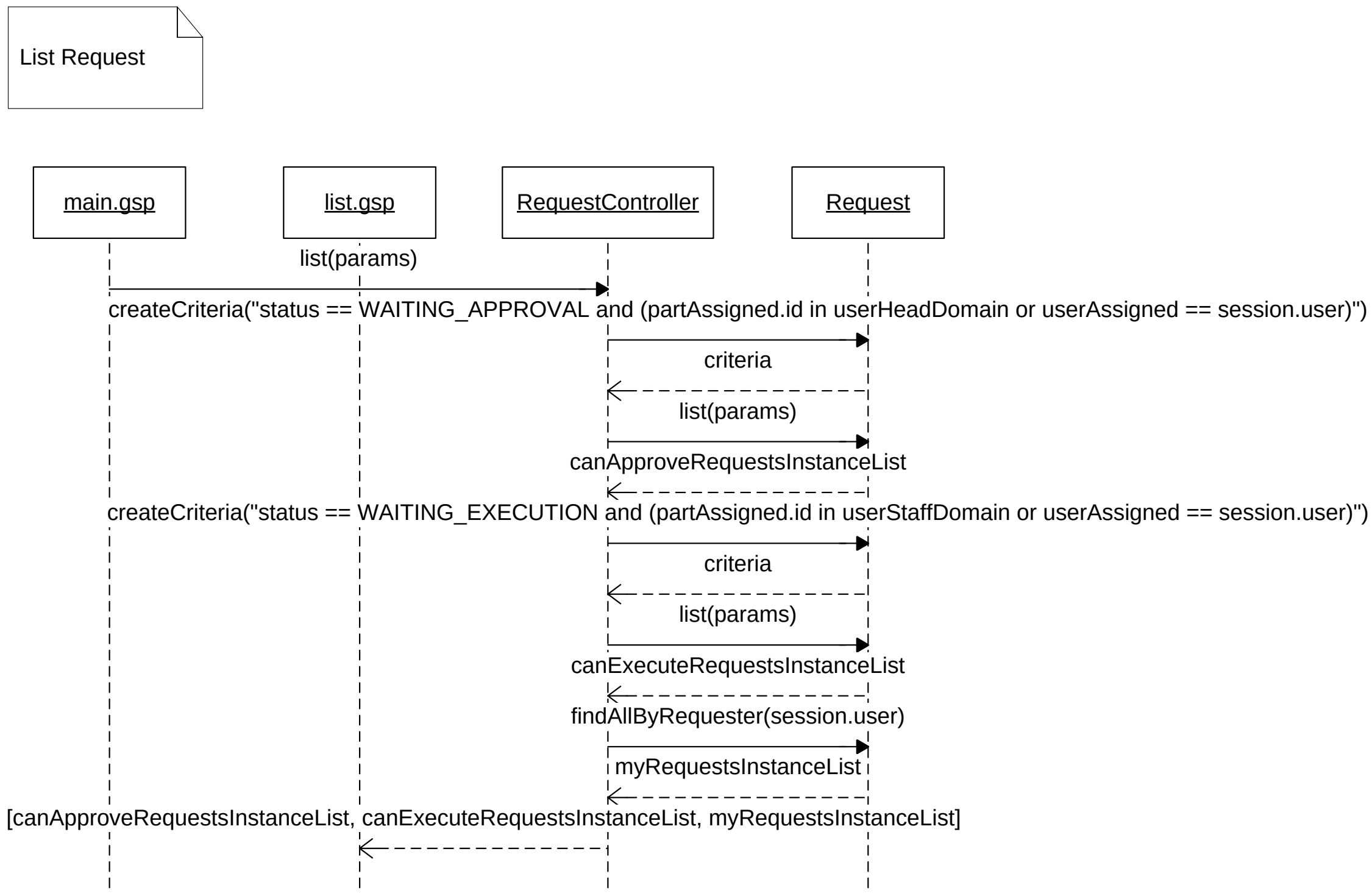

Sequence Diagram: List Request *Figure 7.11*

# Create Request

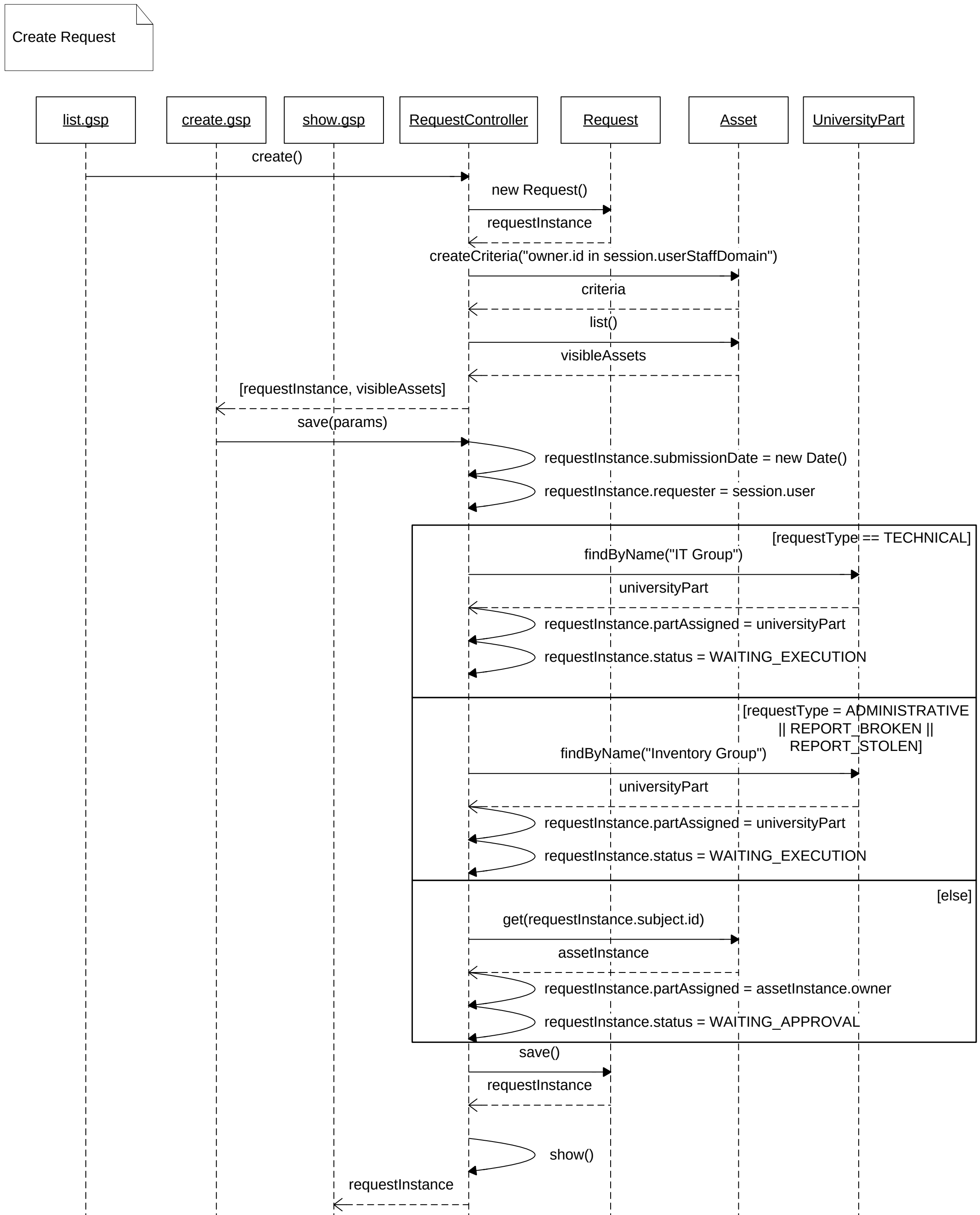

Sequence Diagram: Create Request *Figure 7.12*

## Show Request

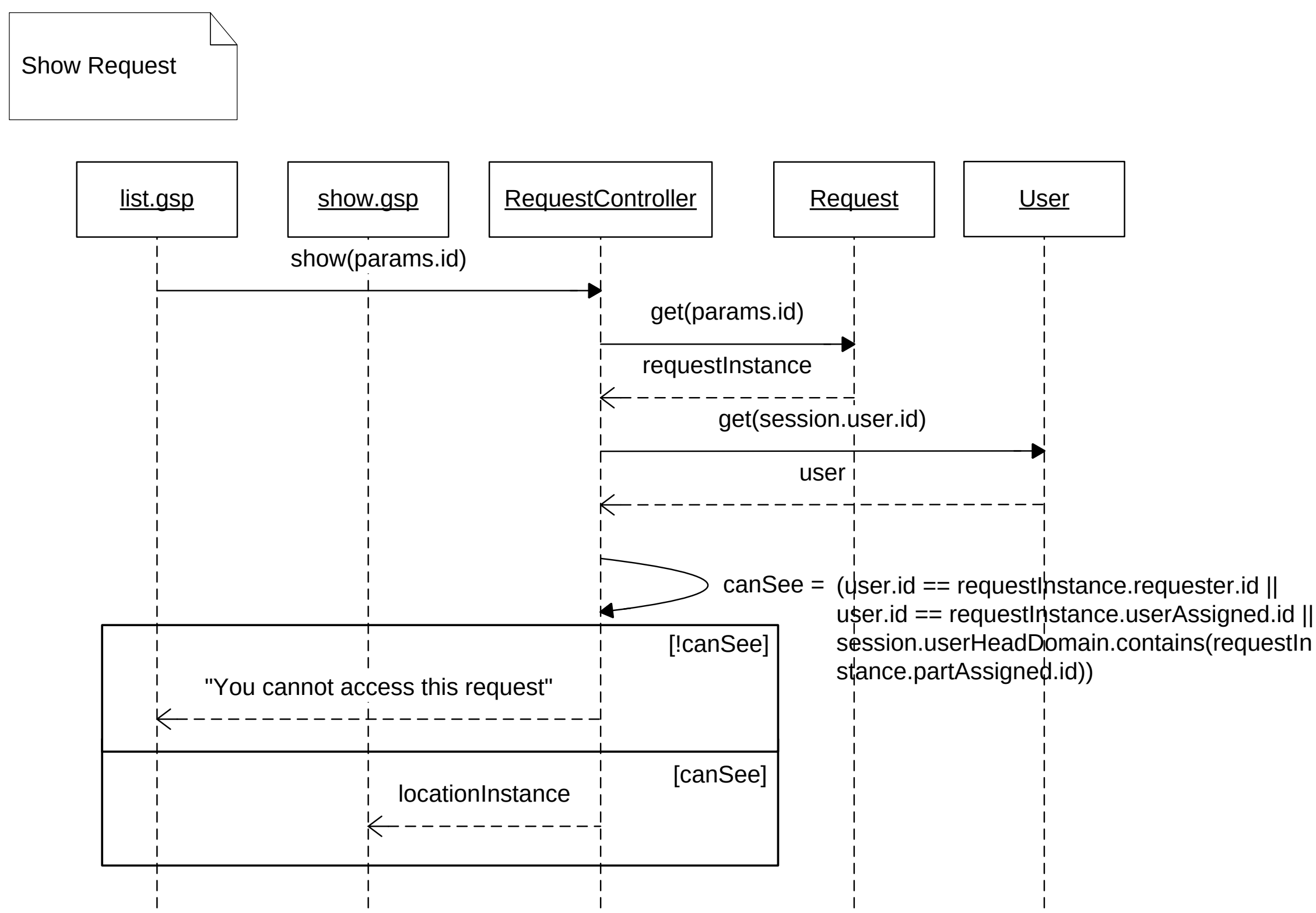

Sequence Diagram: Show Request *Figure 7.13*

## Approve Request

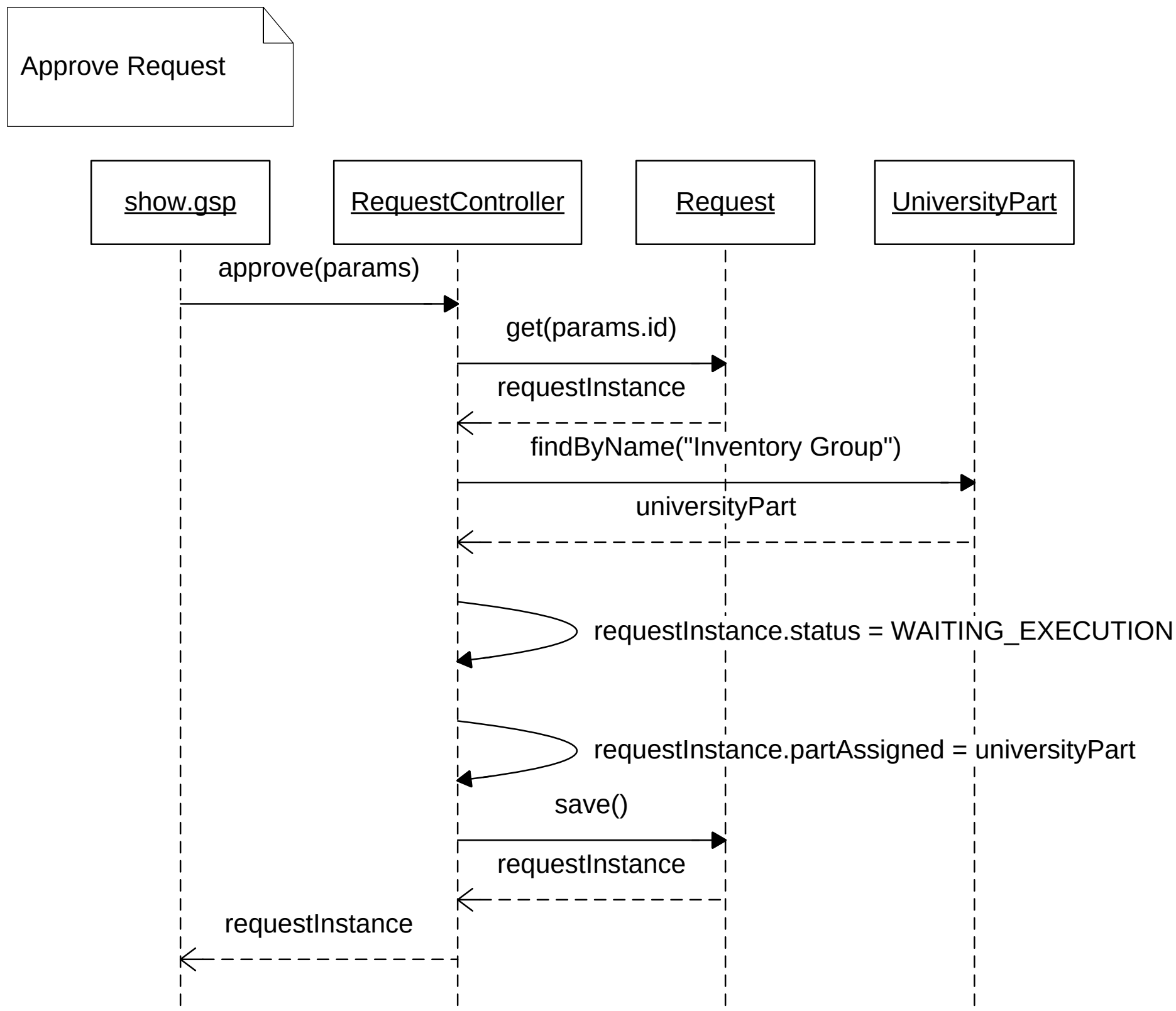

Sequence Diagram: Approve Request *Figure 7.14*

# Reject Request

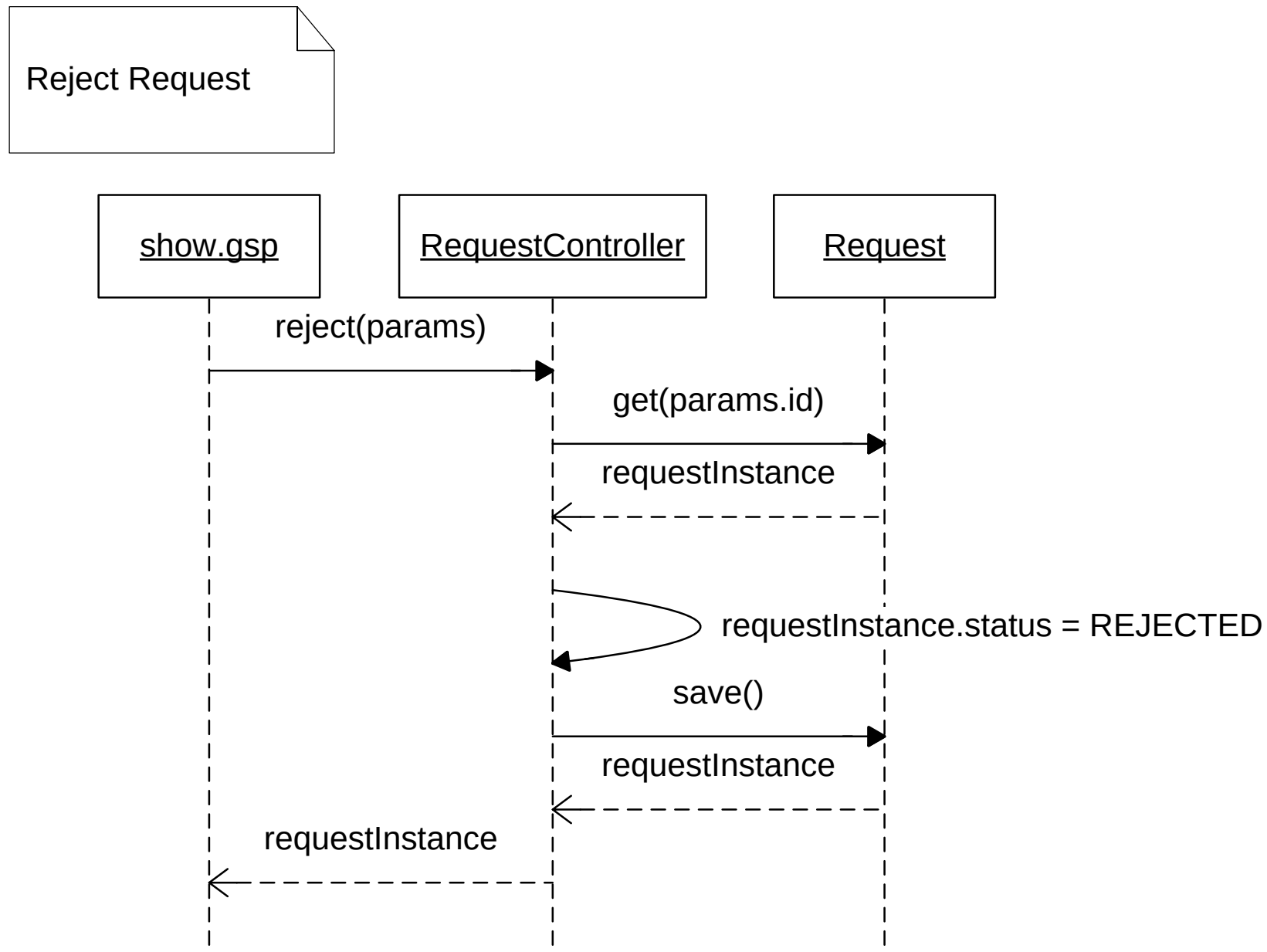

Sequence Diagram: Reject Request *Figure 7.15*

# Execute Request

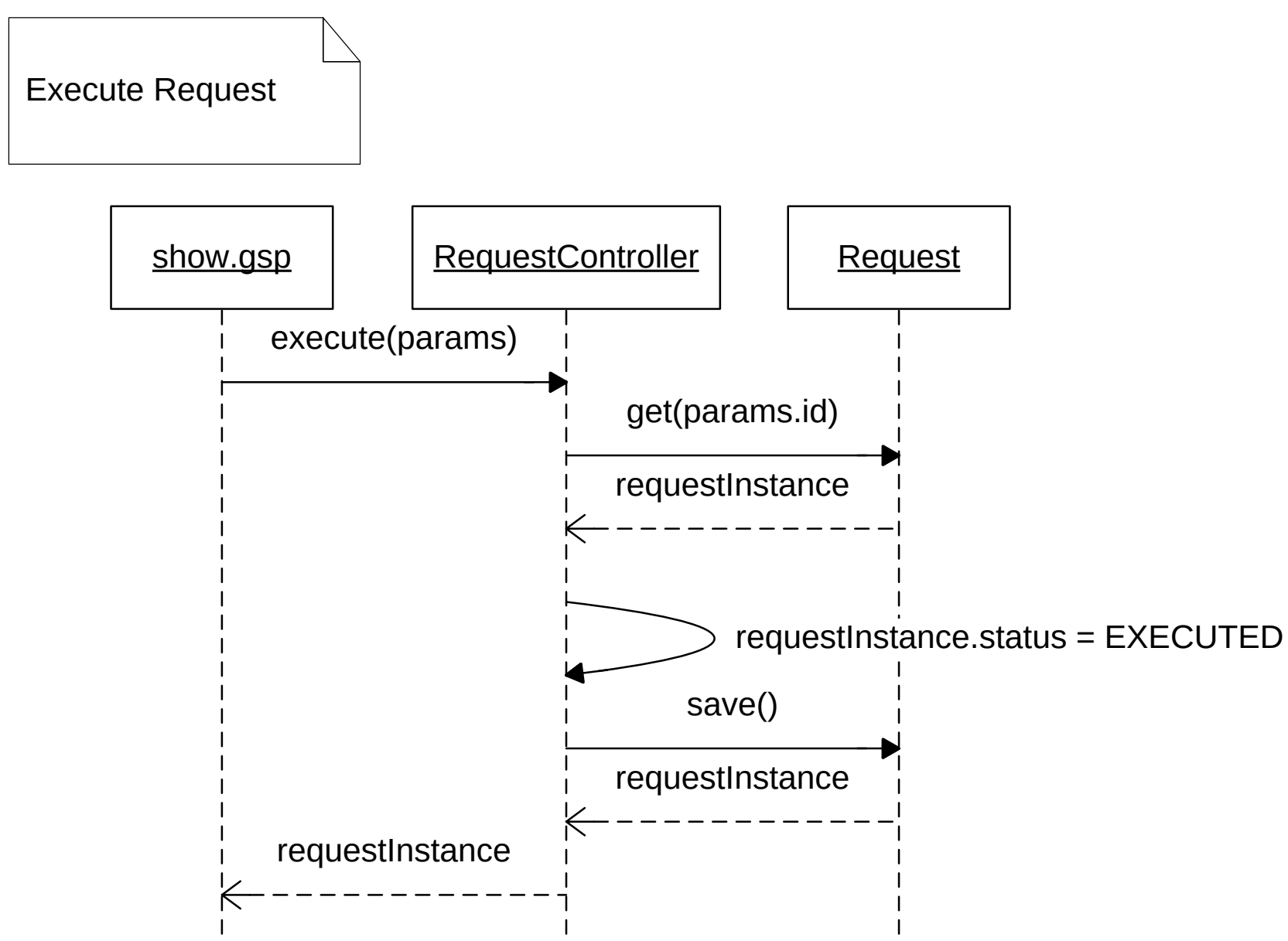

Sequence Diagram: Execute Request *Figure 7.16*

## Not Execute Request

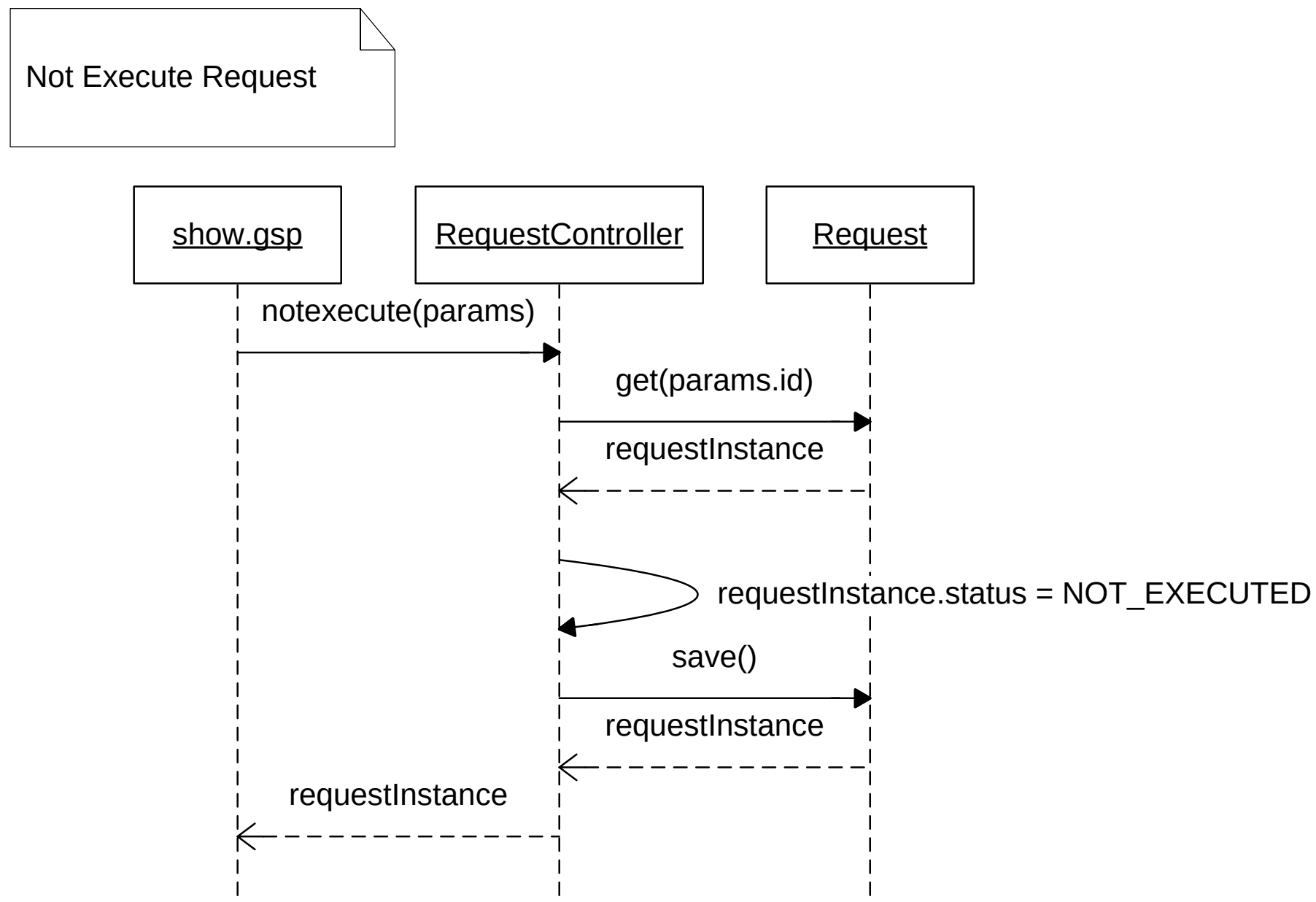

Sequence Diagram: Not Execute Request *Figure 7.17*

# Bulk Insert

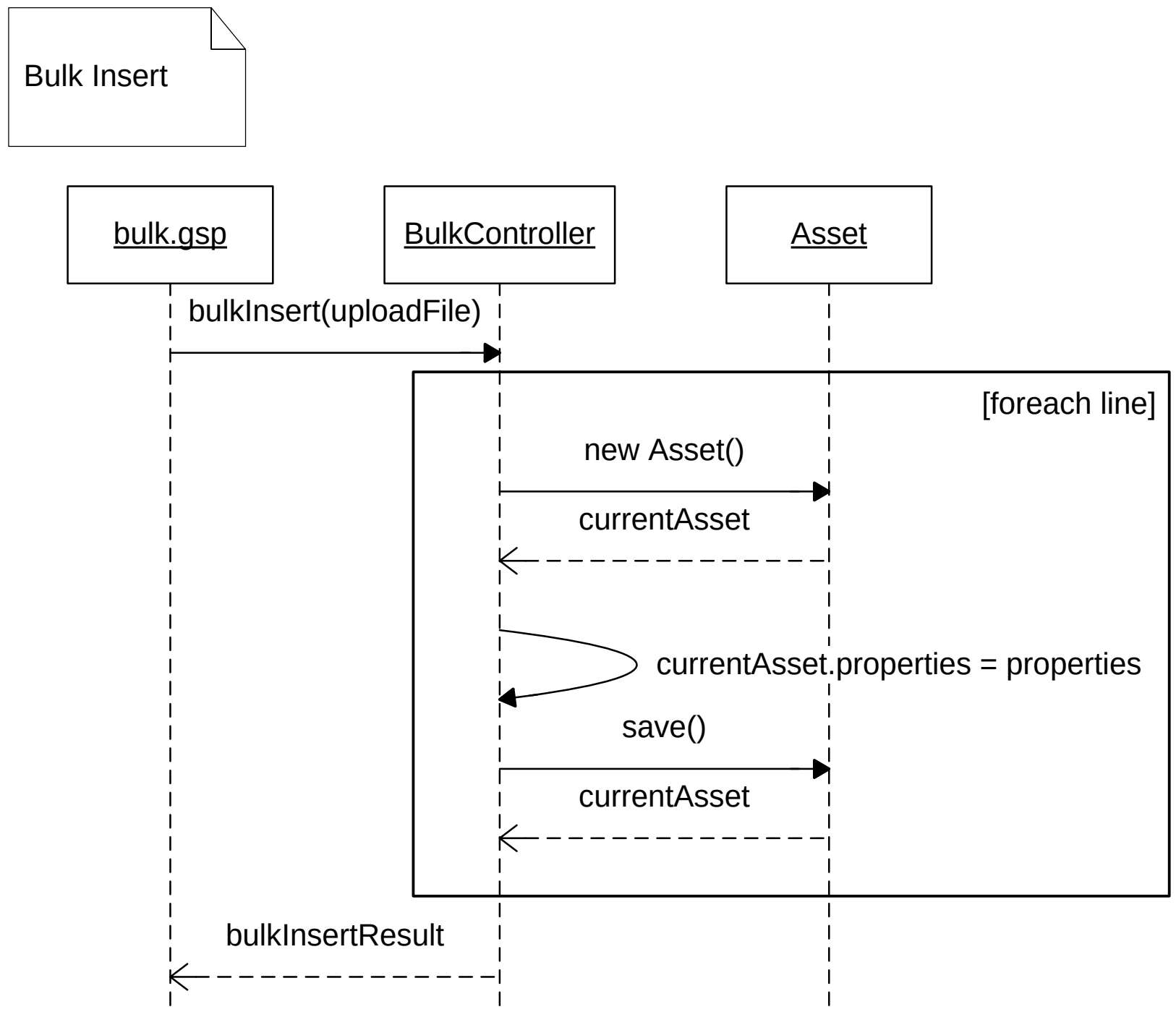

Sequence Diagram: Bulk Insert *Figure 7.18*

# Bulk Update

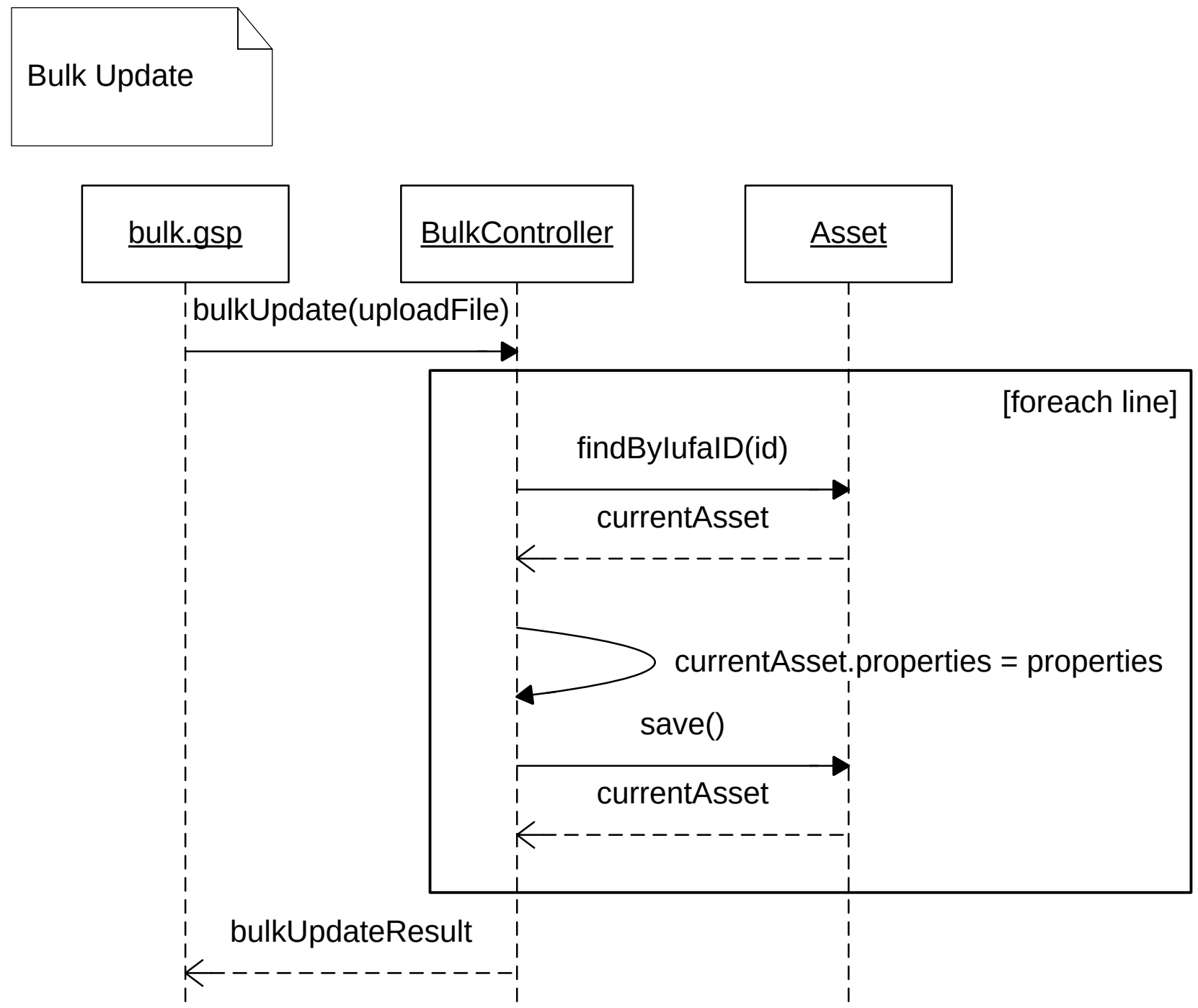

Sequence Diagram: Bulk Update *Figure 7.19*

# Basic Search

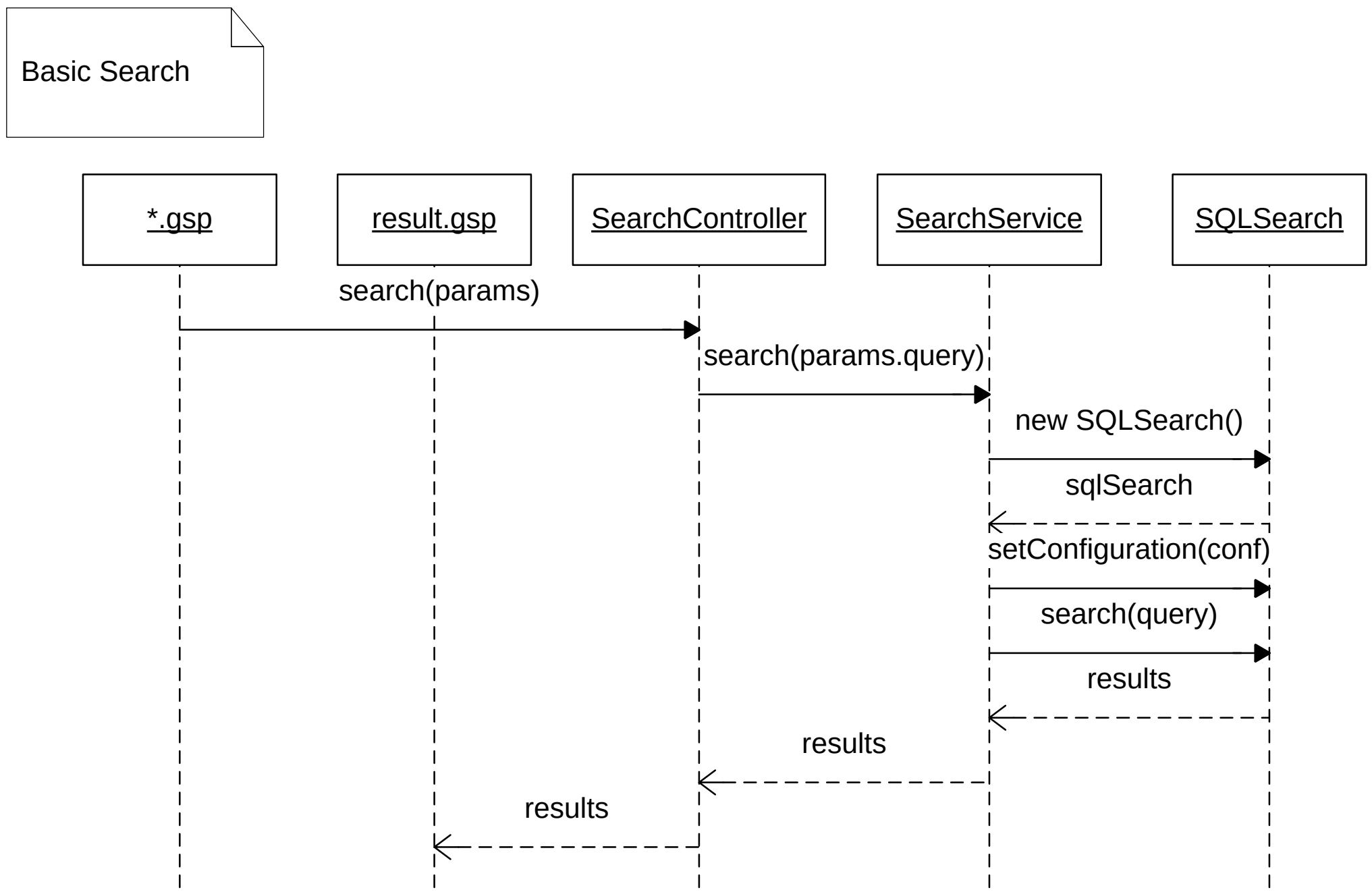

Sequence Diagram: Basic Search *Figure 7.20*

# Advanced Search

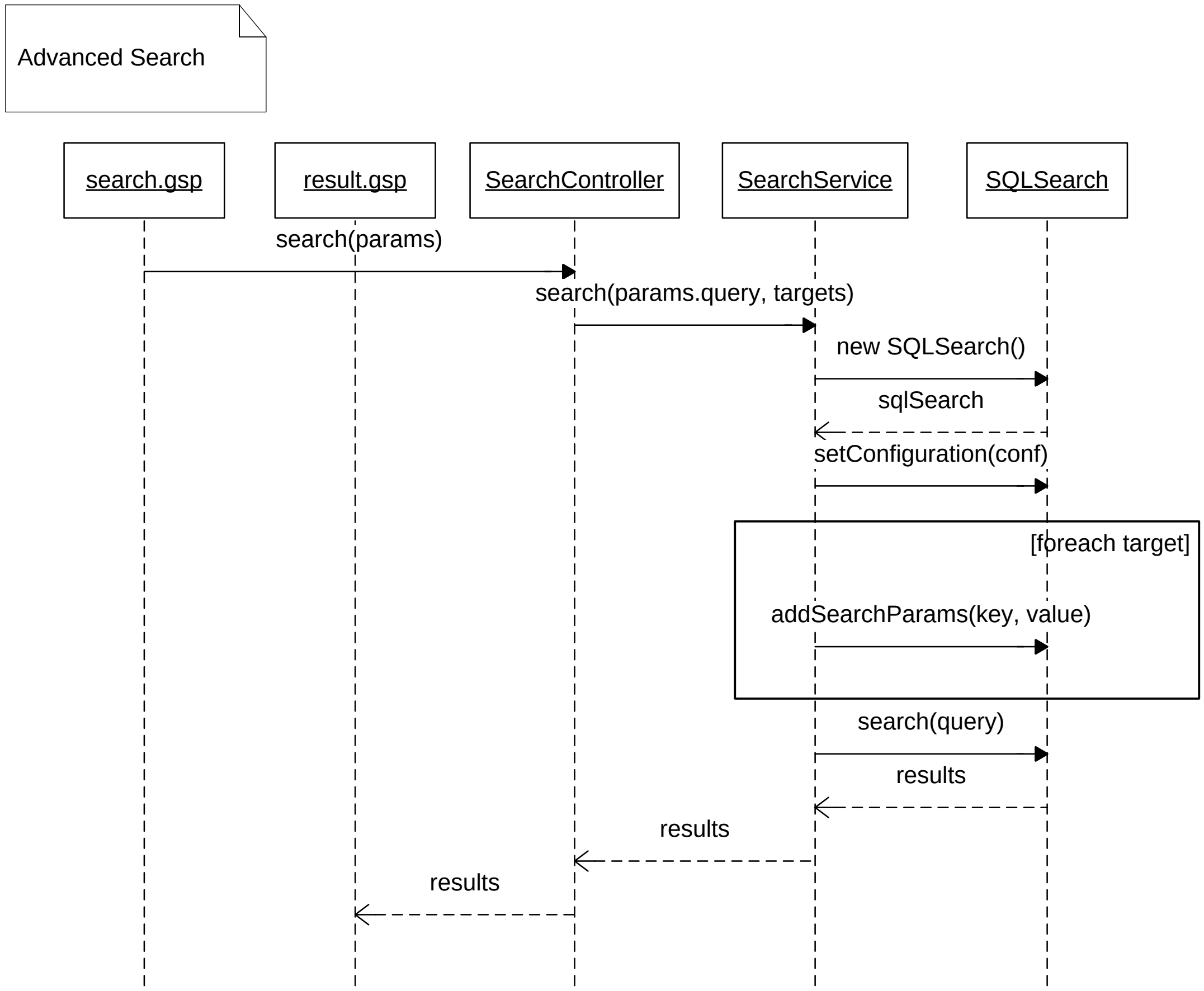

Sequence Diagram: Advanced Search *Figure 7.21*

# 8 Data Dictionary

| Table | Column | Description | Is Required | Maximum Length | Type | Key |
|---|---|---|---|---|---|---|
| user_staff_membership_parts | university_part_id | University/Faculty/Department ID | NO | | bigint(20) | PRI |
| | user_id | User ID | NO | | bigint(20) | PRI |
| user_roles | role_id | Role ID | NO | | bigint(20) | PRI |
| | user_id | User ID | NO | | bigint(20) | PRI |
| user_permissions | user_id | User ID | YES | | bigint(20) | MUL |
| | permissions_string | Action Permissions | YES | 255 | varchar(255) | |
| user_managed_parts | university_part_id | University/Faculty/Department ID | NO | | bigint(20) | PRI |
| | user_id | User ID | NO | | bigint(20) | PRI |
| user | id | User ID | NO | | bigint(20) | PRI |
| | version | Changed veresion | NO | | bigint(20) | |
| | username | User Login Name | NO | 255 | varchar(255) | UNI |
| | name | User First/Lasst Name | NO | 255 | varchar(255) | |
| | password_hash | Encripted Password | NO | 255 | varchar(255) | |
| university_part | id | University/Faculty/Department ID | NO | | bigint(20) | PRI |
| | version | Changed veresion | NO | | bigint(20) | |
| | name | University/Faculty/Department Name | NO | 255 | varchar(255) | |
| | parent_id | University/Faculty/Department Belongs to | YES | | bigint(20) | MUL |
| | type | University/Faculty/Department Type | NO | 255 | varchar(255) | |
| role_permissions | role_id | Role ID | YES | | bigint(20) | MUL |
| | permissions_string | Action Permissions | YES | 255 | varchar(255) | |
| role | id | Role ID | NO | | bigint(20) | PRI |
| | version | Changed veresion | NO | | bigint(20) | |
| | name | Role Name | NO | 255 | varchar(255) | UNI |
| request | id | Request ID | NO | | bigint(20) | PRI |
| | version | Changed veresion | NO | | bigint(20) | |
| | requester_id | Submitted User ID | NO | | bigint(20) | MUL |
| | status | Request Status | NO | 255 | varchar(255) | |
| | part_assigned_id | Assigned to University/Faculty/Department ID | NO | | bigint(20) | MUL |
| | subject_id | Subject ID | YES | | bigint(20) | MUL |
| | request_type | Requeset Type ID | NO | 255 | varchar(255) | |
| | submission_date | Submitted Date | NO | | datetime | |
| | title | Title | NO | 255 | varchar(255) | |
| | user_assigned_id | Assigned to User ID | YES | | bigint(20) | MUL |
| | description | Description | YES | 255 | varchar(255) | |
| | comments | Comments | YES | 255 | varchar(255) | |
| location_type_property | id | Location Type ID | NO | | bigint(20) | PRI |
| | version | Changed veresion | NO | | bigint(20) | |
| | name | Location Type Name | NO | 255 | varchar(255) | UNI |
| | hint | Location Type Hint | YES | 255 | varchar(255) | |
| location_type_location_type_properties | location_type_id | Location Type ID | NO | | bigint(20) | PRI |
| | location_type_property_id | Location Type - Property ID | NO | | bigint(20) | PRI |
| location_type | id | Location Type ID | NO | | bigint(20) | PRI |
| | version | Changed veresion | NO | | bigint(20) | |
| | description | Location Type Description | YES | 255 | varchar(255) | |
| | name | Location Type Name | NO | 255 | varchar(255) | UNI |
| location_property | id | Property ID | NO | | bigint(20) | PRI |
| | version | Changed veresion | NO | | bigint(20) | |
| | location_id | Location ID | NO | | bigint(20) | MUL |
| | value | Property Value | NO | 255 | varchar(255) | |
| | location_type_property_id | Location Type - Property ID | NO | | bigint(20) | MUL |
| location | id | Location ID | NO | | bigint(20) | PRI |
| | version | Changed Veresion | NO | | bigint(20) | |
| | assignee_id | Assiged to User ID | YES | | bigint(20) | MUL |
| | parent_location_id | Location Belongs to | YES | | bigint(20) | MUL |
| | type_id | Location Type | NO | | bigint(20) | MUL |
| | description | Location Description | YES | 255 | varchar(255) | |
| | name | Location Name | NO | 255 | varchar(255) | |
| | map | Location Map | YES | 4294967295 | longblob | |
| | owner_id | University/Faculty/Depamrtment ID | NO | | bigint(20) | MUL |
| | capacity | Capacity | NO | | int(11) | |
| audit_log | id | Audit Log ID | NO | | bigint(20) | PRI |
| | property_name | Property Name | YES | 255 | varchar(255) | |
| | last_updated | Last Updated | NO | | datetime | |
| | old_value | Old Value | YES | 255 | varchar(255) | |
| | actor | Actor | YES | 255 | varchar(255) | |
| | uri | URL | YES | 255 | varchar(255) | |
| | new_value | New Value | YES | 255 | varchar(255) | |
| | persisted_object_version | Persisted Object Version | YES | | bigint(20) | |
| | date_created | Date Created | NO | | datetime | |
| | class_name | Class Name | YES | 255 | varchar(255) | |
| | event_name | Event Name | YES | 255 | varchar(255) | |
| | persisted_object_id | Persisted Object ID | YES | | bigint(20) | |
| | version | Changed veresion | YES | | bigint(20) | |
| asset_type_property | id | Location Type - Property ID | NO | | bigint(20) | PRI |
| | version | Changed veresion | NO | | bigint(20) | |
| | name | Asset Type - Property Name | NO | 255 | varchar(255) | |
| | hint | Hint | NO | 255 | varchar(255) | |
| | asset_type_id | Asset Type ID | YES | | bigint(20) | MUL |
| asset_type_asset_type_properties | asset_type_property_id | Asset Type - Property ID | NO | | bigint(20) | PRI |
| | asset_type_id | Asset Type ID | NO | | bigint(20) | PRI |
| asset_type | id | Asset Type ID | NO | | bigint(20) | PRI |
| | version | Changed veresion | NO | | bigint(20) | |
| | description | Asset Type Description | YES | 255 | varchar(255) | |
| | name | Asset Type Name | NO | 255 | varchar(255) | |
| asset_property | id | Asset Property ID | NO | | bigint(20) | PRI |
| | version | Changed veresion | NO | | bigint(20) | |
| | asset_id | Asset ID | NO | | bigint(20) | MUL |
| | value | Asset Value | NO | 255 | varchar(255) | |
| | asset_type_property_id | Asset Type - Property ID | NO | | bigint(20) | MUL |
| asset | id | Asset ID | NO | | bigint(20) | PRI |
| | version | Changed veresion | NO | | bigint(20) | |
| | iufaid | Unit Asset ID in UUIS (Barcode) | YES | 255 | varchar(255) | UNI |
| | status | Asset Status | NO | 255 | varchar(255) | |
| | legacyid | Legacy ID | YES | 255 | varchar(255) | UNI |
| | location_id | Location ID | NO | | bigint(20) | MUL |
| | assignee_id | Assiged to User ID | YES | | bigint(20) | MUL |
| | parent_id | Group Asset ID | YES | | bigint(20) | MUL |
| | serial_number | Serial Number | YES | 255 | varchar(255) | |
| | type_id | Asset Type ID | NO | | bigint(20) | MUL |
| | details | Asset Details | YES | 255 | varchar(255) | |
| | name | Asset Name | NO | 255 | varchar(255) | |
| | owner_id | University/Faculty/Depa rmtment ID | NO | | bigint(20) | MUL |

# 9 Time logs

| | Bing | Robin | Deyvisson | Abdulrahman | Ali | Kanj | Max | Yuming |
|---|---|---|---|---|---|---|---|---|
| Week 1 | 0 | 0 | 0 | 0 | 0 | 0 | 0 | 0 |
| Week 2 | 1 | 2 | 1 | 2 | 1 | 2 | 2 | 2 |
| Week 3 | 2 | 7 | 3 | 2 | 2 | 8 | 3 | 3 |
| Week 4 | 5 | 4 | 1 | 5 | 6 | 8 | 2 | 3 |
| Week 5 | 5 | 5 | 1 | 6 | 2 | 6 | 5 | 3 |
| Week 6 | 7 | 6 | 4 | 6 | 1 | 6 | 4 | 5 |
| Week 7 | 7 | 6 | 2 | 7 | 4 | 5 | 4 | 4 |
| Week 8 | 6 | 6 | 5 | 15 | 15 | 5 | 5 | 10 |
| Week 9 | 15 | 7 | 10 | 4 | 3 | 7 | 10 | 5 |
| Week 10 | 6 | 10 | 20 | 10 | 15 | 5 | 10 | 5 |
| Week 11 | 11 | 16 | 15 | 12 | 15 | 13 | 16 | 10 |
| Week 12 | 8 | 6 | 9 | 8 | 5 | 4 | 4 | 5 |
| Week 13 | 9 | 4 | 6 | 5 | 10 | 8 | 7 | 7 |
| Week 14 | 6 | 3 | 9 | 5 | 6 | 9 | 4 | 6 |
| Week 15 | 7 | 8 | 9 | 8 | 6 | 5 | 6 | 8 |
| Week 16 | 4 | 5 | 5 | 0 | 5 | 5 | 1 | 5 |
| Week 17 | 5 | 5 | 5 | 5 | 1 | 6 | 1 | 1 |
| Total | 104 | 100 | 105 | 101 | 97 | 102 | 84 | 82 |

Time log of the team (in hours)

# Appendix I: User Guild Document

## System Requirements

You may run the system web interface application any OS including Windows XP, Windows Vista, Windows 7 MAC OS, or Unix using IE7 or higher Firefox ,Chrome, Opera or Safari

## Access to the web application

Any user can having internet connection and using one the listed browsers can access the web interface application using the following link: http://spec109.encs.concordia.ca/uuis/

## Login

The access to the application is restricted to authorized users only, each user should know his user name and password
Any user must be authenticated to be able to use the application
Login is the first page displayed when you access the application

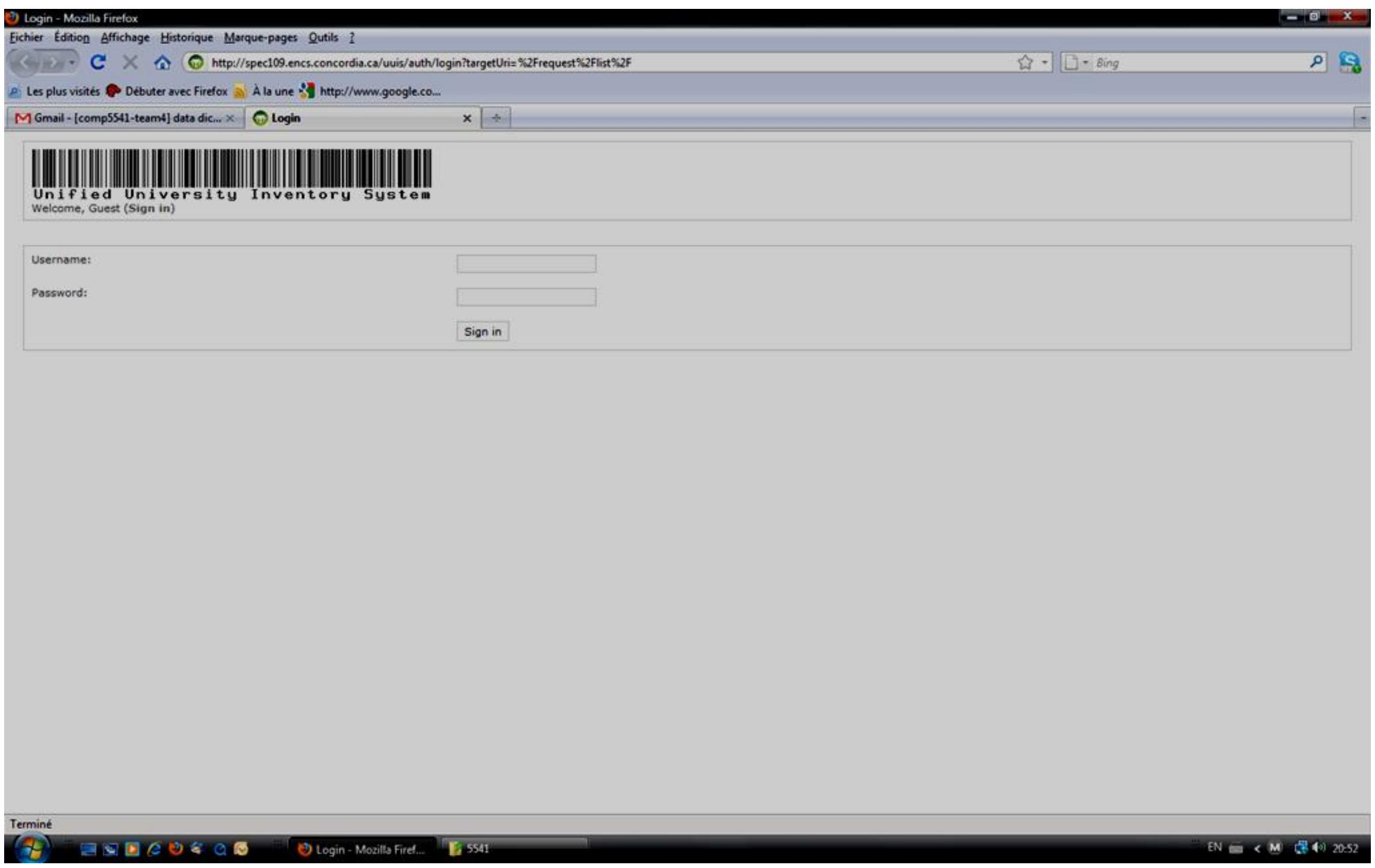

## Application main page

After authentication this page change according to user permissions

Student and professor can only create requests

For any administrative level user this page is divided into 3 sections the main menu at the left side, sub menu at the middle and search menu at the right side

By default the assets page is displayed it contains 3 lists:

1. Requests waiting for approval: includes any request that this administrator can approve
2. Requests waiting for execution: any approved requests pass from waiting for approval to waiting for execution when it is approved, when an asset is picked up the data update is made
3. My Request contains the request that he use made

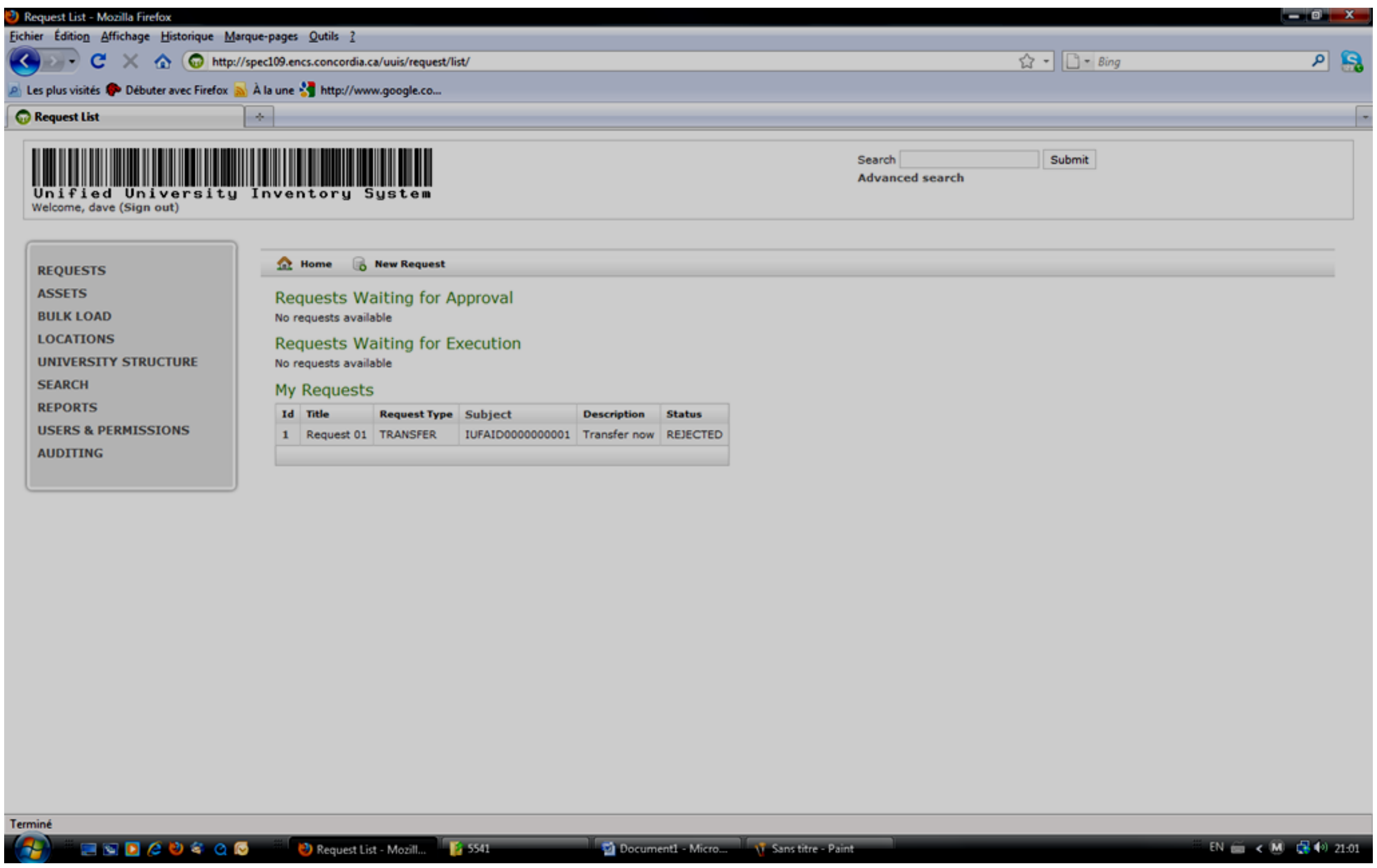

## <u>Approving a request</u>

An inventory admin having the permission to approve or reject requests

1. Click the request details are displayed in as shown in the following screen
2. Click Approve

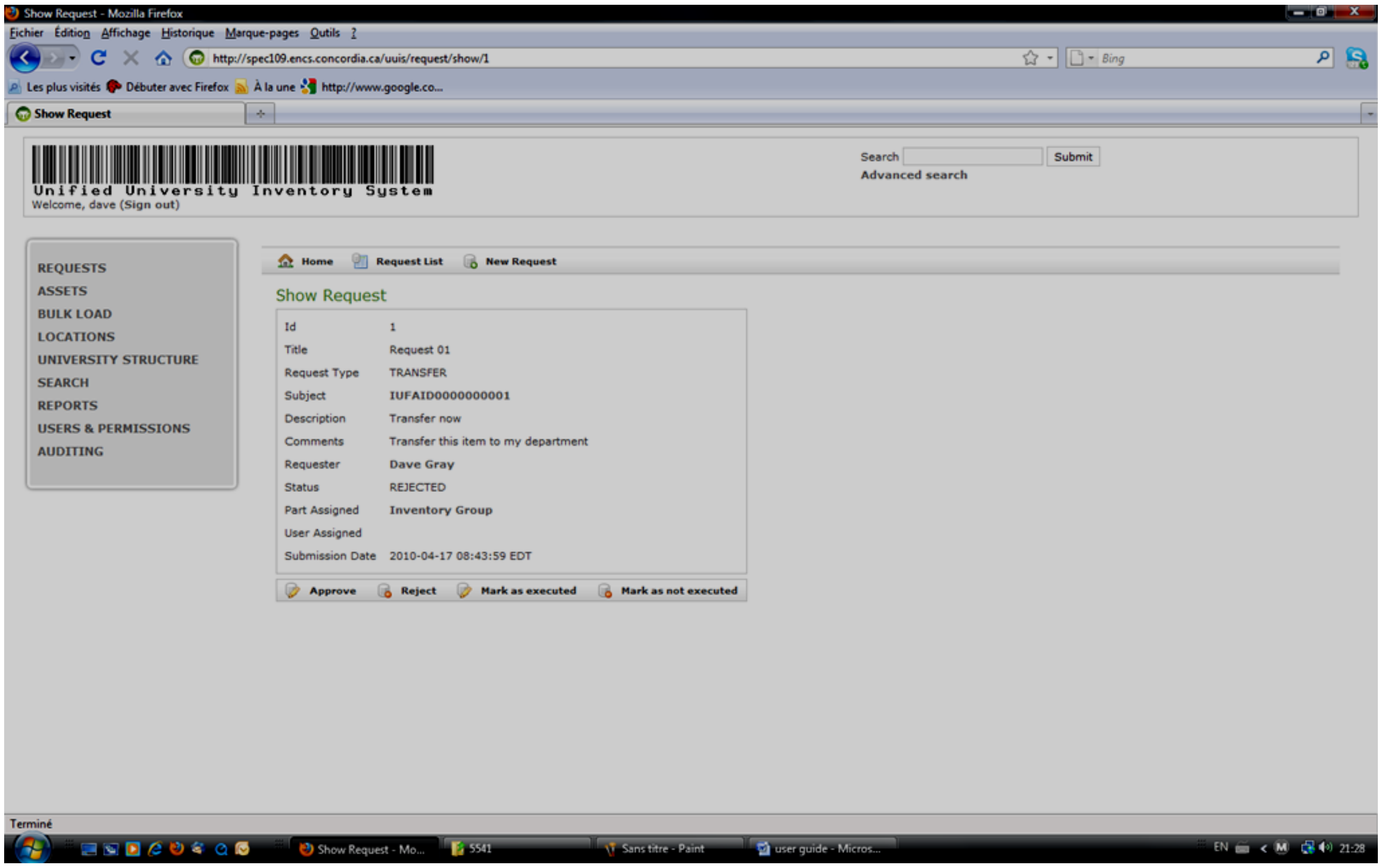

The request is transferred to waiting for execution list

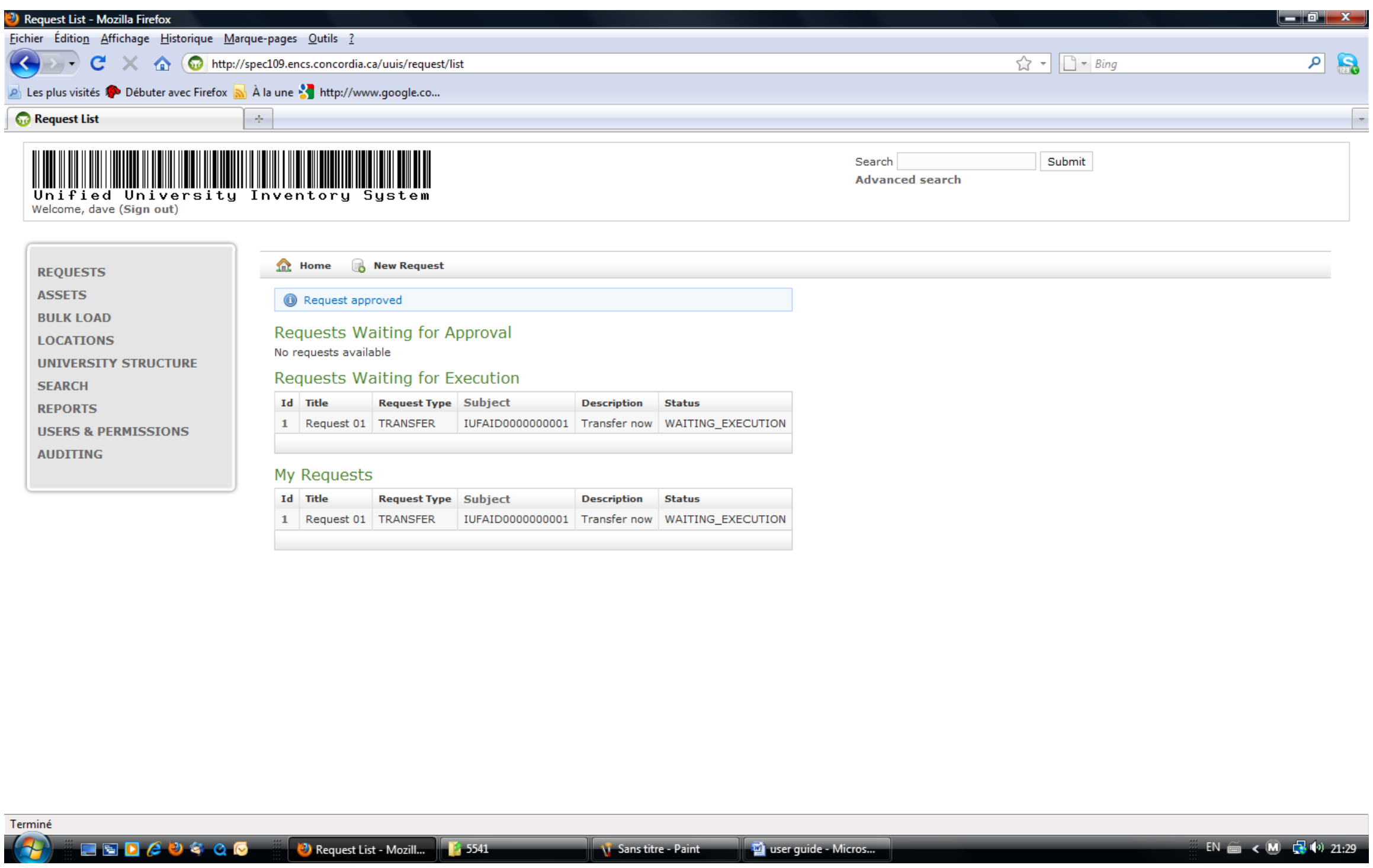

## Rejecting Request

1. Click on the request
2. Click Reject

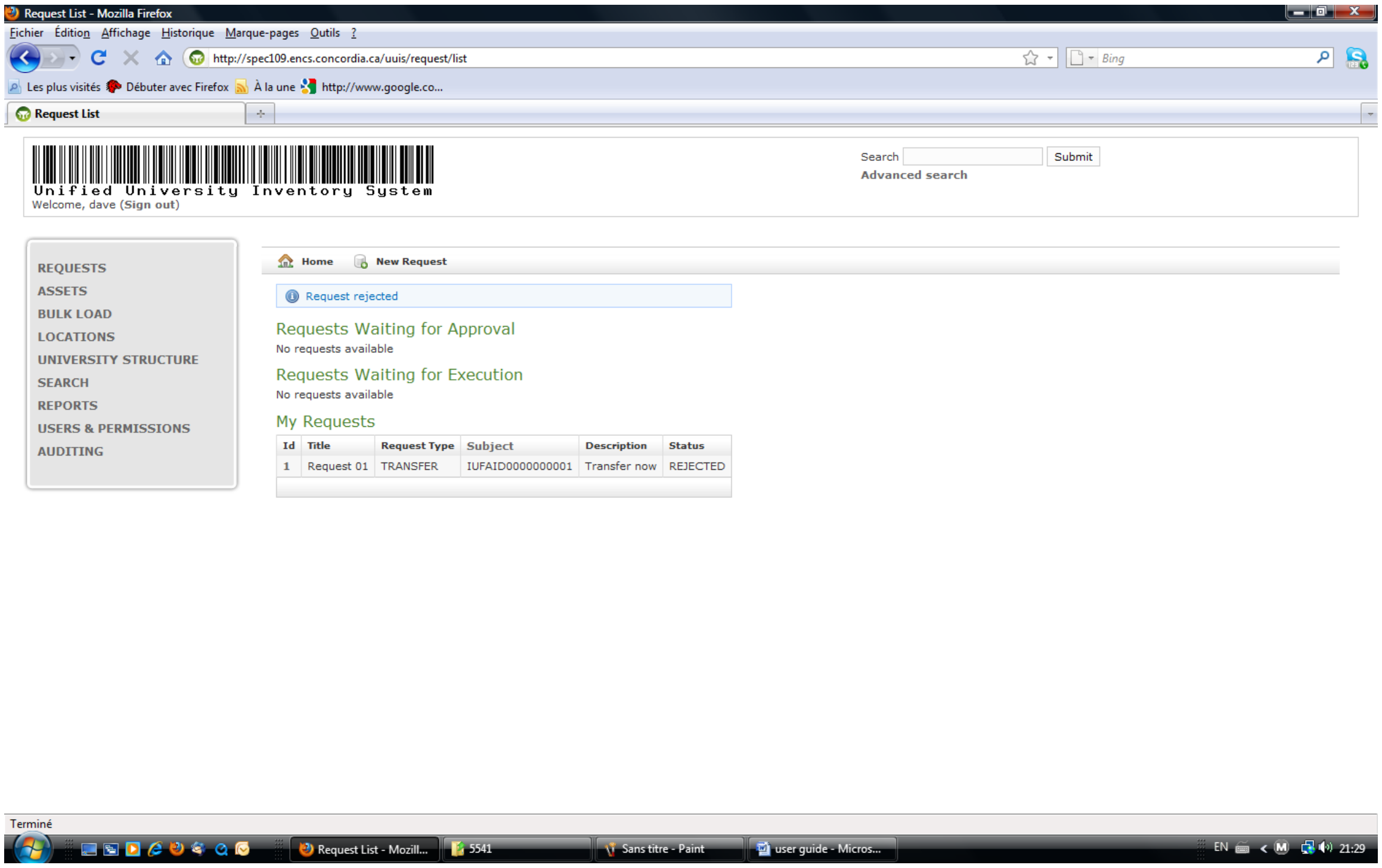

## Executed Request Form Waiting For Execution List

1. Click the request
2. Click the reject mark as executed

Request is removed from the list

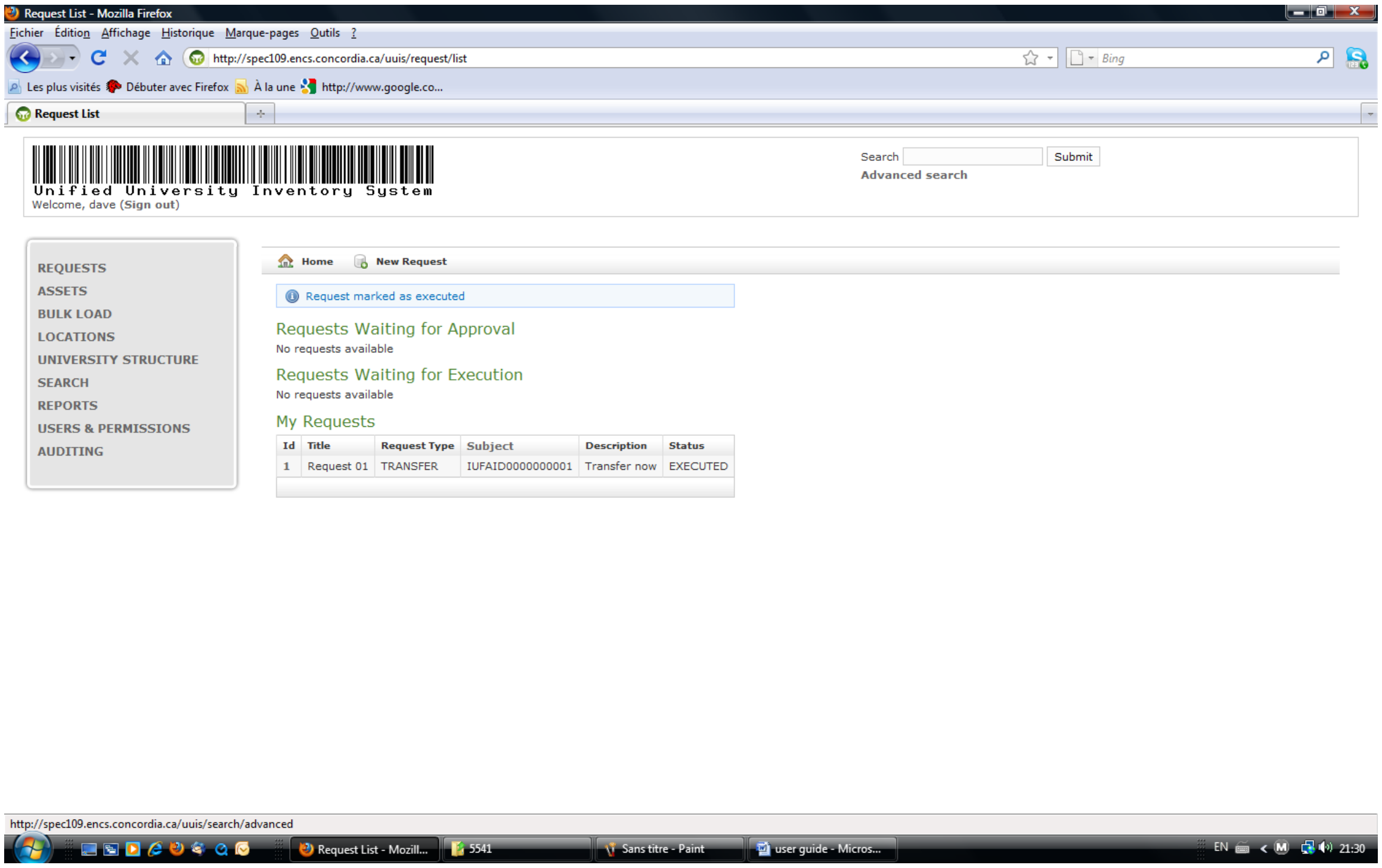

## Creating a New Basic Request New Request

1. Click the button New request
2. Type a request description in the comment box
3. Click create

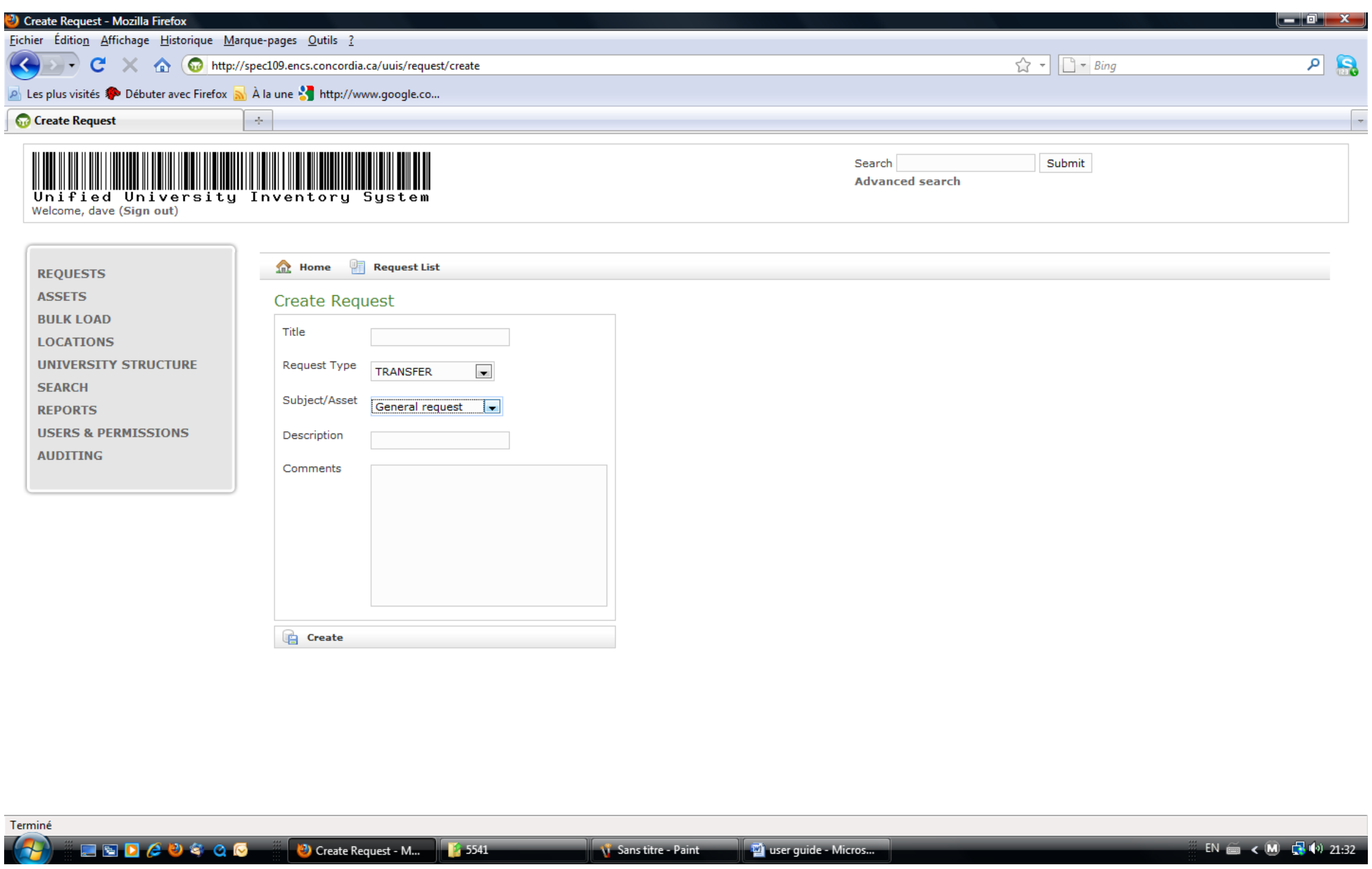

## Creating a New Advanced Request

1. Click the button New request
2. Type a request description in the comment box
3. Select request type
4. Select required assets
5. Click Create

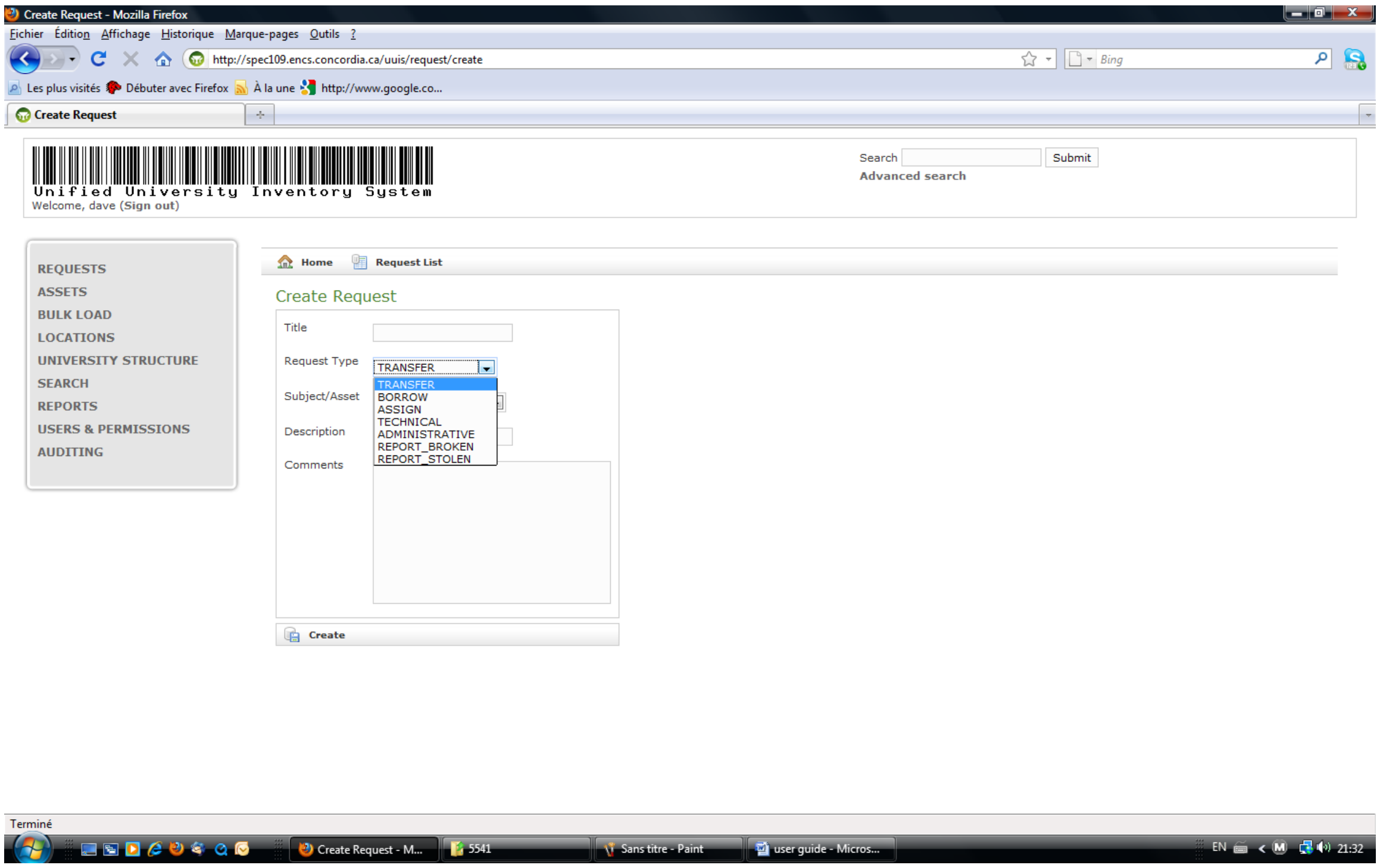

## Display Assets List

1. Click Asset on left menu

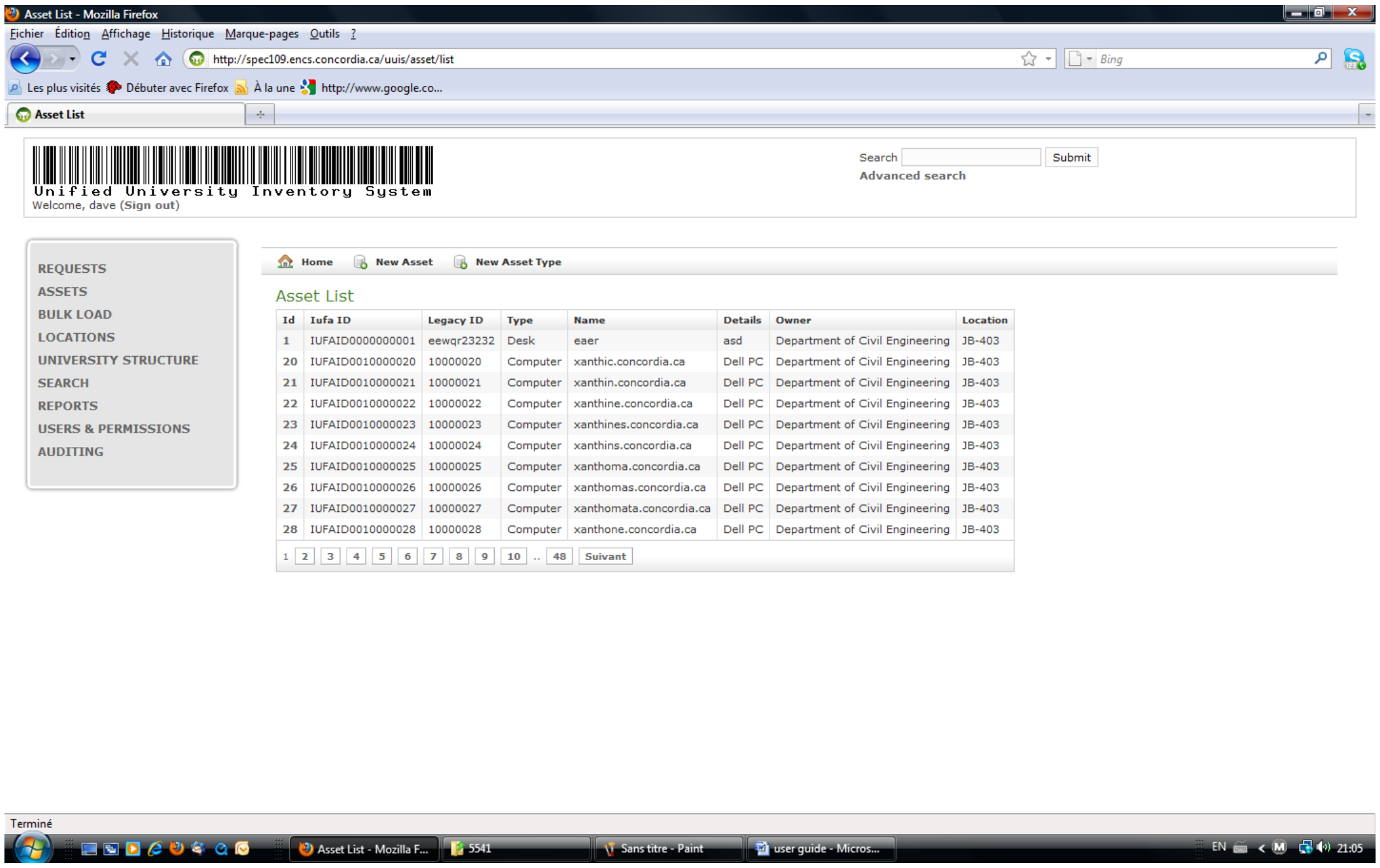

## Create New Asset

1. Click Asset on left menu
2. Click New Asset
3. Fill all the required data and click Create

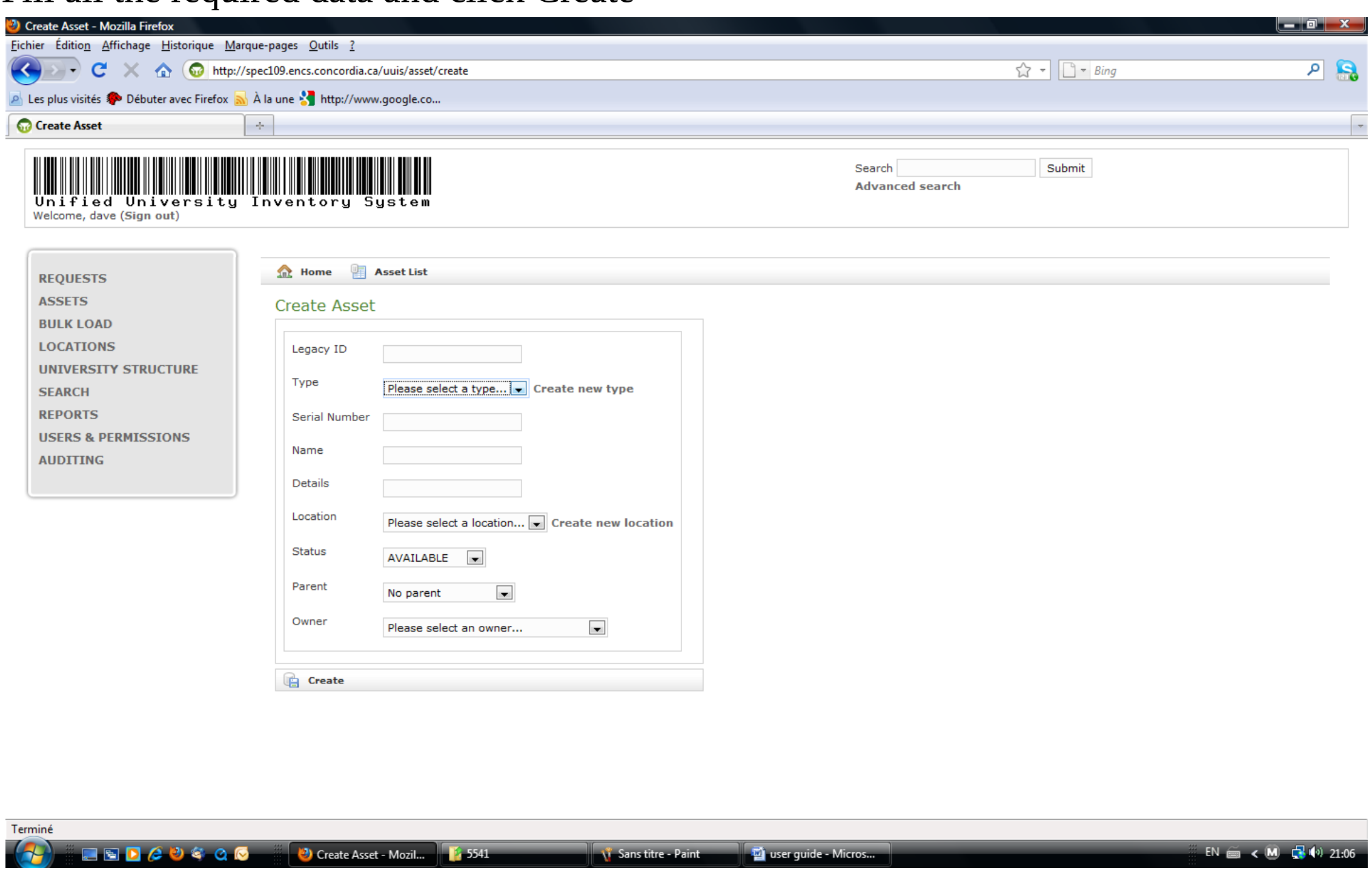

## Create New Asset Type (only IT Administrator)

1. Click Asset on left menu
2. Click New Asset Type
3. Fill all the required data and click Create

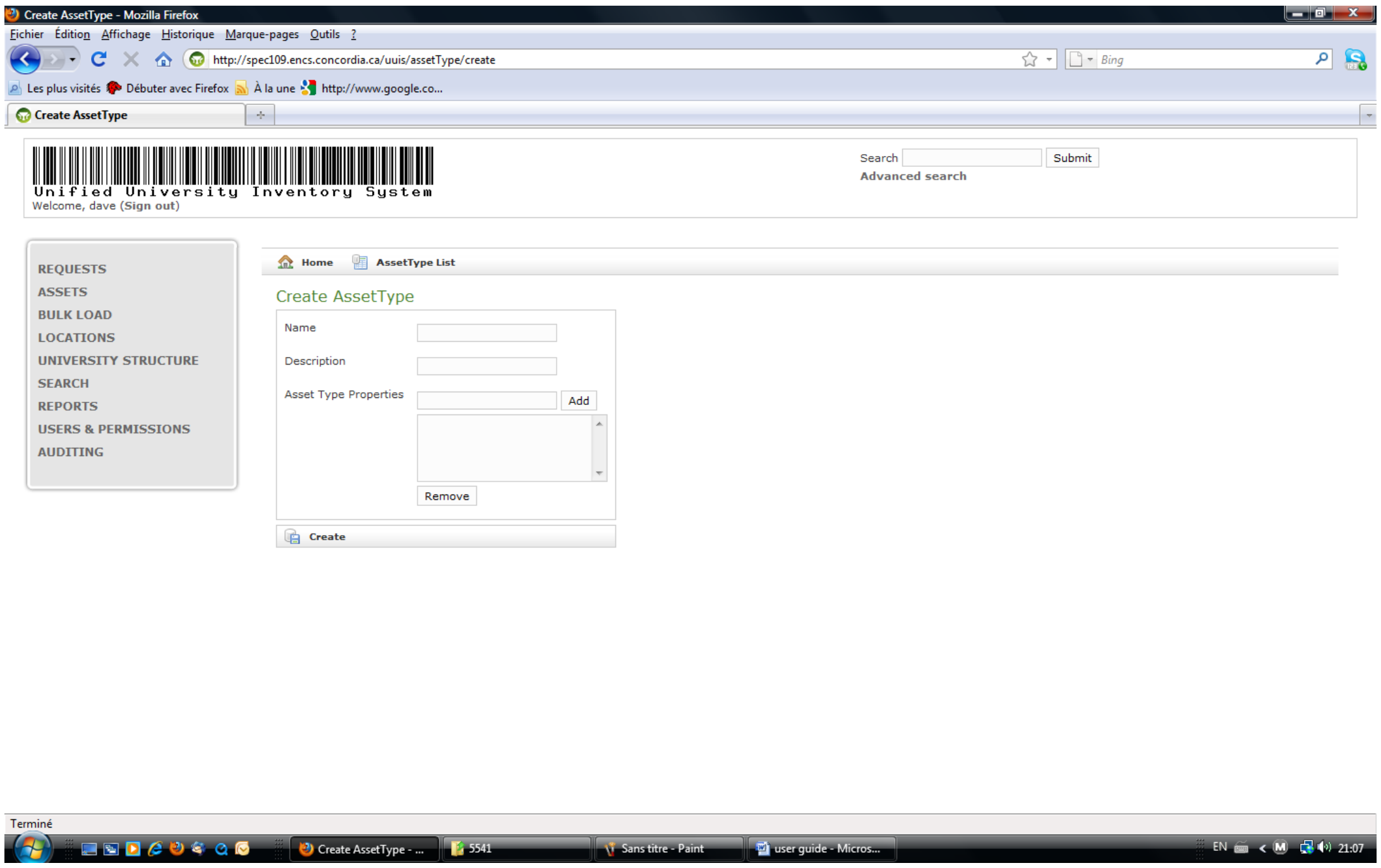

## Displaying Asset properties

1. Click the Asset from left Menu
2. Click the Asset from the Assets list

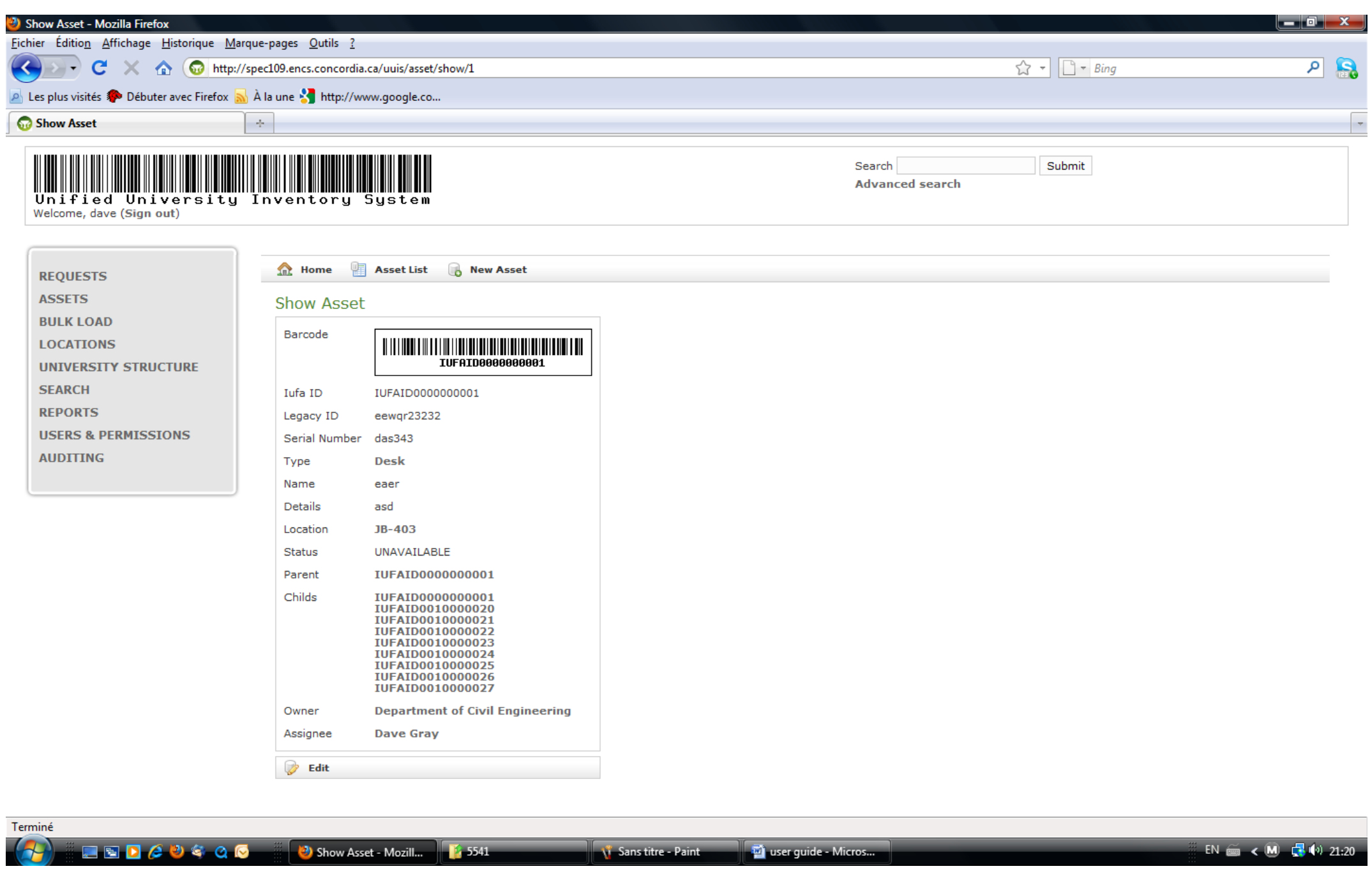

## Change Asset properties

1. Click the Asset from left Menu
2. Click the Asset from the Assets list
3. Click Edit asset
4. Change properties and click Update

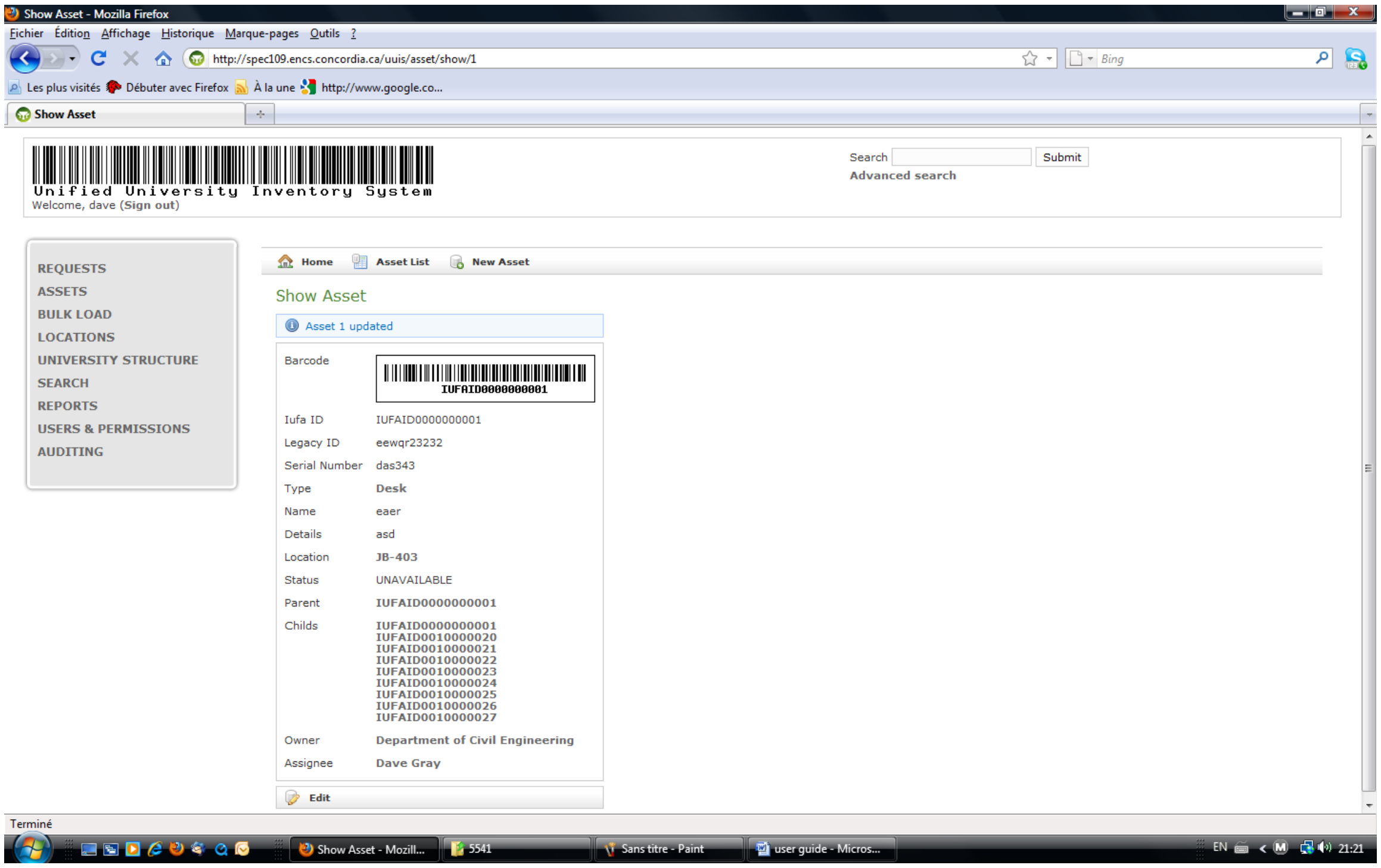

## Bulk load

Allow the a faster creation or updating of the Assets or locationsby using bulk entry form a file that must contains a header in the first line specifying data order and data records arranged according to the header

Bulk Insert: used to add or create new assets

1. Click browse
2. Select the file
3. Click upload file

Bulk update: used to update assets

1. Click browse
2. Select the file
3. Click Upload file

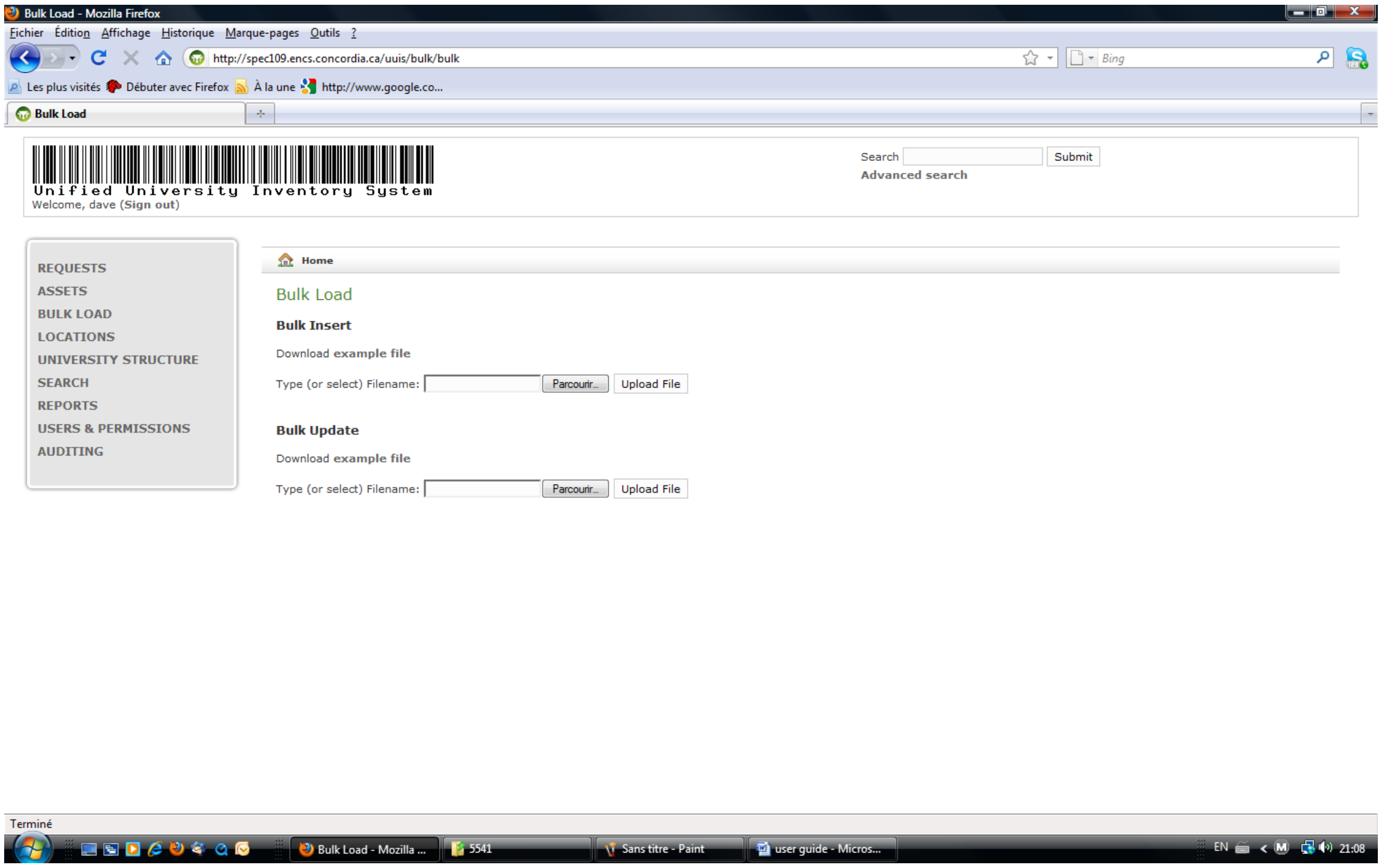

## Display Location list

1. Click Location left menu

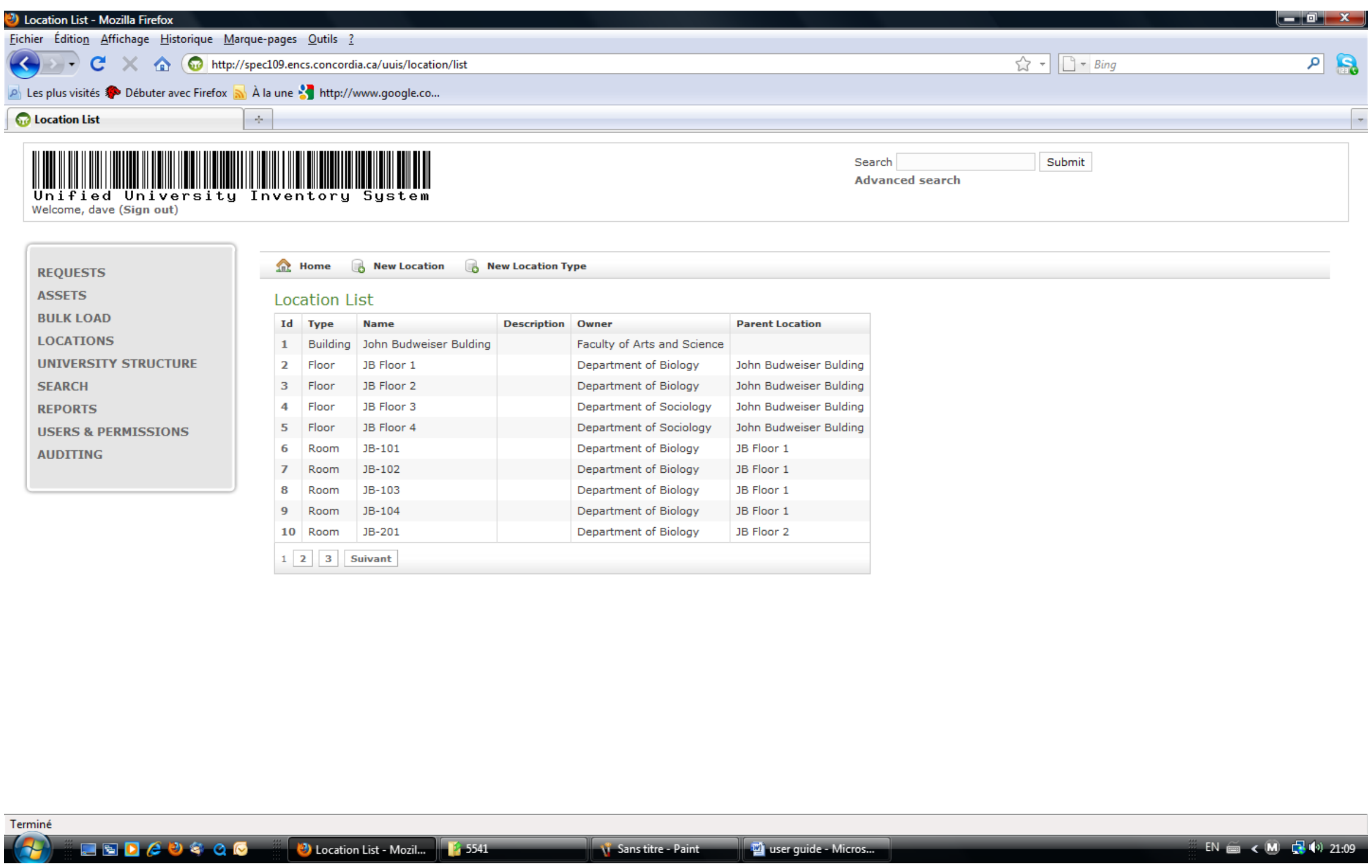

# Create New Location

1. Click Location on left menu
2. Click New Location
3. Fill all the required data and click Create

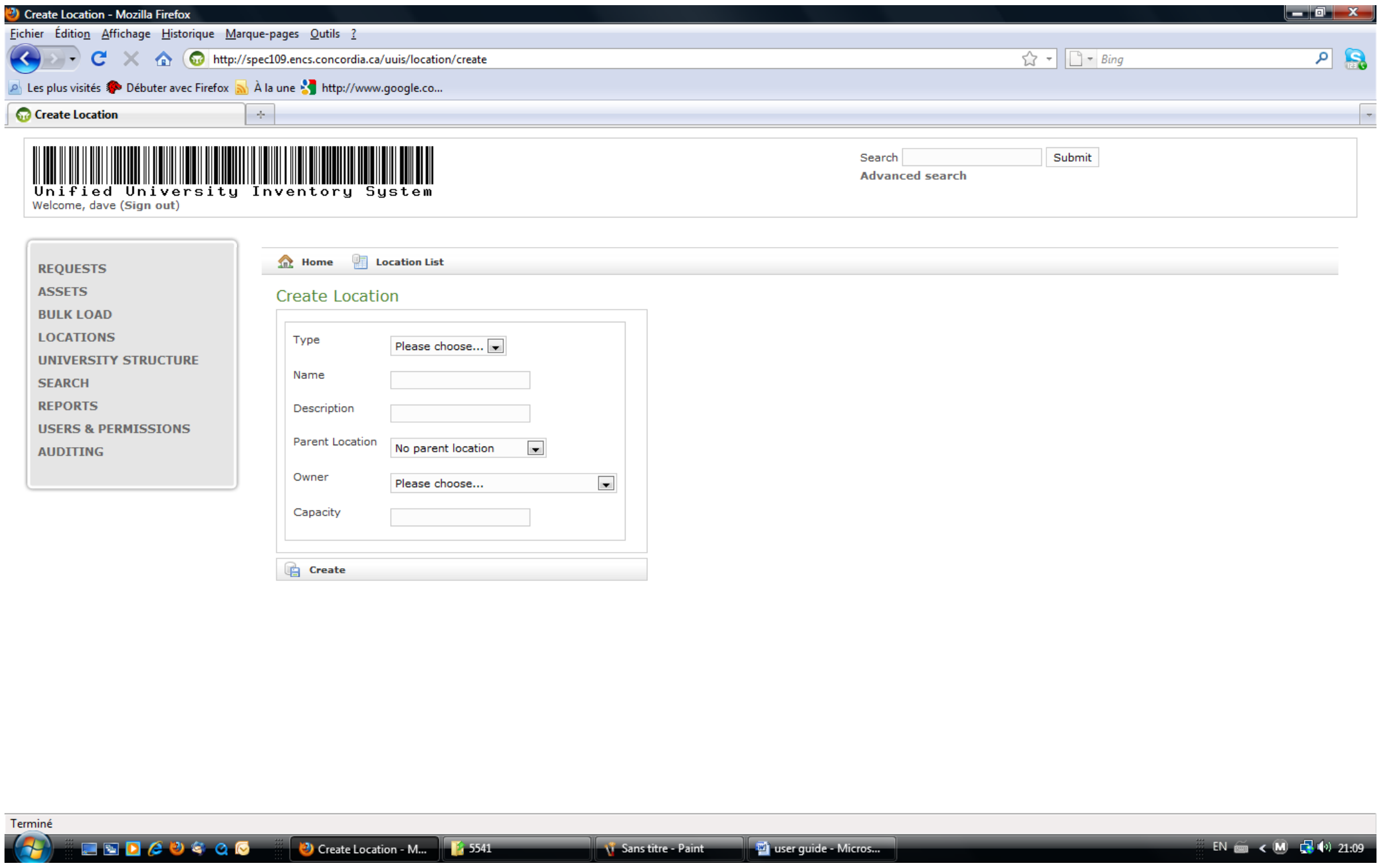

# Create New Location Type (IT group only)

1. Click Location on left menu
2. Click New Location Type
3. Fill all the required data and click on Create

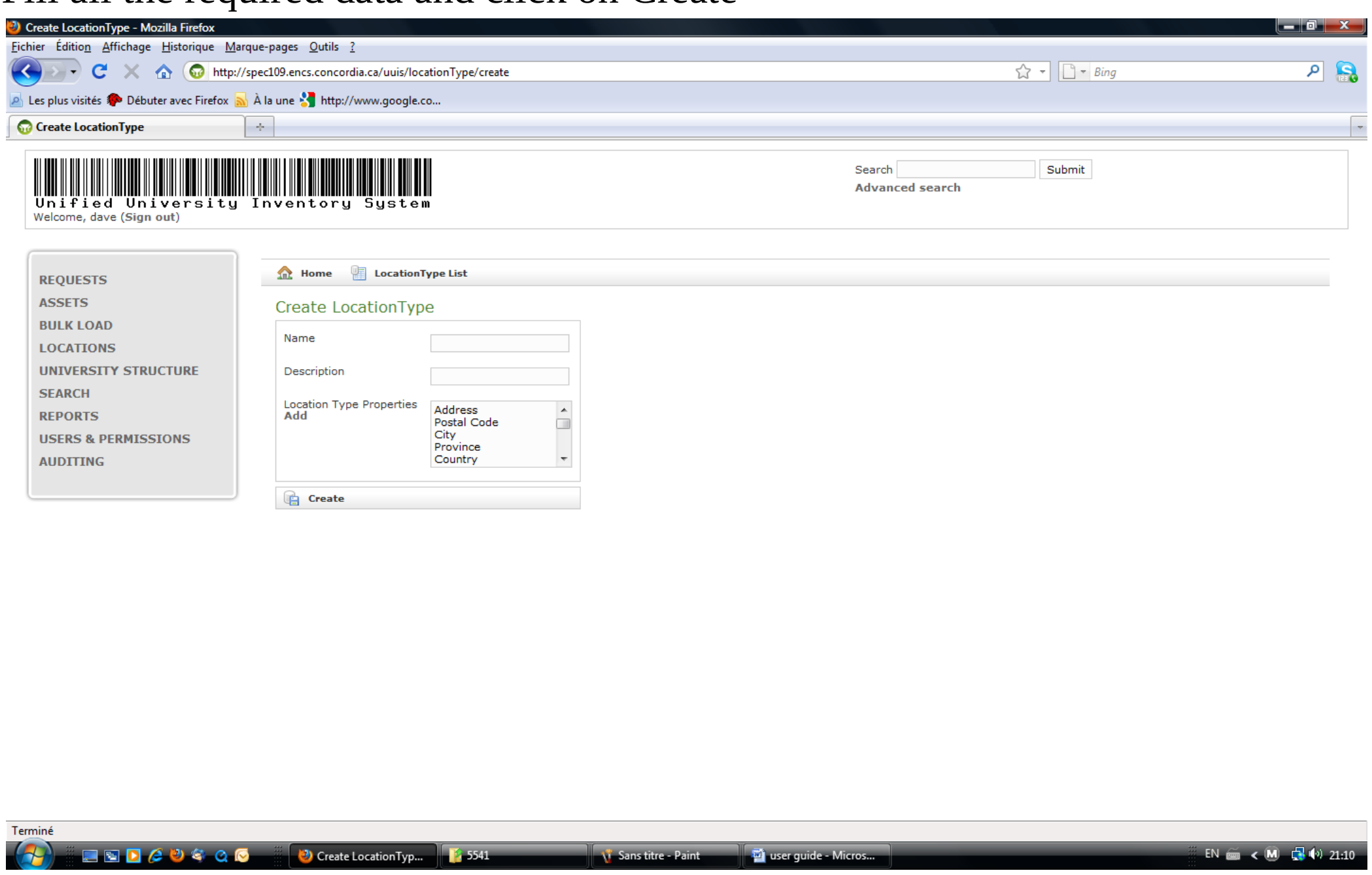

## Displaying Location Properties

1. Click the Location from left Menu
2. Click the Location from the Location list

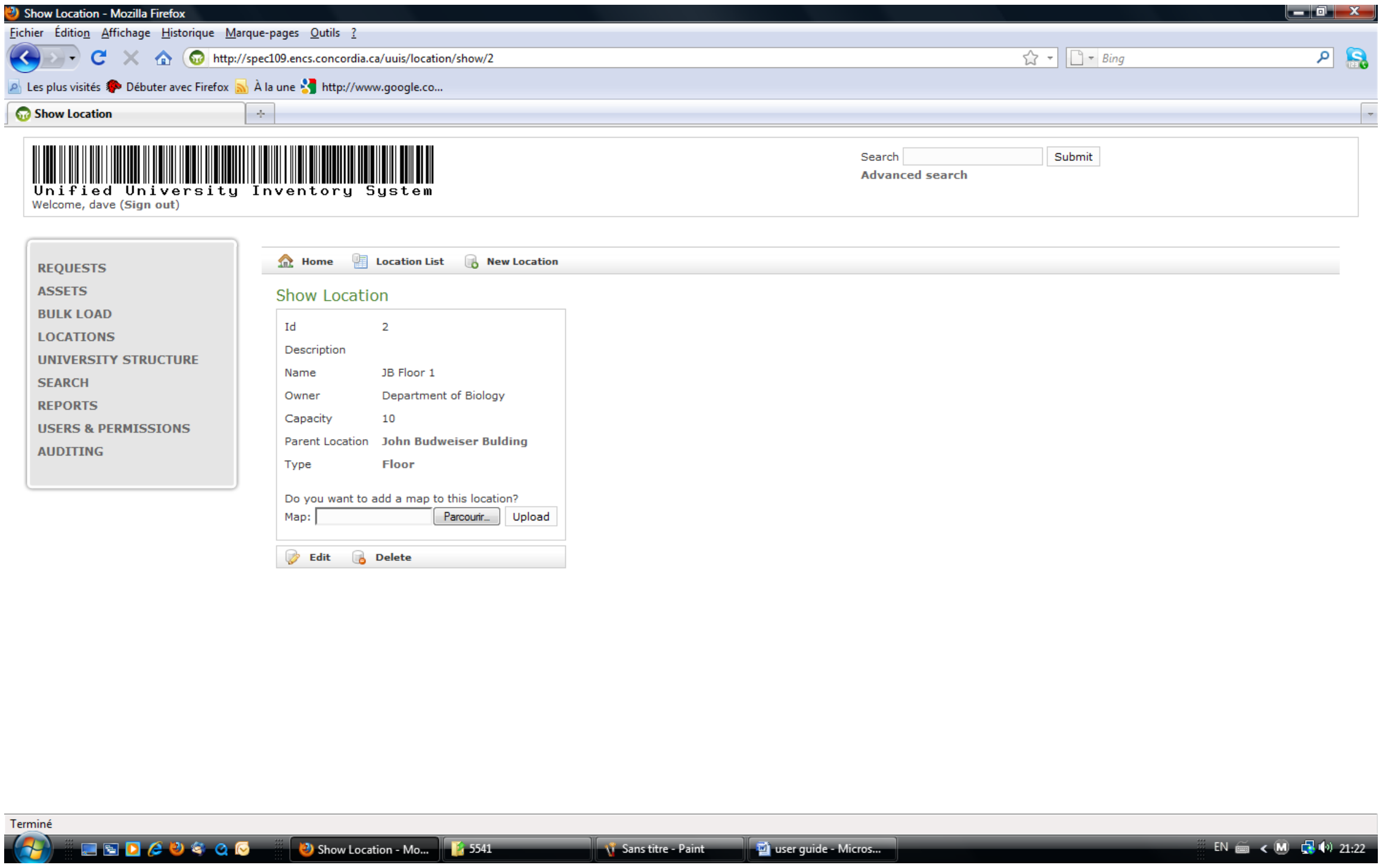

## Change Location Properties

1. Click the Location from left Menu
2. Click the Location from the Location list
3. Click Edit Location
4. Change properties and click Update

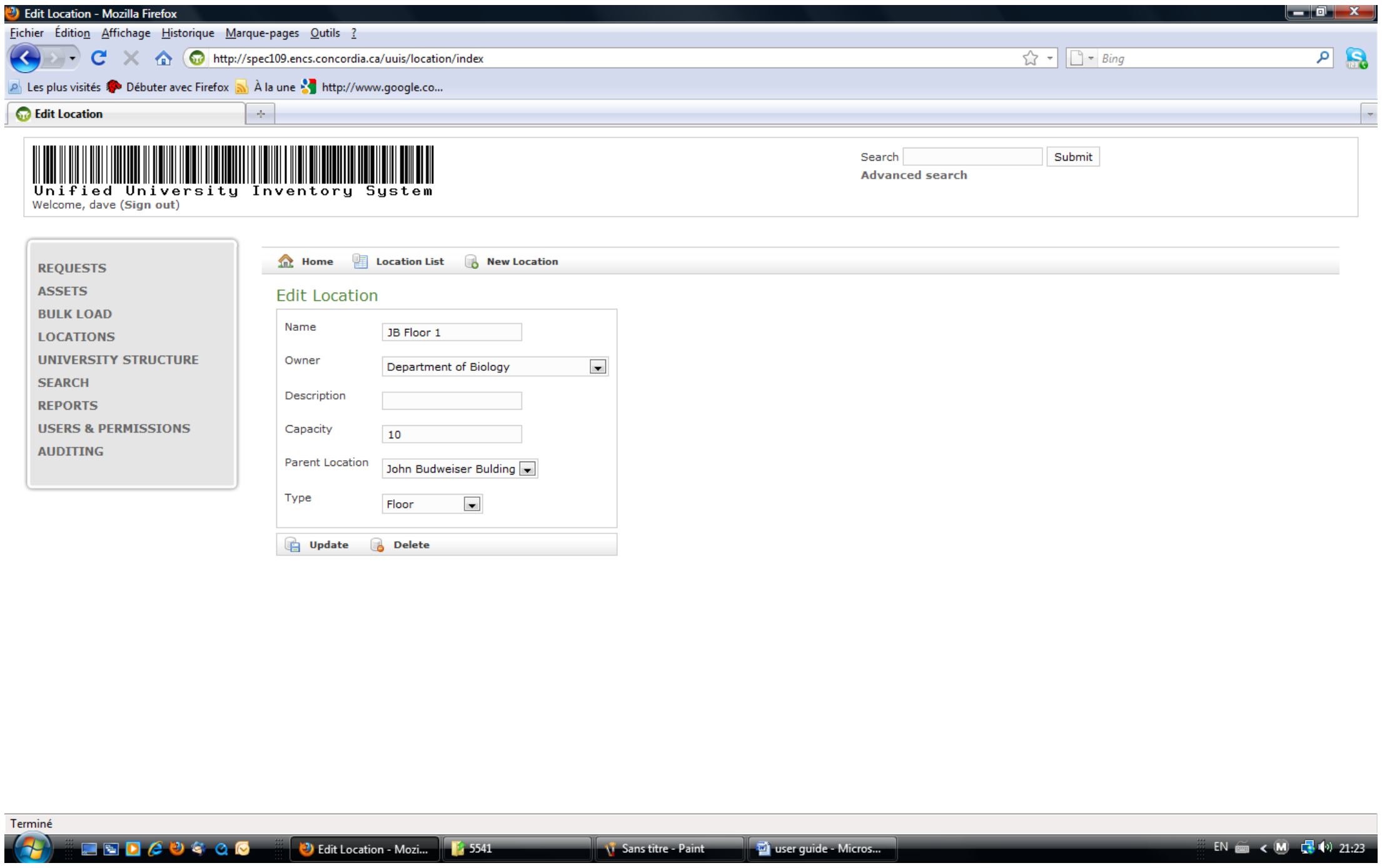

## Delete Location (IT Group Only)

1. Click the Location from left Menu
2. Click the Location from the Location list
3. Click Edit Location
4. Click Delete Location
5. Click ok to confirm

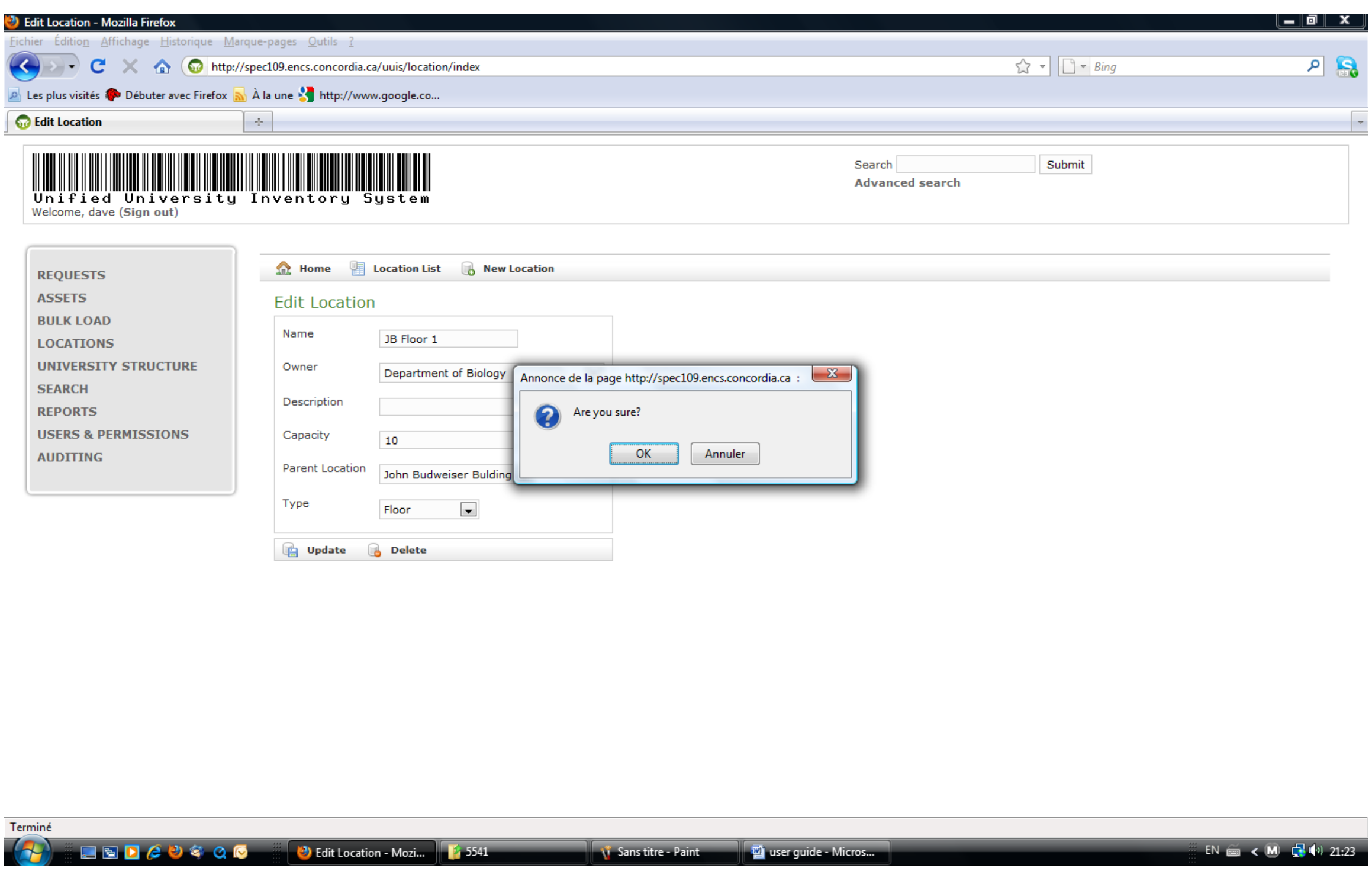

## Displaying University Structure

Permit to display administrators list groups and levels

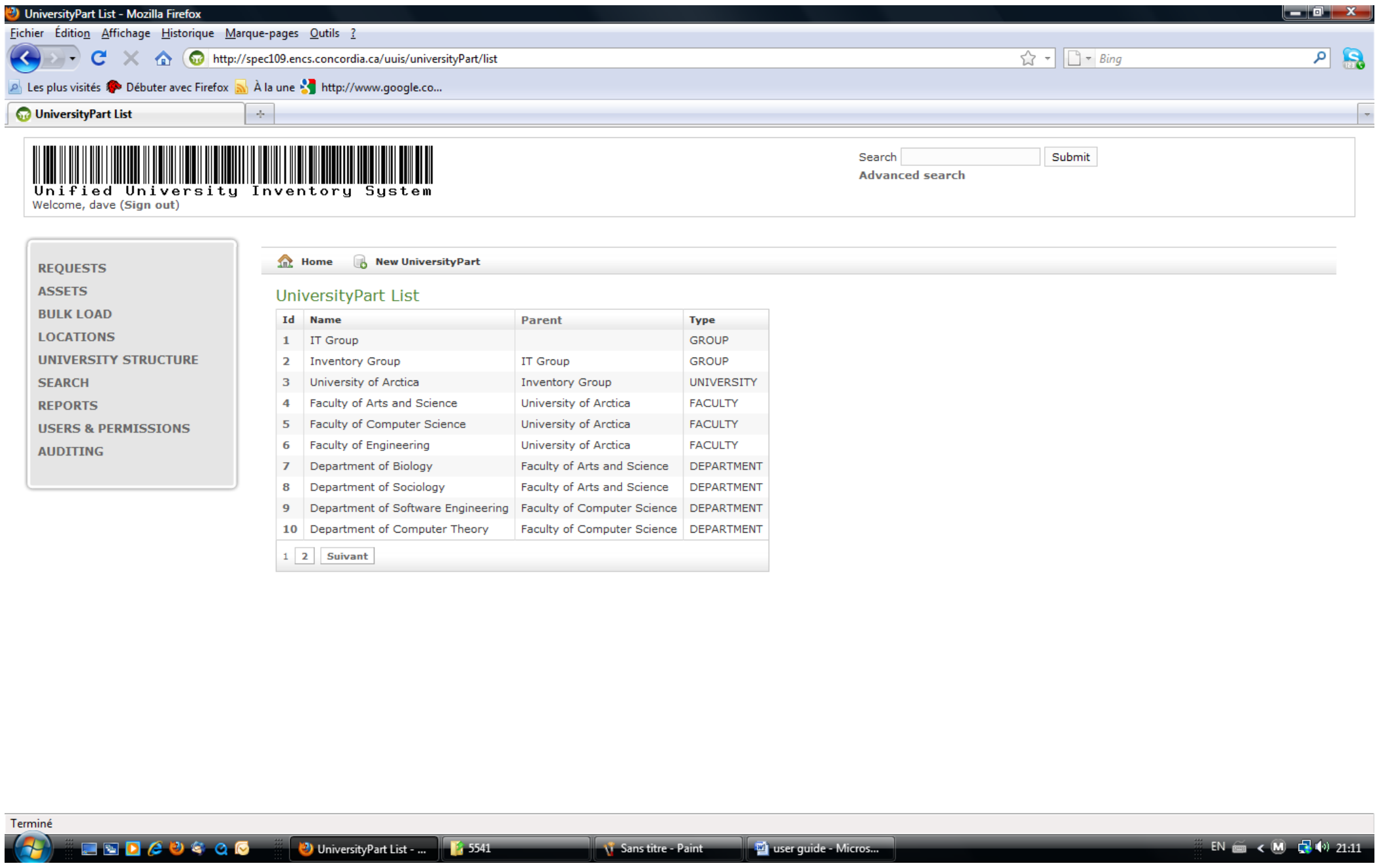

## Adding New Entity to University Structure (IT Group Only)

1. Click University Structure from the left Menu
2. Click Create

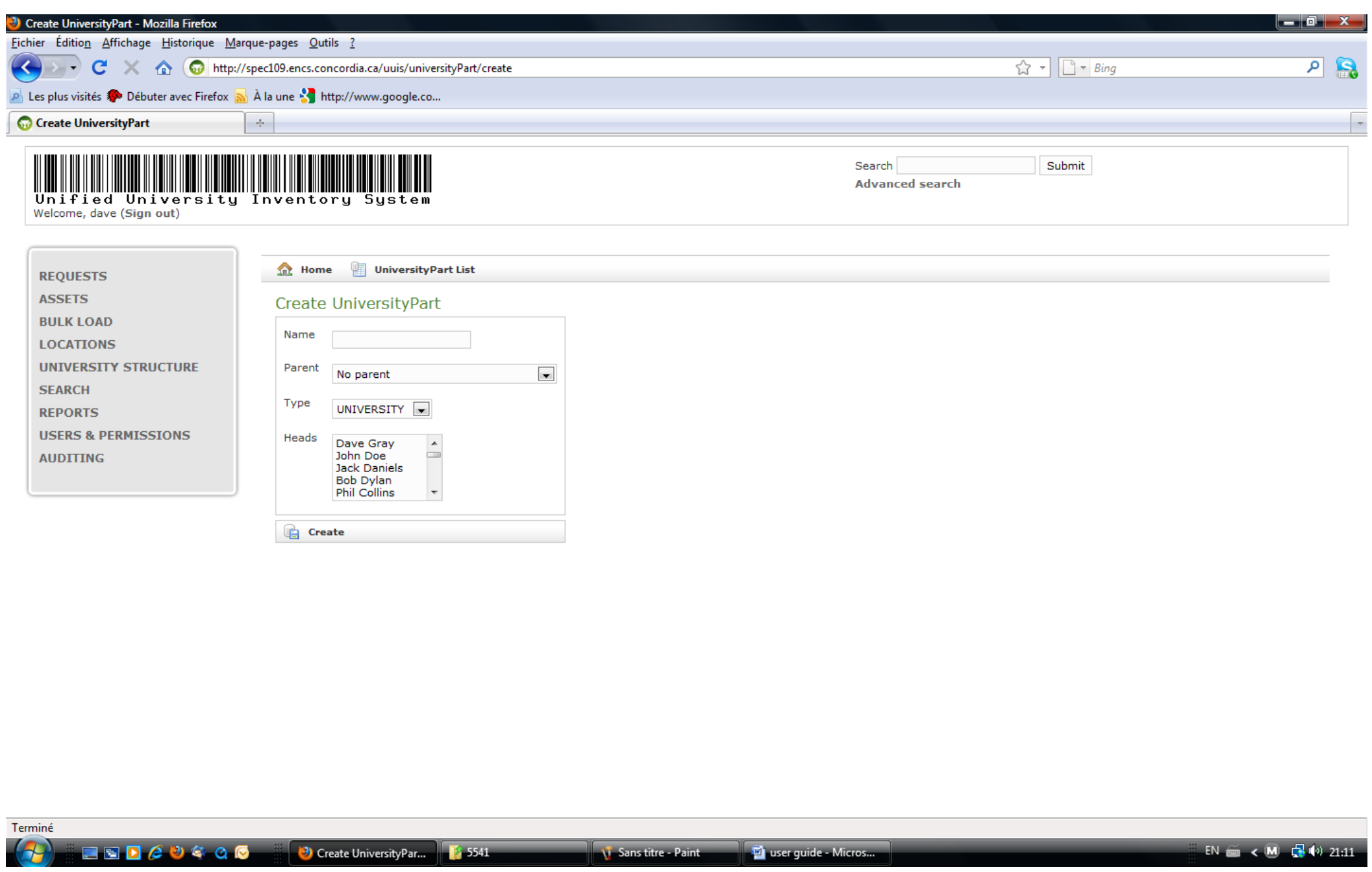

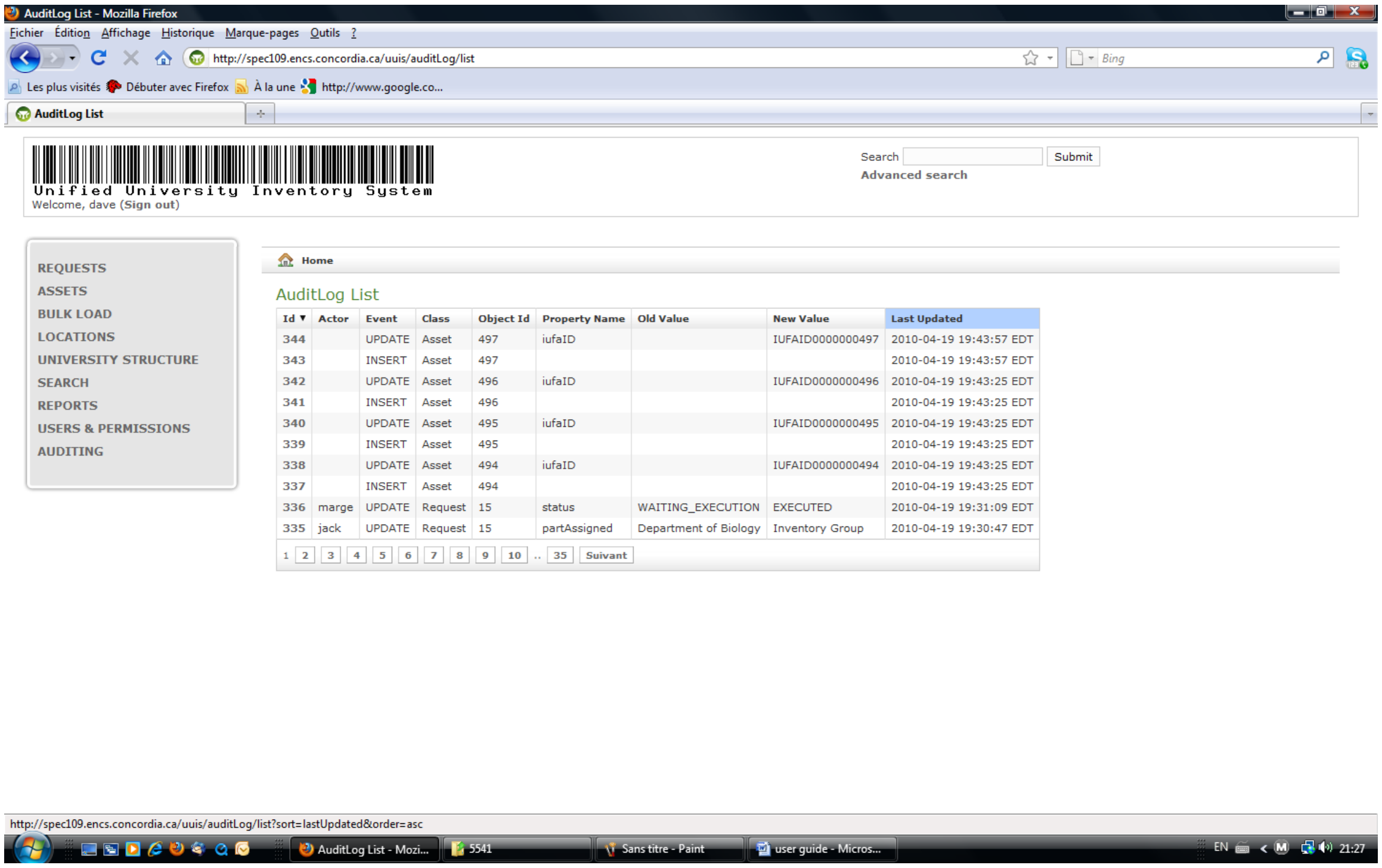

## <u>Search</u>

Search box is always displayed at right side of the screen

To display advance search page:

1. Click Search form the left Menu
2. Fill the search box
3. Using drop down list specify search  attributes

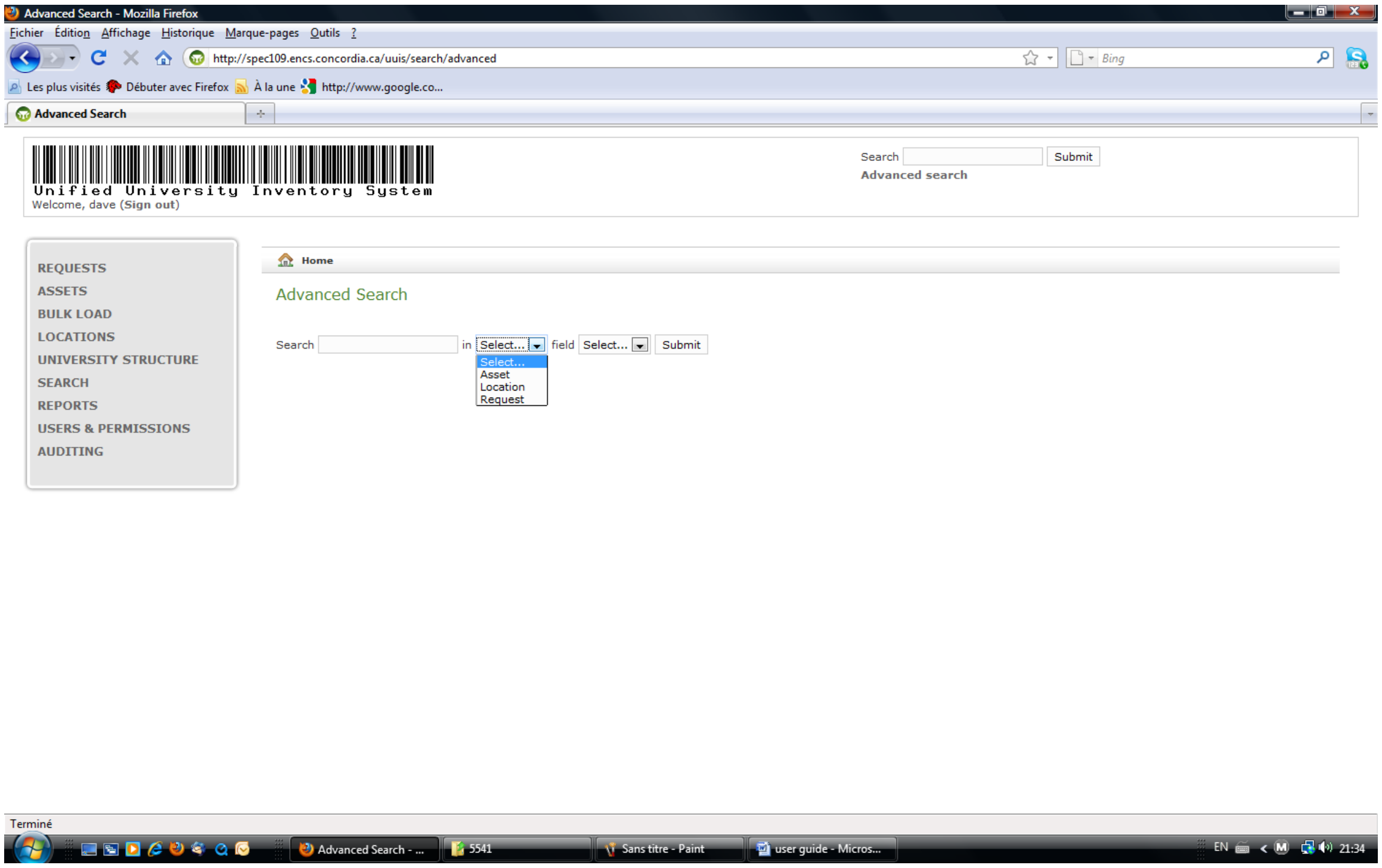

## Displaying Reports

The application can display three kinds of reports

1. User Permission Report
2. Requests Report
3. Assets by location report

To display a report:

1. Click Report from the left Menu
2. Click the Report you want to display

Report page allows to filter and sort data

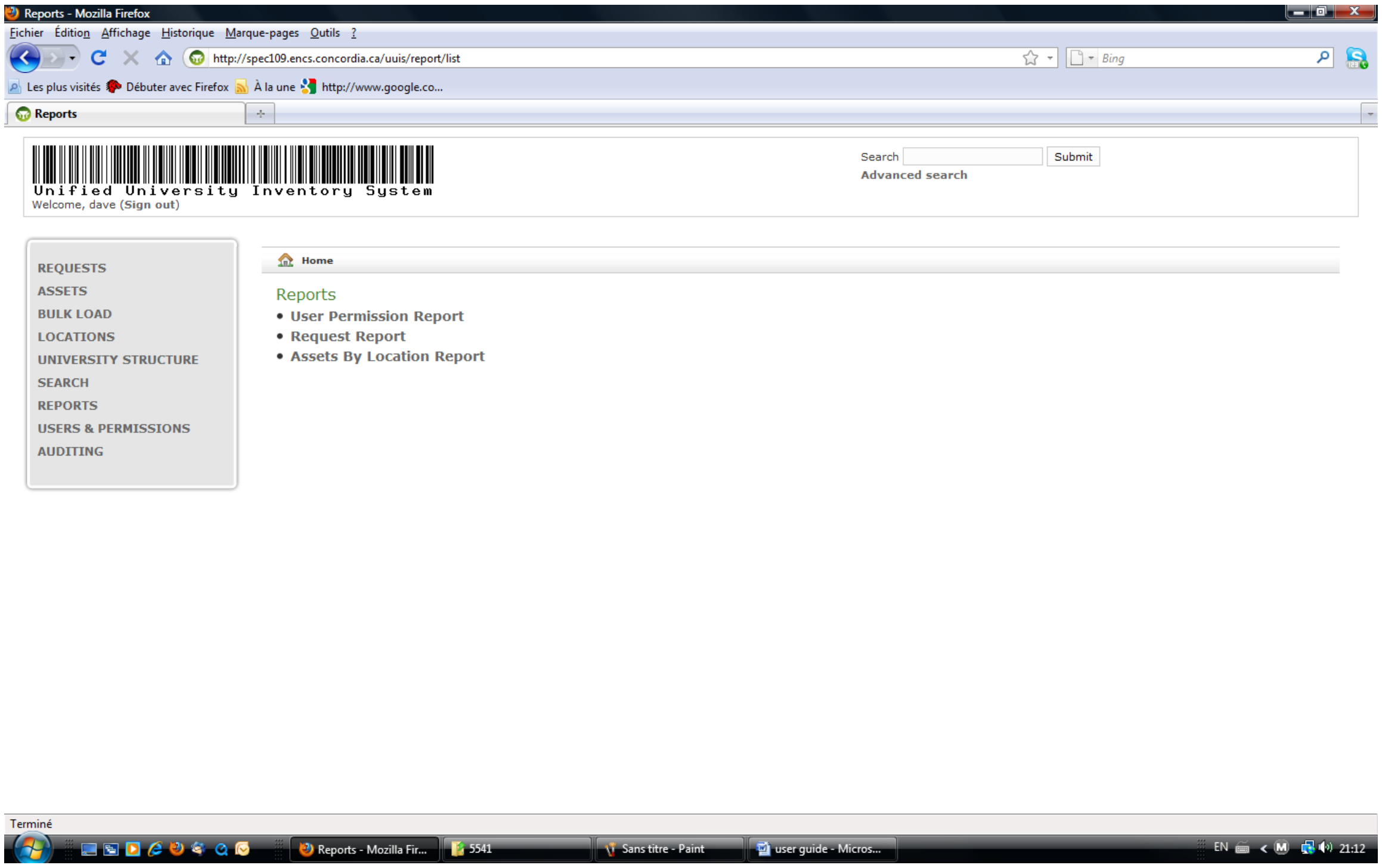

# User Permission Report

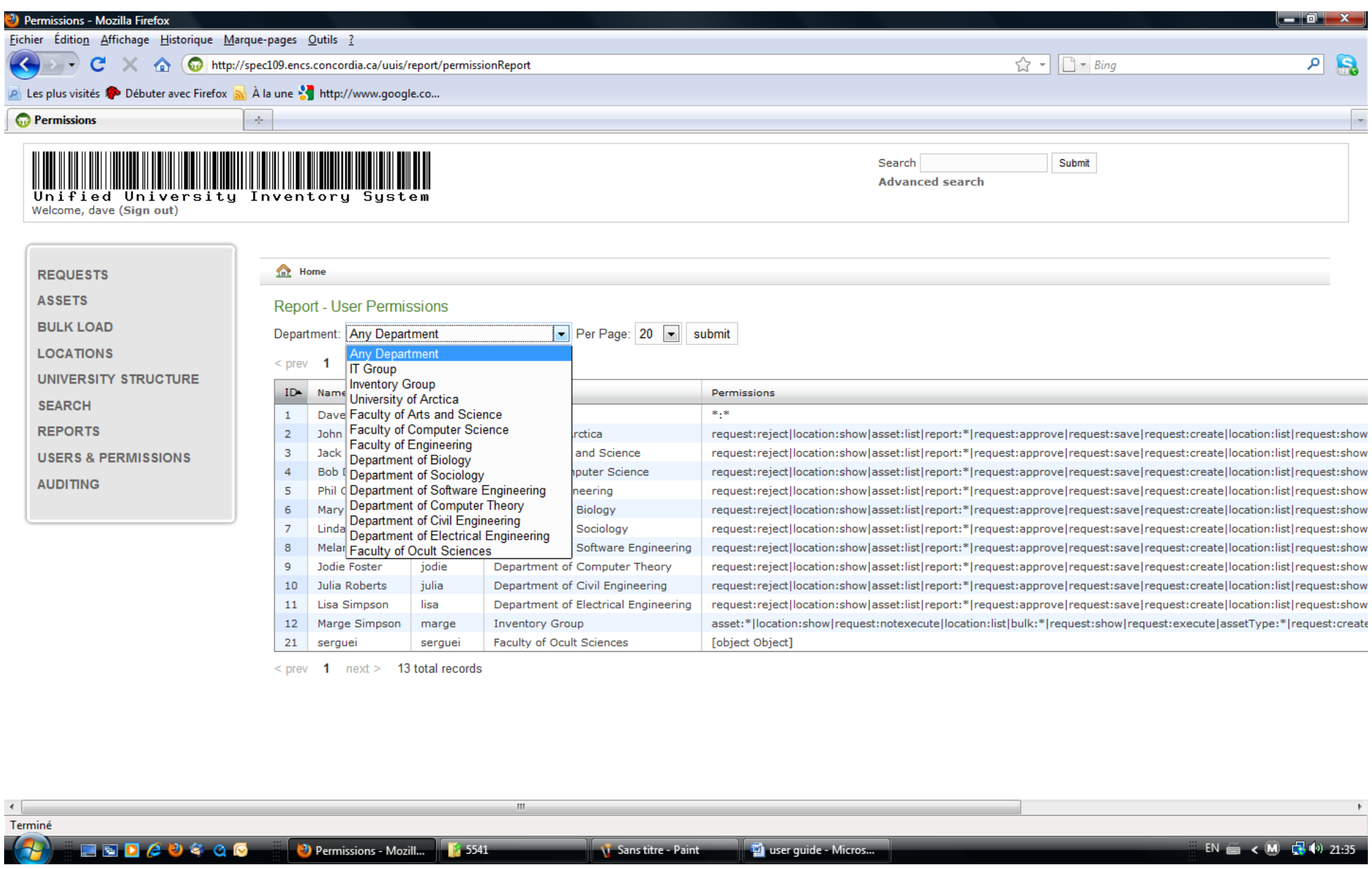

# Request Report

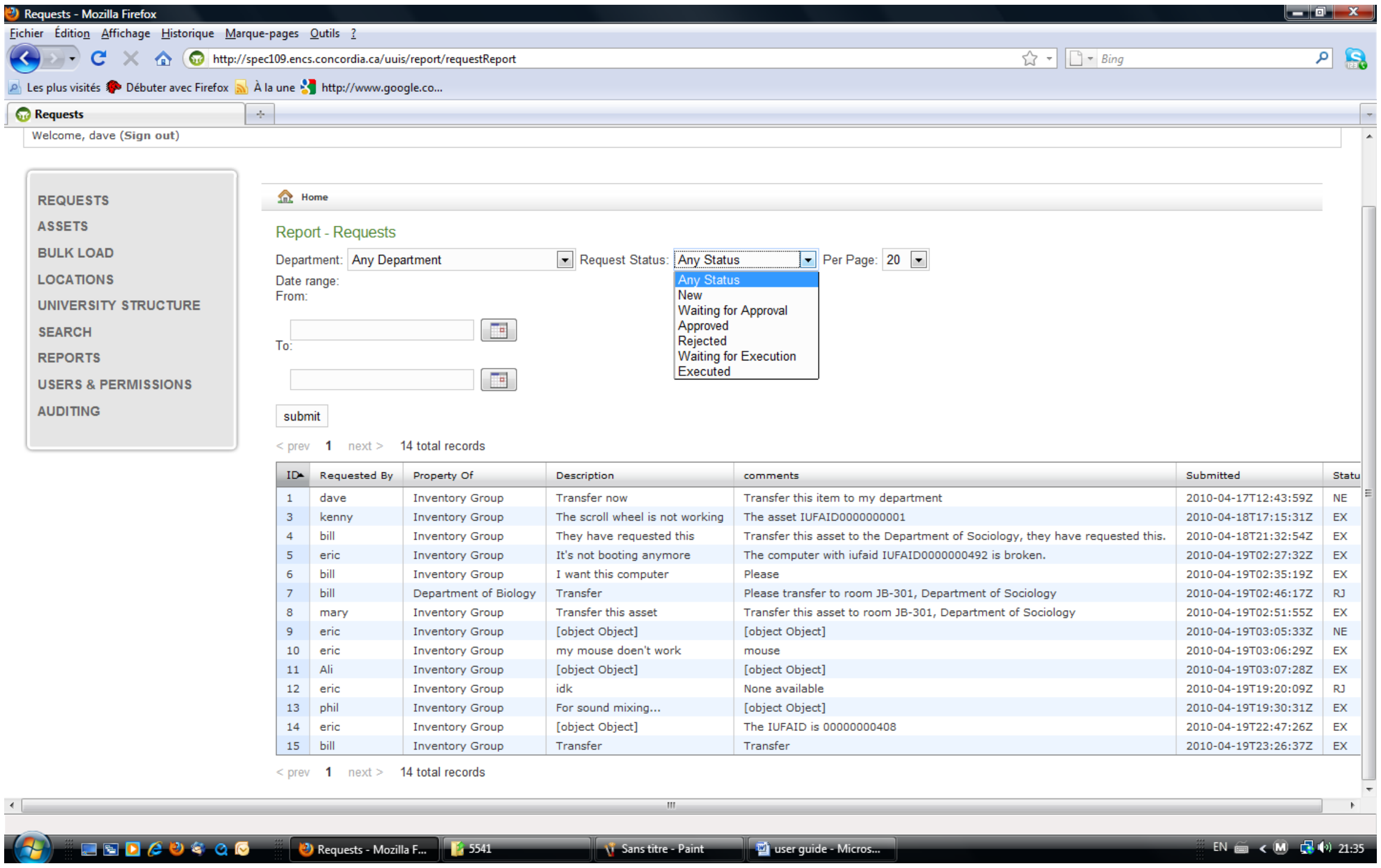

Requests - Mozilla Firefox

http://spec109.encs.concordia.ca/uuis/report/requestReport

Welcome, dave (Sign out)

REQUESTS
ASSETS
BULK LOAD
LOCATIONS
UNIVERSITY STRUCTURE
SEARCH
REPORTS
USERS & PERMISSIONS
AUDITING

Home

Report - Requests

Department: Any Department Request Status: Any Status Per Page: 20

Any Status
New
Waiting for Approval
Approved
Rejected
Waiting for Execution
Executed

Date range:
From:
To:

submit

< prev 1 next > 14 total records

| ID | Requested By | Property Of | Description | comments | Submitted | Statu |
|---|---|---|---|---|---|---|
| 1 | dave | Inventory Group | Transfer now | Transfer this item to my department | 2010-04-17T12:43:59Z | NE |
| 3 | kenny | Inventory Group | The scroll wheel is not working | The asset IUFAID0000000001 | 2010-04-18T17:15:31Z | EX |
| 4 | bill | Inventory Group | They have requested this | Transfer this asset to the Department of Sociology, they have requested this. | 2010-04-18T21:32:54Z | EX |
| 5 | eric | Inventory Group | It's not booting anymore | The computer with iufaid IUFAID0000000492 is broken. | 2010-04-19T02:27:32Z | EX |
| 6 | bill | Inventory Group | I want this computer | Please | 2010-04-19T02:35:19Z | EX |
| 7 | bill | Department of Biology | Transfer | Please transfer to room JB-301, Department of Sociology | 2010-04-19T02:46:17Z | RJ |
| 8 | mary | Inventory Group | Transfer this asset | Transfer this asset to room JB-301, Department of Sociology | 2010-04-19T02:51:55Z | EX |
| 9 | eric | Inventory Group | [object Object] | [object Object] | 2010-04-19T03:05:33Z | NE |
| 10 | eric | Inventory Group | my mouse doen't work | mouse | 2010-04-19T03:06:29Z | EX |
| 11 | Ali | Inventory Group | [object Object] | [object Object] | 2010-04-19T03:07:28Z | EX |
| 12 | eric | Inventory Group | idk | None available | 2010-04-19T19:20:09Z | RJ |
| 13 | phil | Inventory Group | For sound mixing... | [object Object] | 2010-04-19T19:30:31Z | EX |
| 14 | eric | Inventory Group | [object Object] | The IUFAID is 00000000408 | 2010-04-19T22:47:26Z | EX |
| 15 | bill | Inventory Group | Transfer | Transfer | 2010-04-19T23:26:37Z | EX |

< prev 1 next > 14 total records

# Assets by Location Report

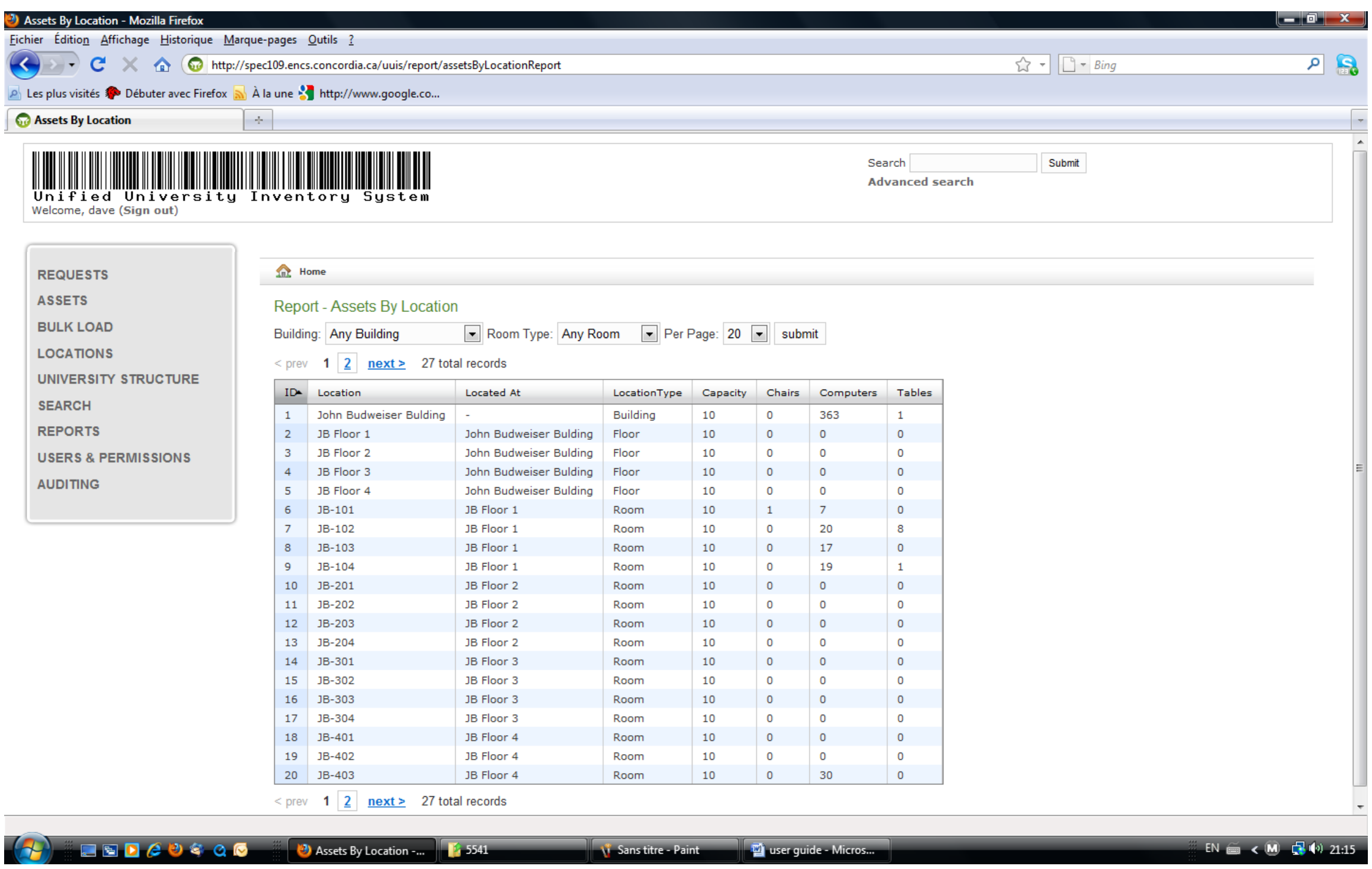

Assets By Location - Mozilla Firefox

http://spec109.encs.concordia.ca/uuis/report/assetsByLocationReport

Unified University Inventory System

Welcome, dave (Sign out)

Search Submit

Advanced search

REQUESTS
ASSETS
BULK LOAD
LOCATIONS
UNIVERSITY STRUCTURE
SEARCH
REPORTS
USERS & PERMISSIONS
AUDITING

Home

Report - Assets By Location

Building: Any Building Room Type: Any Room Per Page: 20 submit

< prev 1 2 next > 27 total records

| ID | Location | Located At | LocationType | Capacity | Chairs | Computers | Tables |
|---|---|---|---|---|---|---|---|
| 1 | John Budweiser Bulding | - | Building | 10 | 0 | 363 | 1 |
| 2 | JB Floor 1 | John Budweiser Bulding | Floor | 10 | 0 | 0 | 0 |
| 3 | JB Floor 2 | John Budweiser Bulding | Floor | 10 | 0 | 0 | 0 |
| 4 | JB Floor 3 | John Budweiser Bulding | Floor | 10 | 0 | 0 | 0 |
| 5 | JB Floor 4 | John Budweiser Bulding | Floor | 10 | 0 | 0 | 0 |
| 6 | JB-101 | JB Floor 1 | Room | 10 | 1 | 7 | 0 |
| 7 | JB-102 | JB Floor 1 | Room | 10 | 0 | 20 | 8 |
| 8 | JB-103 | JB Floor 1 | Room | 10 | 0 | 17 | 0 |
| 9 | JB-104 | JB Floor 1 | Room | 10 | 0 | 19 | 1 |
| 10 | JB-201 | JB Floor 2 | Room | 10 | 0 | 0 | 0 |
| 11 | JB-202 | JB Floor 2 | Room | 10 | 0 | 0 | 0 |
| 12 | JB-203 | JB Floor 2 | Room | 10 | 0 | 0 | 0 |
| 13 | JB-204 | JB Floor 2 | Room | 10 | 0 | 0 | 0 |
| 14 | JB-301 | JB Floor 3 | Room | 10 | 0 | 0 | 0 |
| 15 | JB-302 | JB Floor 3 | Room | 10 | 0 | 0 | 0 |
| 16 | JB-303 | JB Floor 3 | Room | 10 | 0 | 0 | 0 |
| 17 | JB-304 | JB Floor 3 | Room | 10 | 0 | 0 | 0 |
| 18 | JB-401 | JB Floor 4 | Room | 10 | 0 | 0 | 0 |
| 19 | JB-402 | JB Floor 4 | Room | 10 | 0 | 0 | 0 |
| 20 | JB-403 | JB Floor 4 | Room | 10 | 0 | 30 | 0 |

< prev 1 2 next > 27 total records

## Displaying Users List (IT Group Only)

1. Click Users &Permissions from Left side Menu

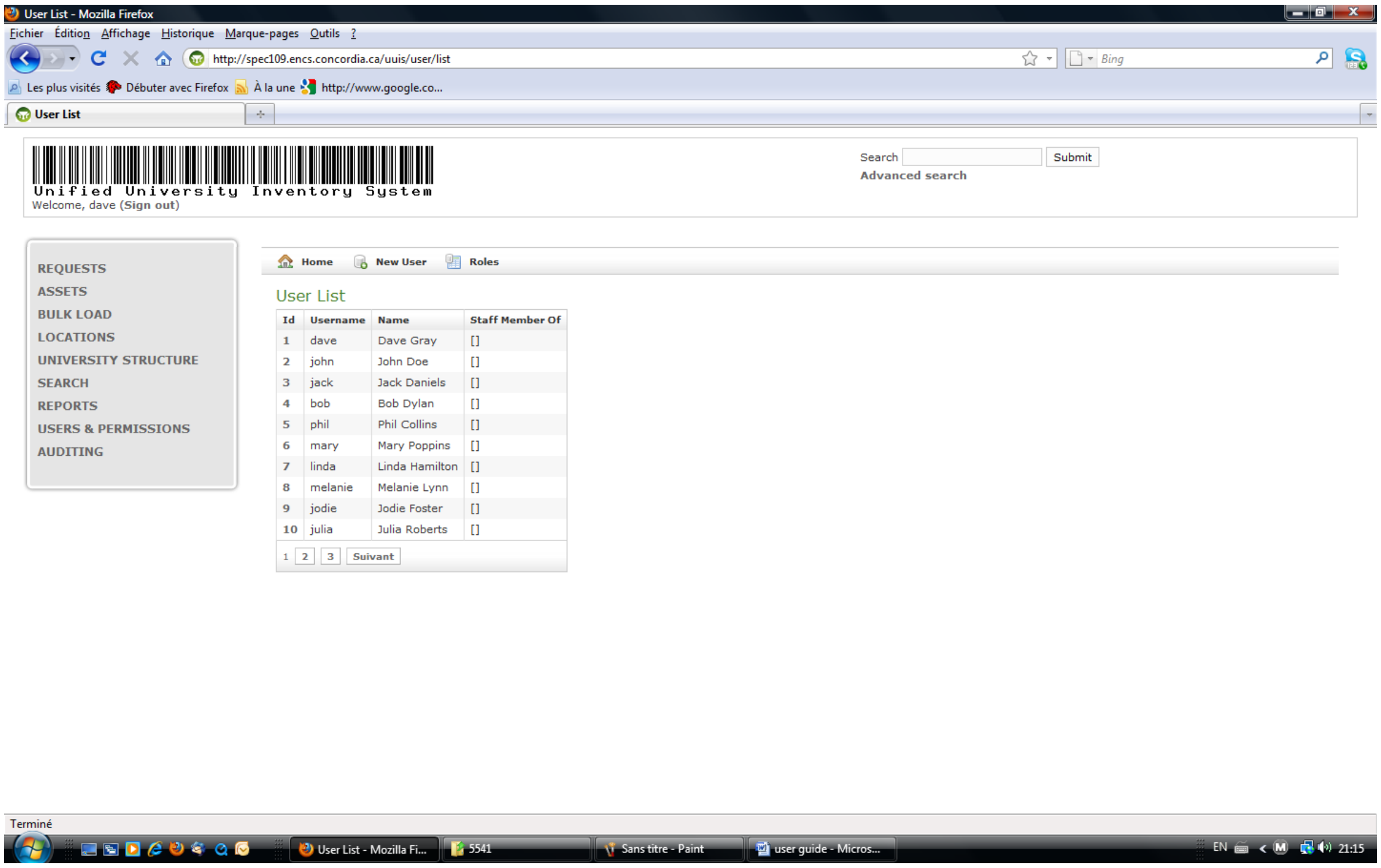

## Create a New user (IT Group only)

1. Click Users &Permissions from Left side Menu
2. Click New User
3. Fill user info and click Create

   **It is important to pay attention to the user assigned role because by assigning roles we assign permission to the users**

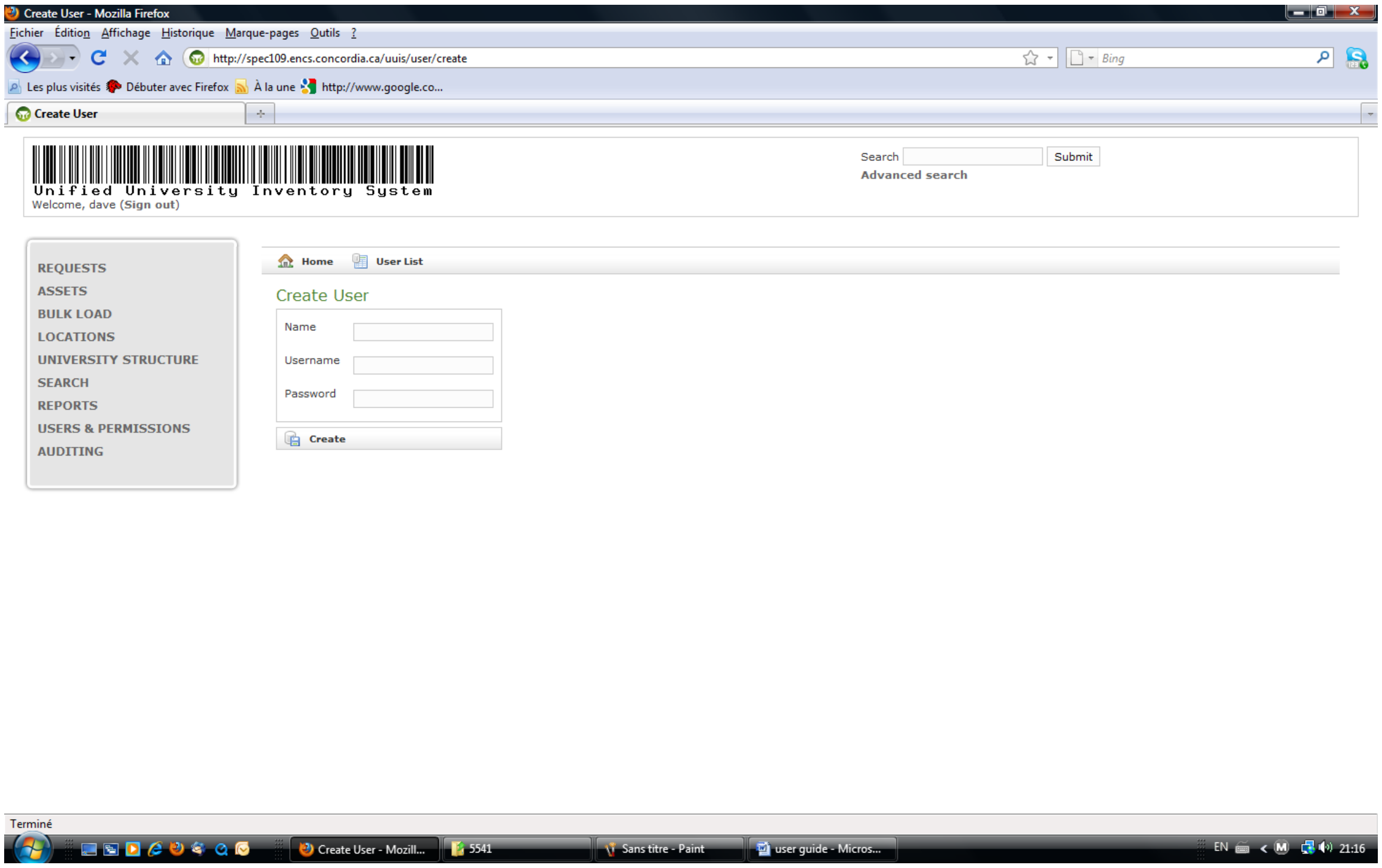

## Display User properties (IT Group only)

1. Click Users &Permissions from Left side Menu
2. Click on the User

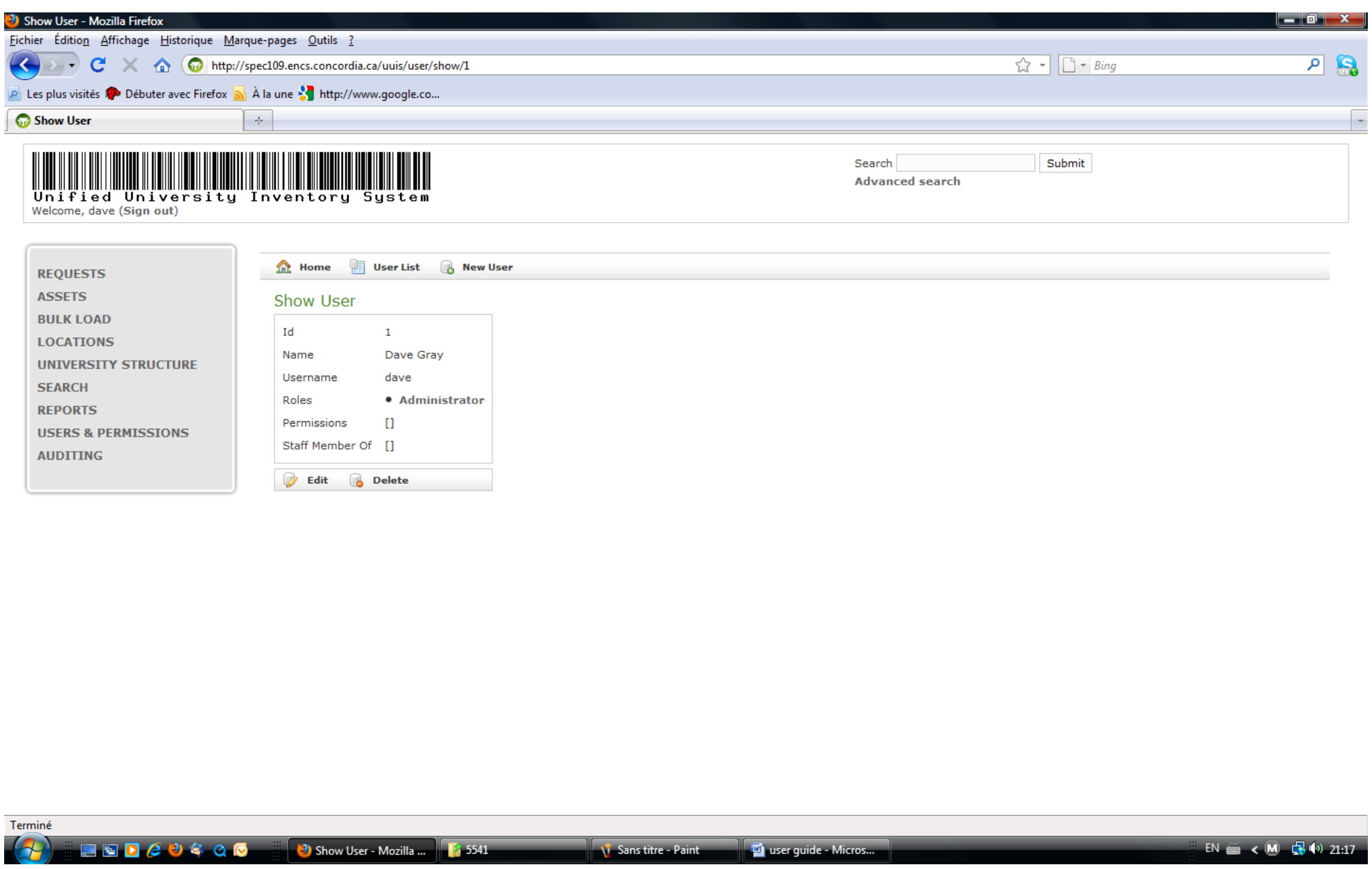

## Modifying user properties (IT Group only)

1. Click Users &Permissions from Left side Menu
2. Click the User
3. Fill user info and click Edit user
4. Change properties and click Update

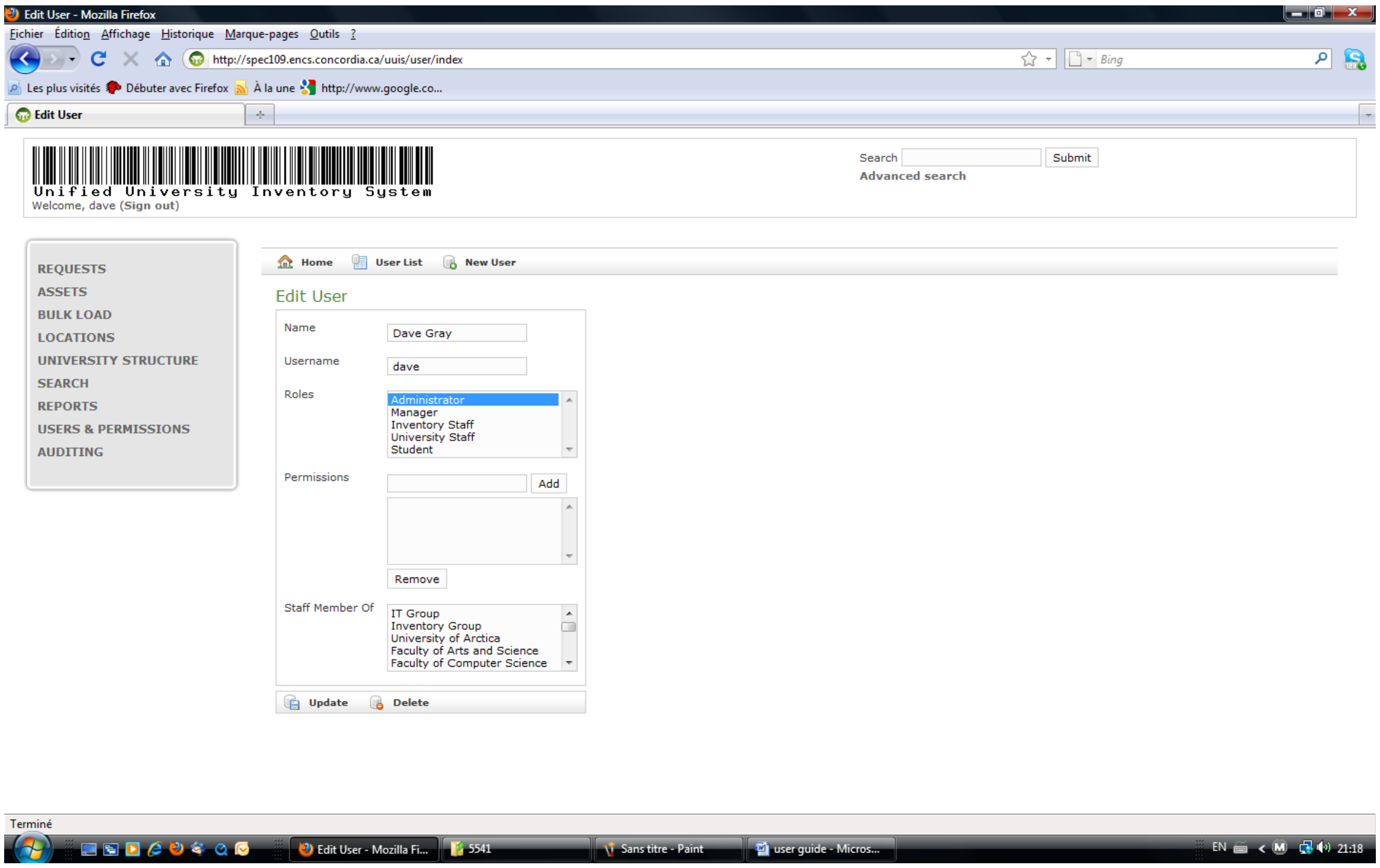

## Delete user (IT Group only)

1. Click Users &Permissions from Left side Menu
2. Click the User
3. Fill user info and click on Delete user
4. Confirm

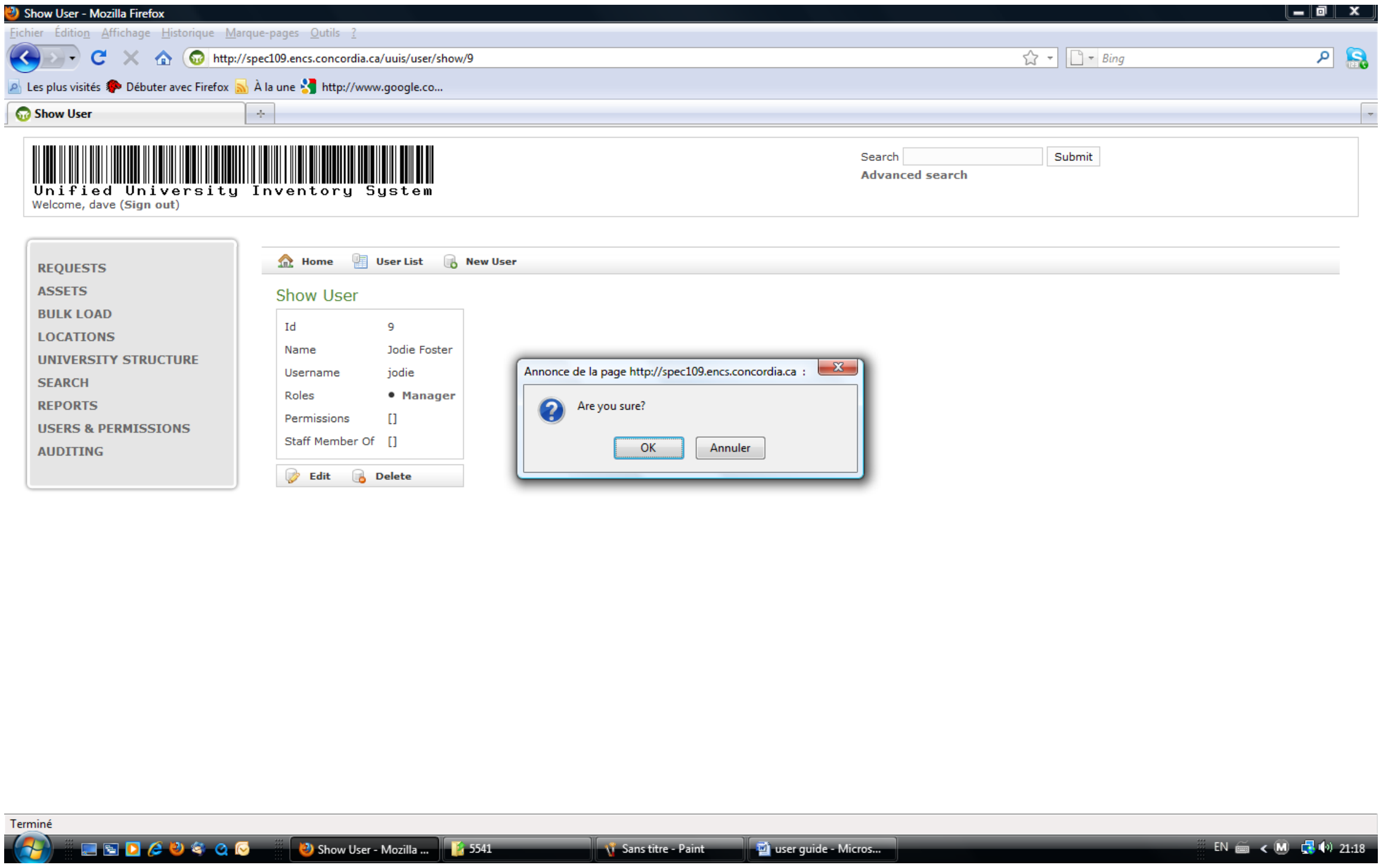

## Displaying the List of Roles List (IT Group Only)

1. Click Users &Permissions from Left side Menu
2. Click Roles

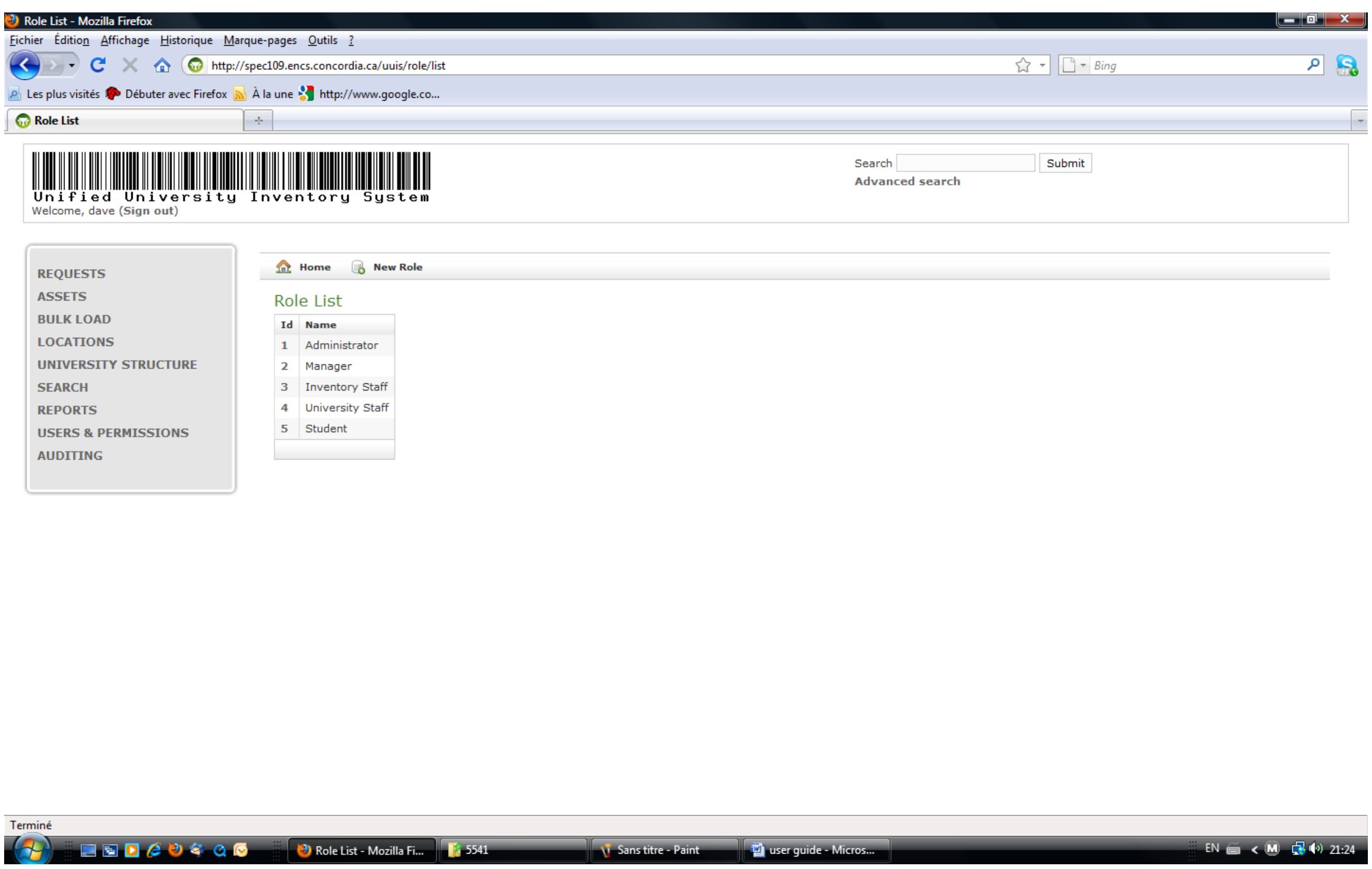

## Create a New Role (IT Group Only)

1. Click Users &Permissions from Left side Menu
2. Click Roles
3. Click create
4. Assign permissions
5. Click Create

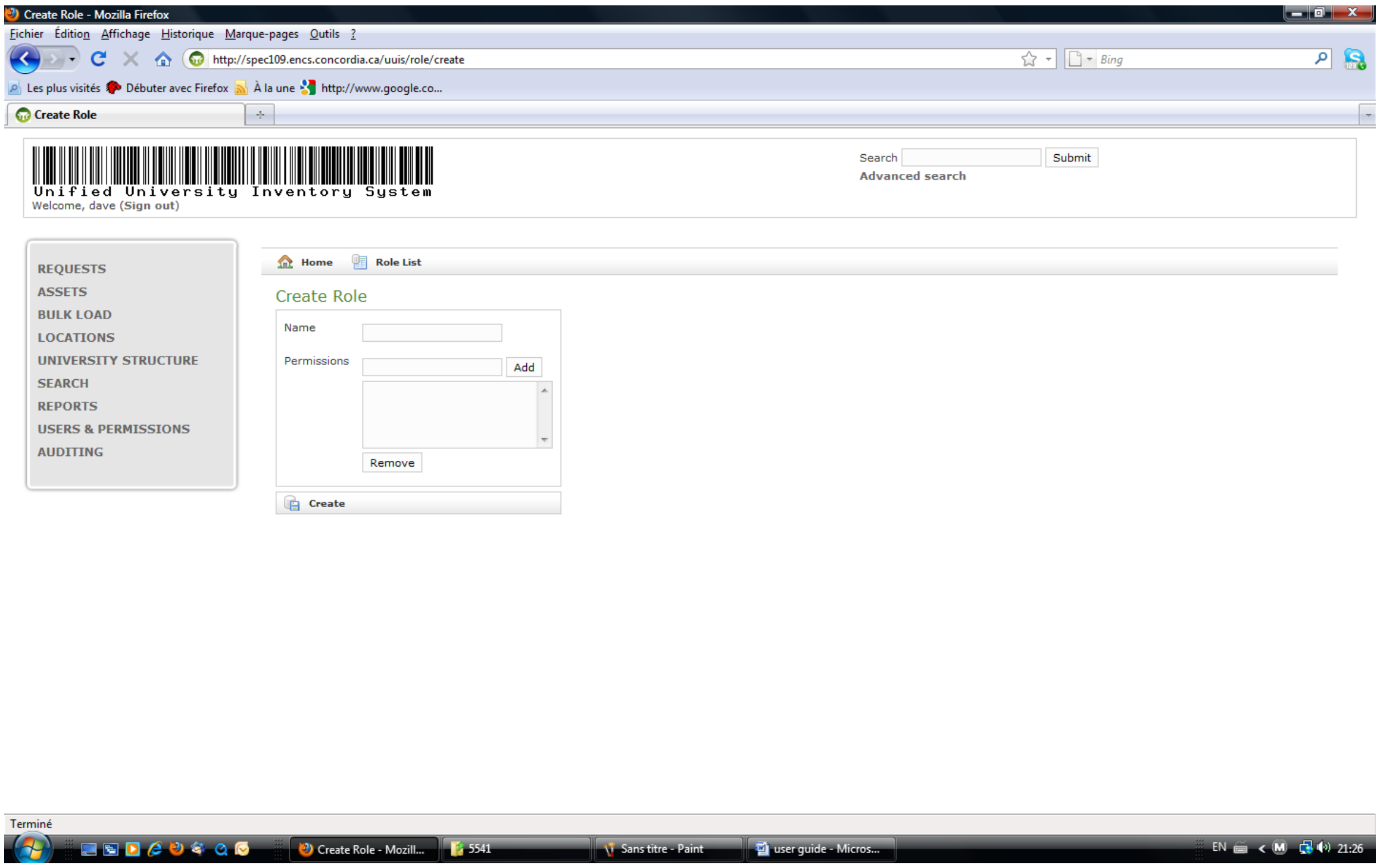

## Display List of Users By Role (IT Group Only)

You can display all the administrators for examples

1. Click Users &Permissions from Left side Menu
2. Click Roles
3. Click Role List

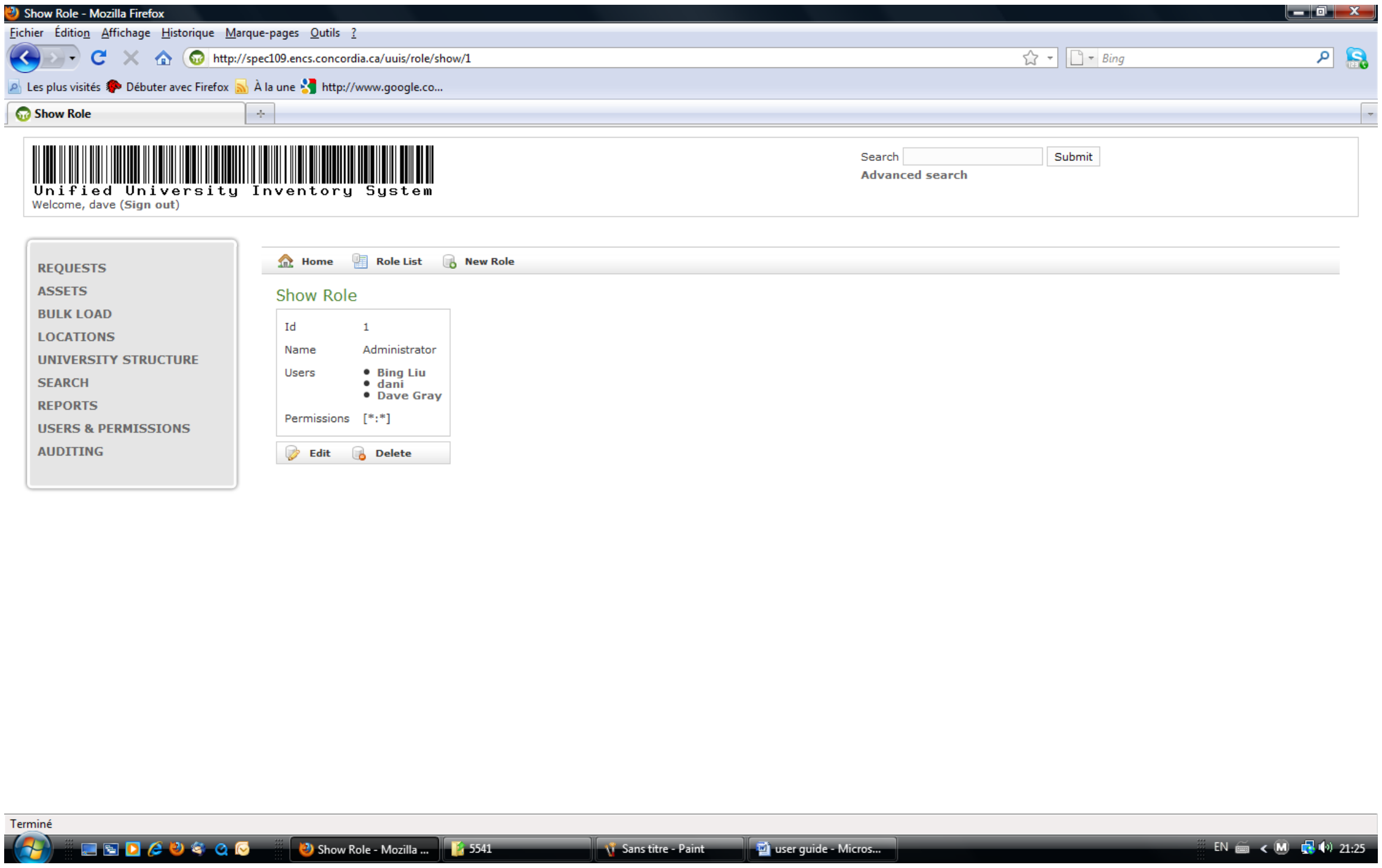

## Modify Role Properties (IT Group Only)

1. Click Users &Permissions from Left side Menu
2. Click Roles
3. Edit role
4. Change properties and click  Update

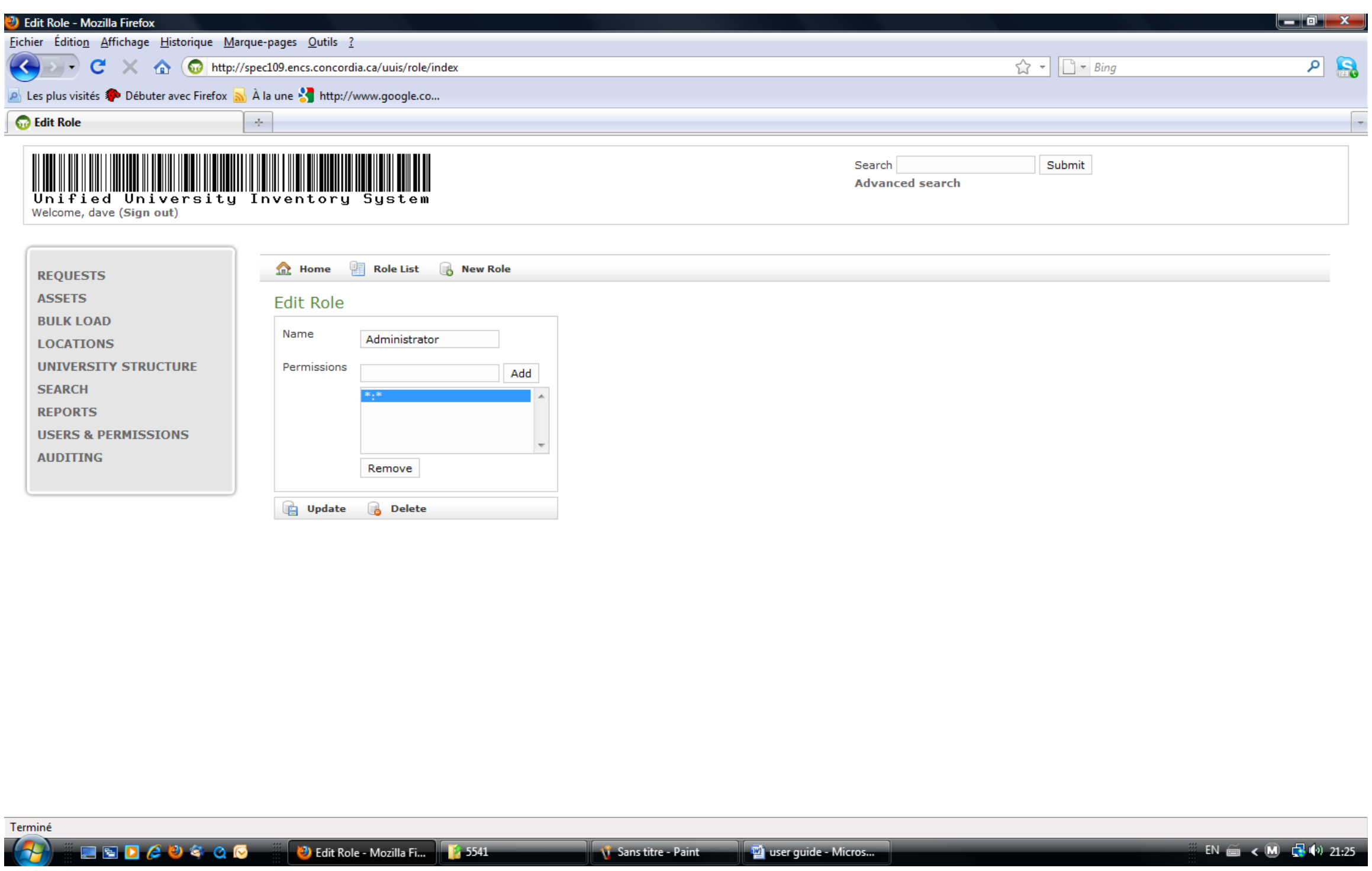

## Delete Role (IT Group only)

1. Click Users &Permissions from Left side Menu
2. Click Roles
3. Click Delete and confirm

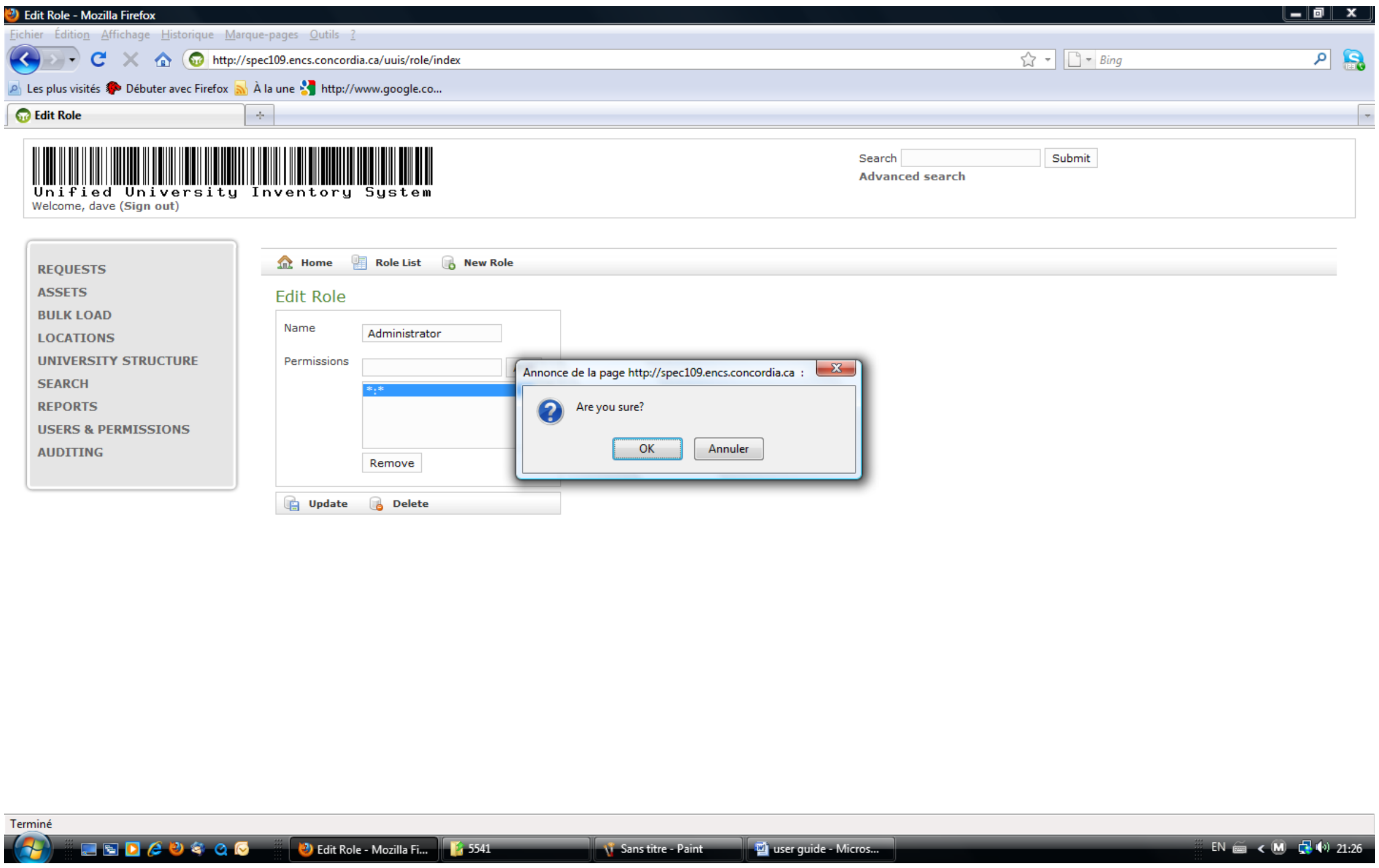

## Display Audit Report

1. Click Auditing from the left Menu

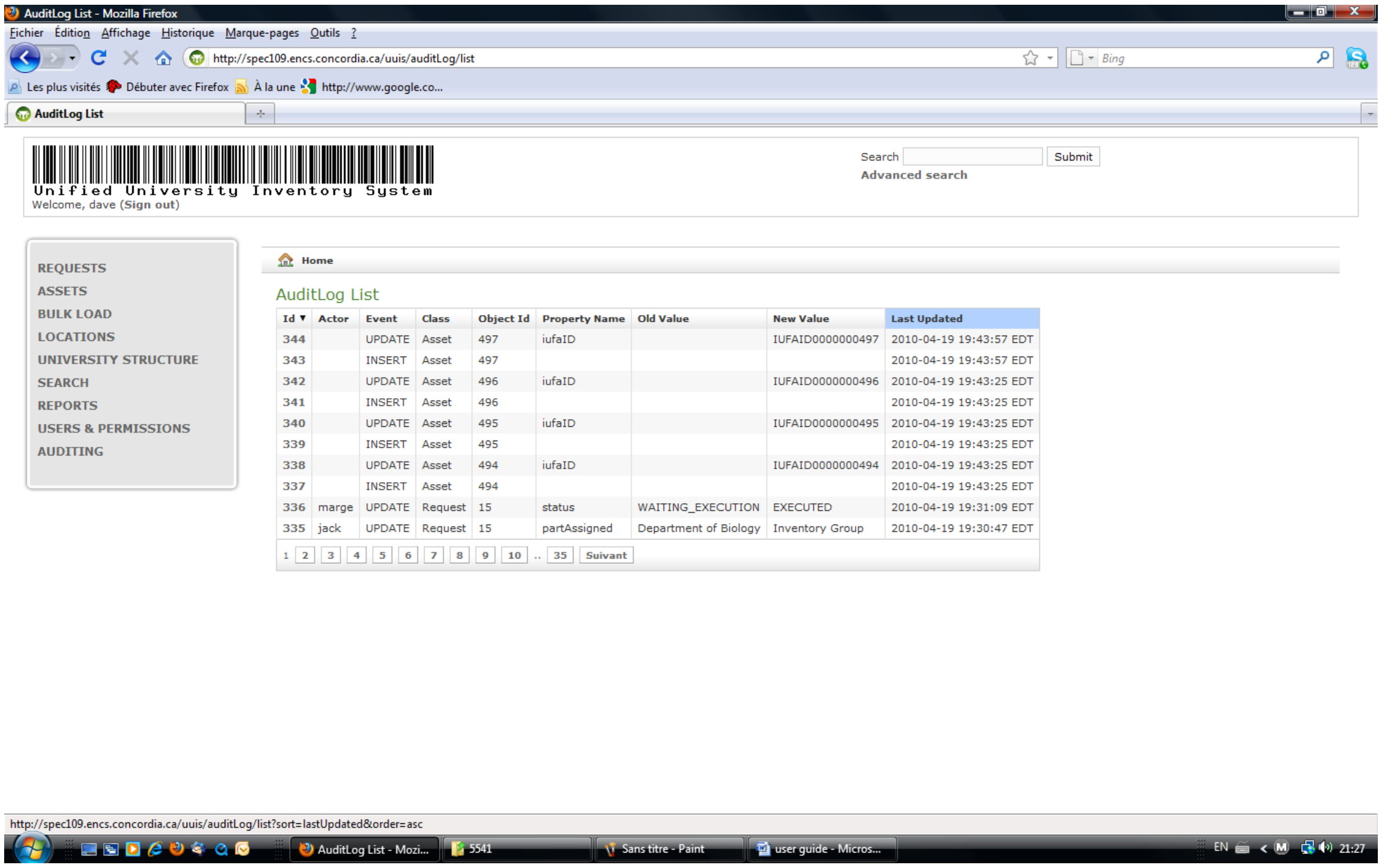
AuditLog List - Mozilla Firefox
Fichier Édition Affichage Historique Marque-pages Outils ?
http://spec109.encs.concordia.ca/uuis/auditLog/list
Les plus visités Débuter avec Firefox À la une http://www.google.co...
AuditLog List
Unified University Inventory System
Welcome, dave (Sign out)
Search Submit
Advanced search
REQUESTS
ASSETS
BULK LOAD
LOCATIONS
UNIVERSITY STRUCTURE
SEARCH
REPORTS
USERS & PERMISSIONS
AUDITING
Home
AuditLog List
Id Actor Event Class Object Id Property Name Old Value New Value Last Updated
344 UPDATE Asset 497 iufaID IUFAID0000000497 2010-04-19 19:43:57 EDT
343 INSERT Asset 497 2010-04-19 19:43:57 EDT
342 UPDATE Asset 496 iufaID IUFAID0000000496 2010-04-19 19:43:25 EDT
341 INSERT Asset 496 2010-04-19 19:43:25 EDT
340 UPDATE Asset 495 iufaID IUFAID0000000495 2010-04-19 19:43:25 EDT
339 INSERT Asset 495 2010-04-19 19:43:25 EDT
338 UPDATE Asset 494 iufaID IUFAID0000000494 2010-04-19 19:43:25 EDT
337 INSERT Asset 494 2010-04-19 19:43:25 EDT
336 marge UPDATE Request 15 status WAITING_EXECUTION EXECUTED 2010-04-19 19:31:09 EDT
335 jack UPDATE Request 15 partAssigned Department of Biology Inventory Group 2010-04-19 19:30:47 EDT
1 2 3 4 5 6 7 8 9 10 .. 35 Suivant
http://spec109.encs.concordia.ca/uuis/auditLog/list?sort=lastUpdated&order=asc
AuditLog List - Mozi... 5541 Sans titre - Paint user guide - Micros...
EN 21:27

# Appendix II: Configuration and Deployment Document

## Windows Server Configuration

Install Windows Server 2008 using default configuration.

Open the Windows Firewall Configuration and select Inbound Rules.

Click New Rule. In the window that opens select "Port" and click Next.

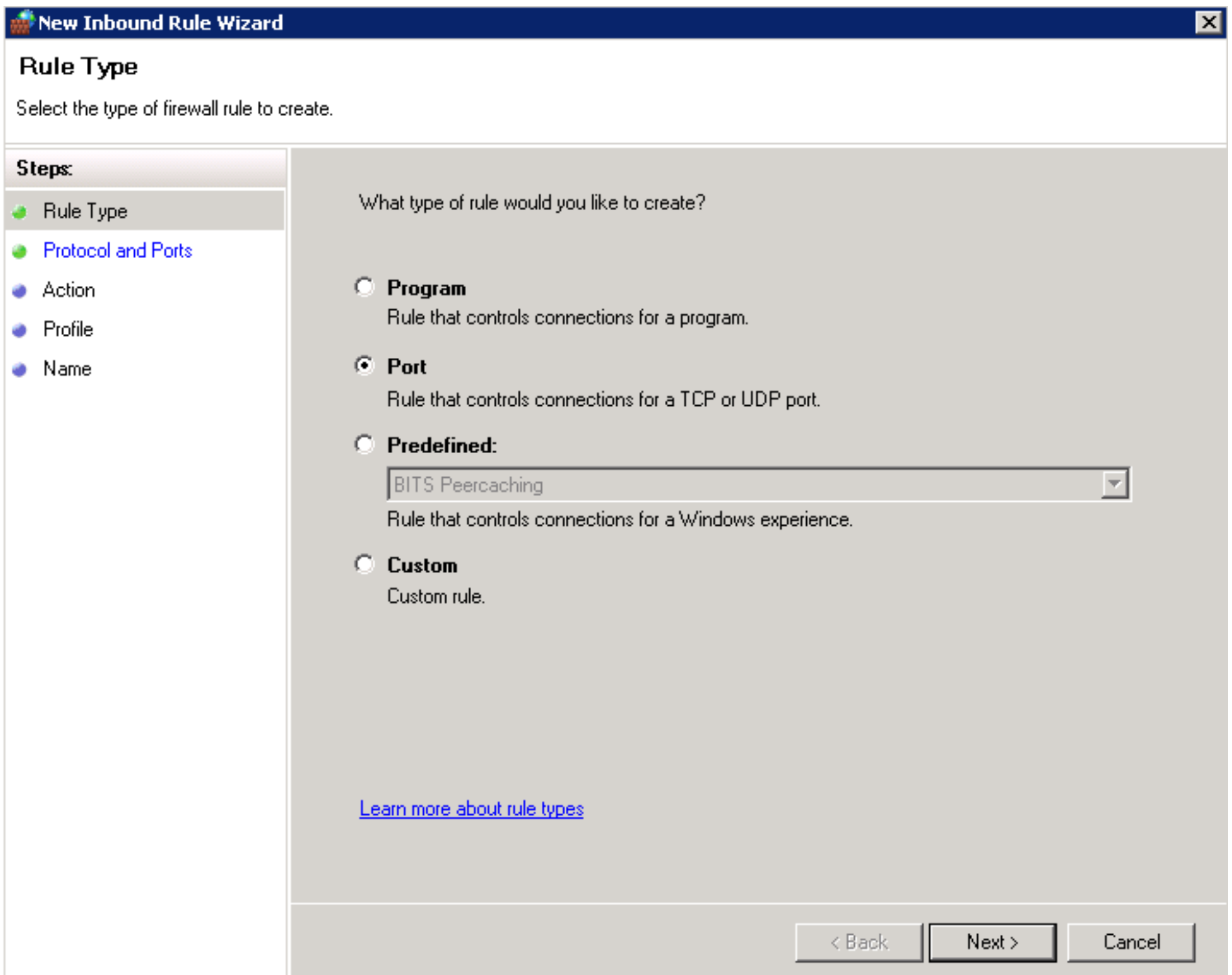

Select "Specific local ports" and enter 80 in the text box.

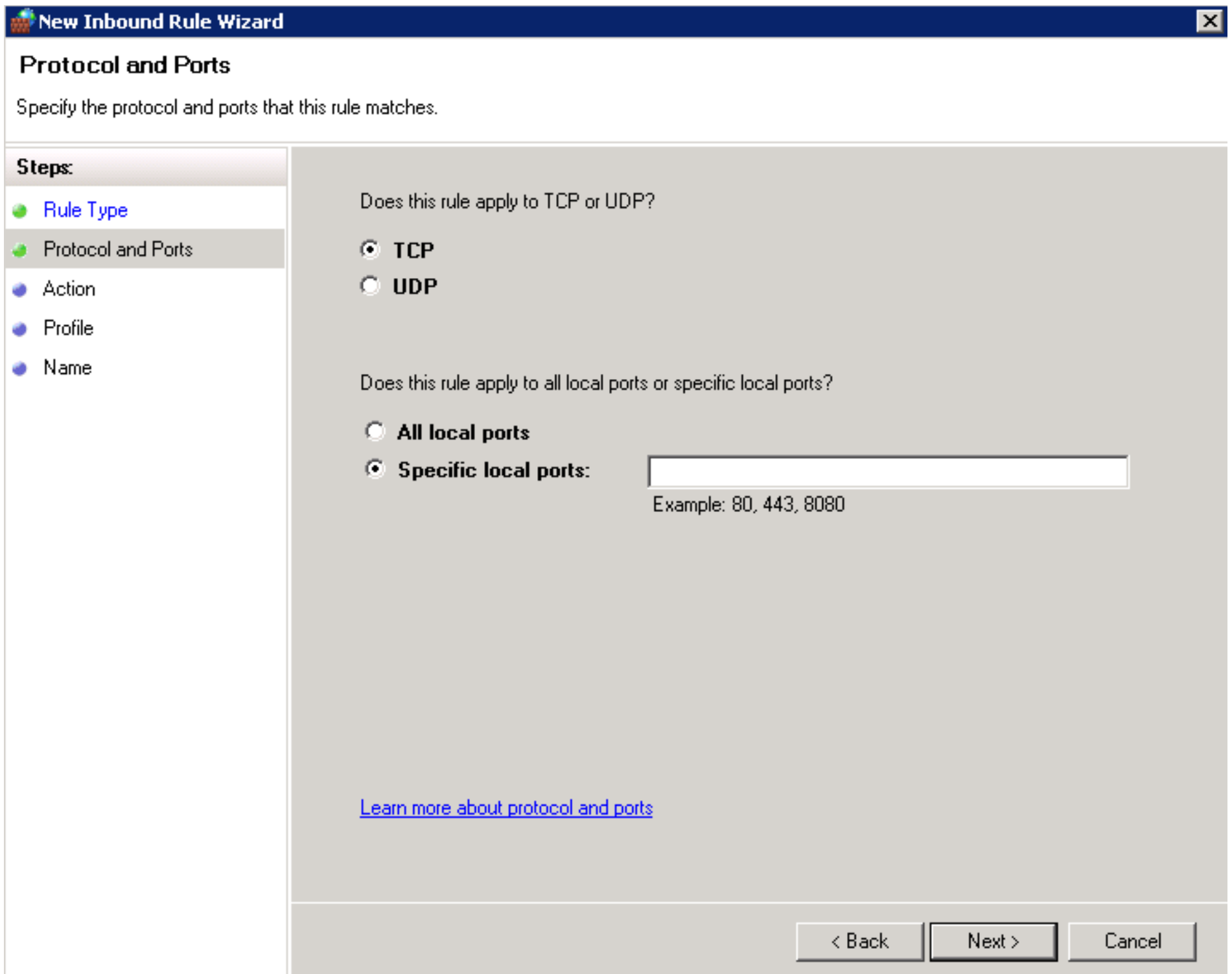

Leave the default "Allow the connection" enabled and click Next.

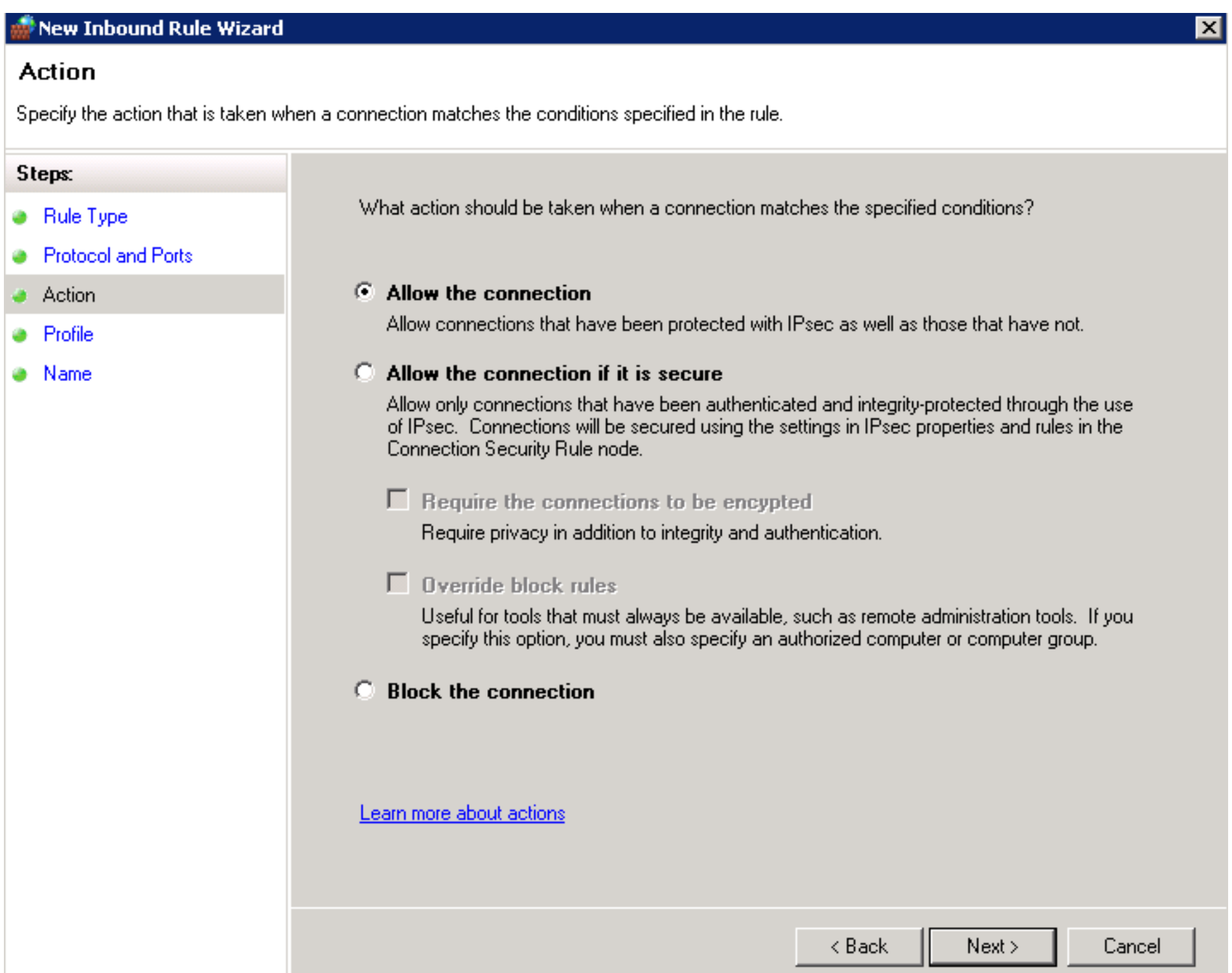

Apply the rule to all three areas.

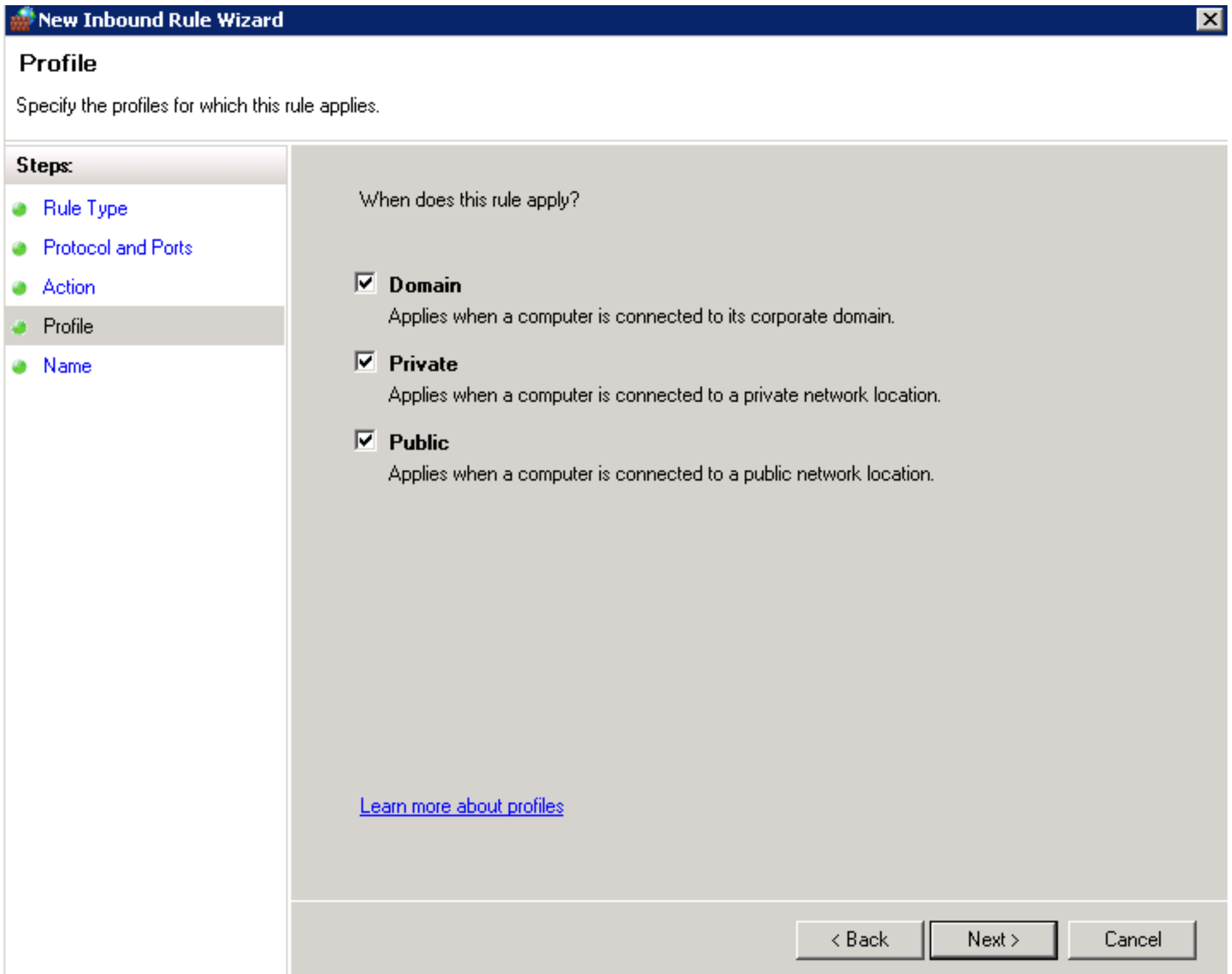

Give the rule a name like "Web Server"  and click Finish to complete the firewall setup.

## Apache Configuration

Download XAMP v1.7.3 from here. Install with default options. Also install the Tomcat 6.0.20 plugin available from here. Default options work but you may specify an alternate install path if you prefer.

After installation XAMPP should run automatically. Click the Start button beside Apache, MySql, and Tomcat.

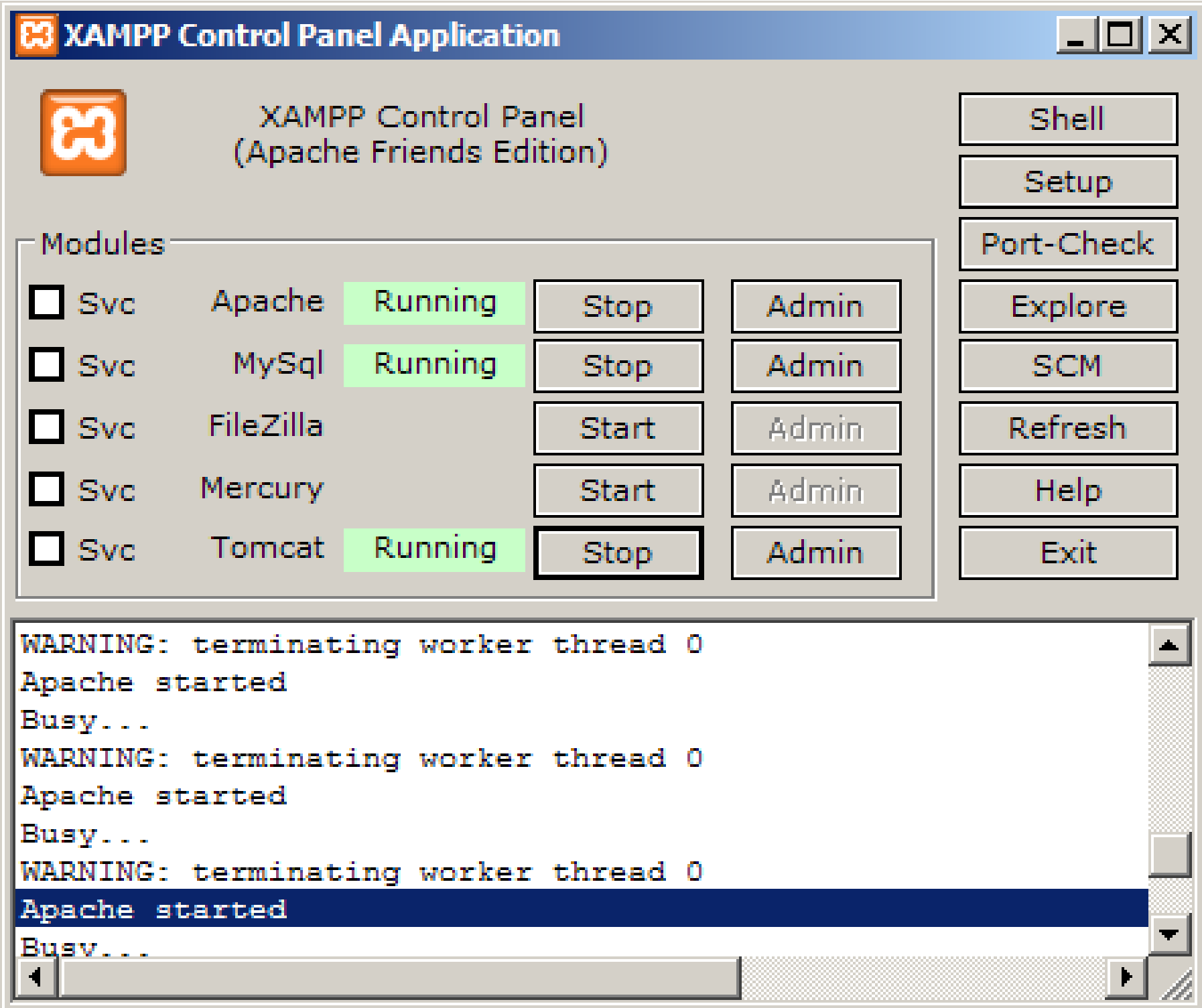

To run these servers as Windows services, click the empty checkboxes on the left side under the Modules heading.

## MySQL Configuration

MySQL and MySQL WorkBench were installed using default configuration.

1. Create default connection (here, we use localhost)

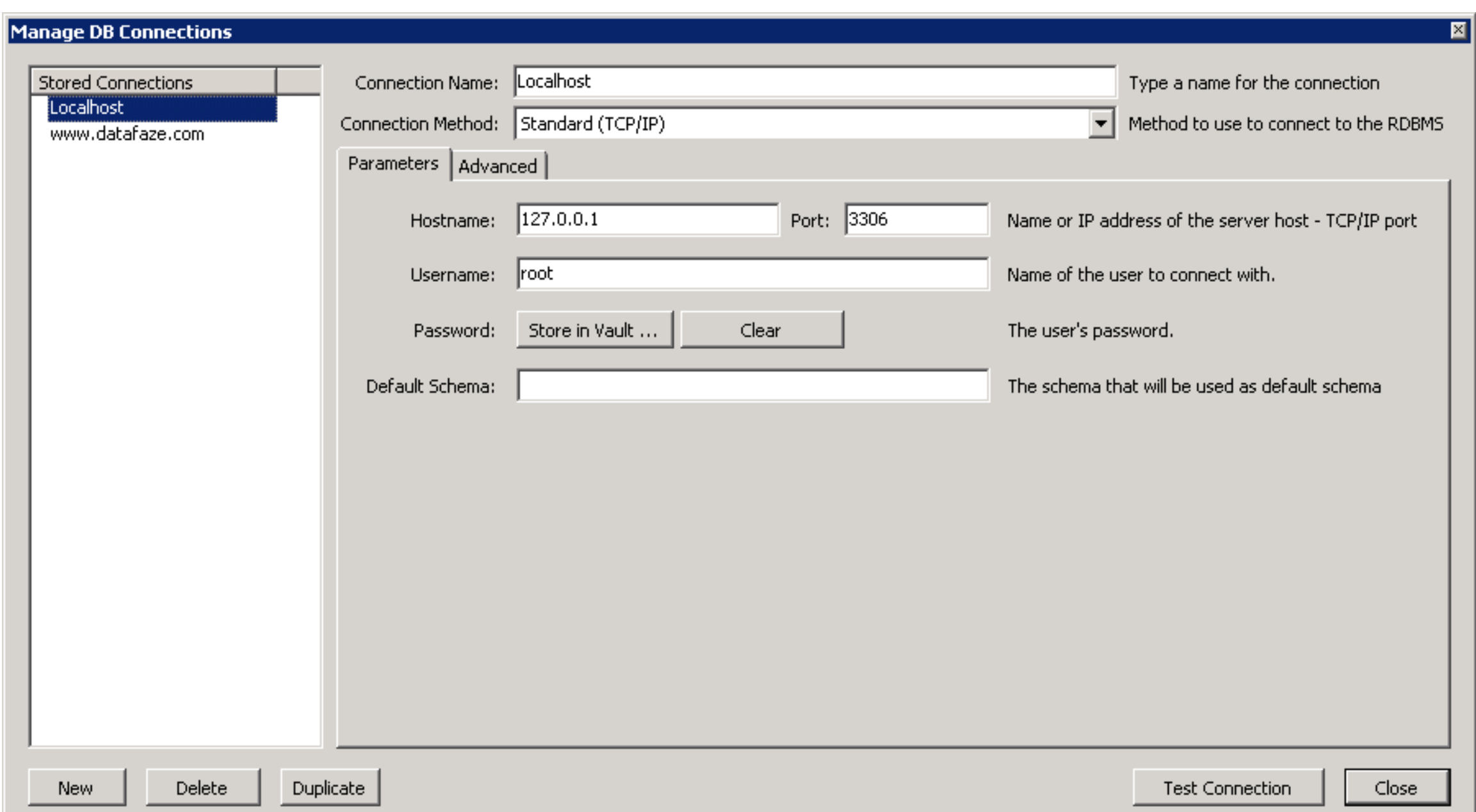

2. Create a new database: uuis_team4, click on apply

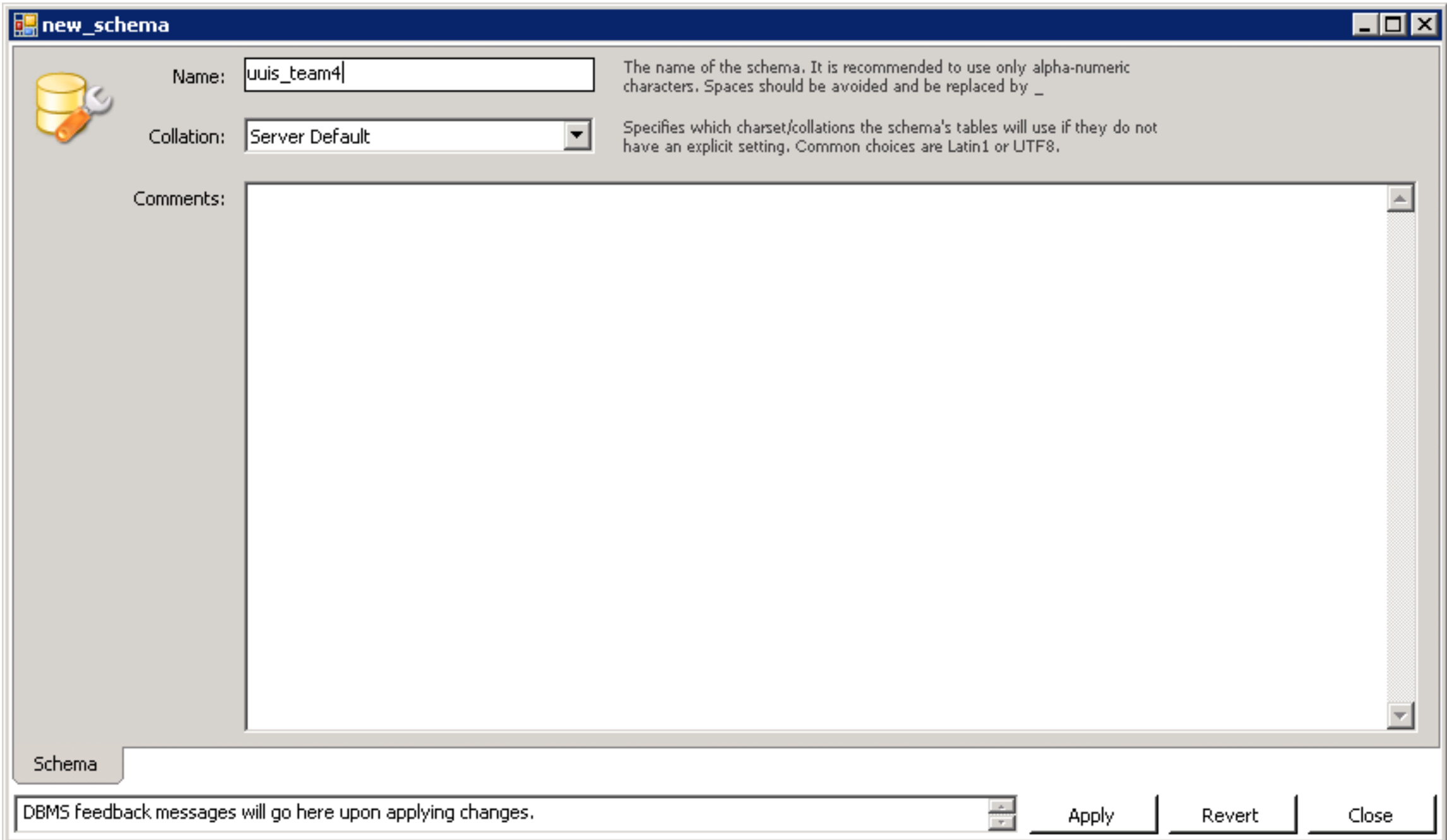

3. Open the database schema script at <source path root>\uuis\scripts\DBSchema.sql from the query editor window and run it. The tables and some initial data will be generated.

## Grails Configuration

1. Download Grails 1.2.1. Do it by clicking here.
2. Unpack the zip file to your computer, I'm unpacking it to G:\ as I'm in one of the labs now, but you can put it in you C:\
3. Add the Grails path to your environment variables. Right click My Computer, then click Properties, then the Advanced tab, then the Environment Variables button. Under System variables (if you're in the labs, under User variables) click New and add the values:

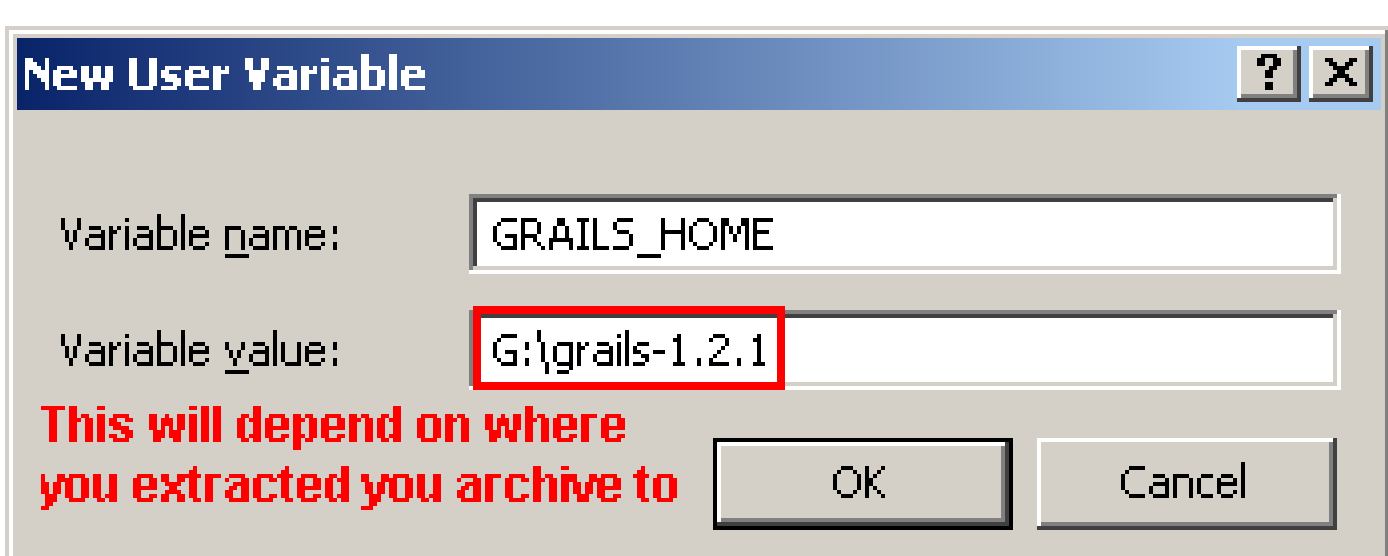

4. Still under System variables (or User variables), find Path in the list, and edit it adding %GRAILS_HOME%\bin

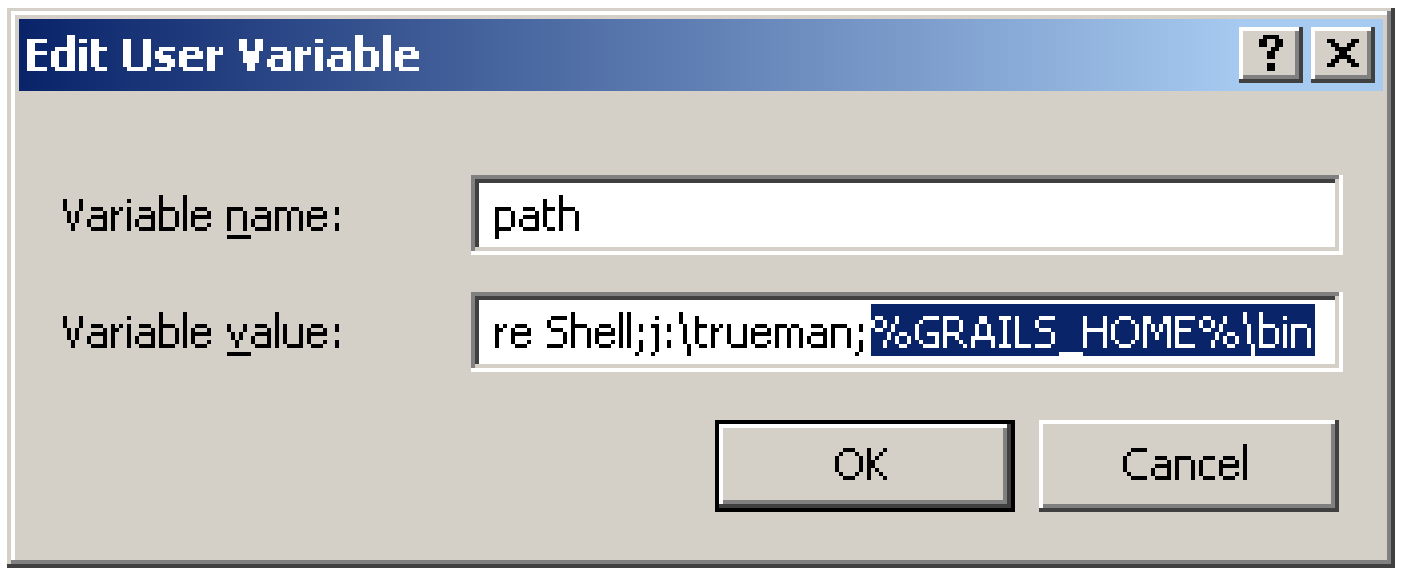

Click OK to close the windows.

5. Let's test if it works, open a command line and type "grails -version". You should see something like this:

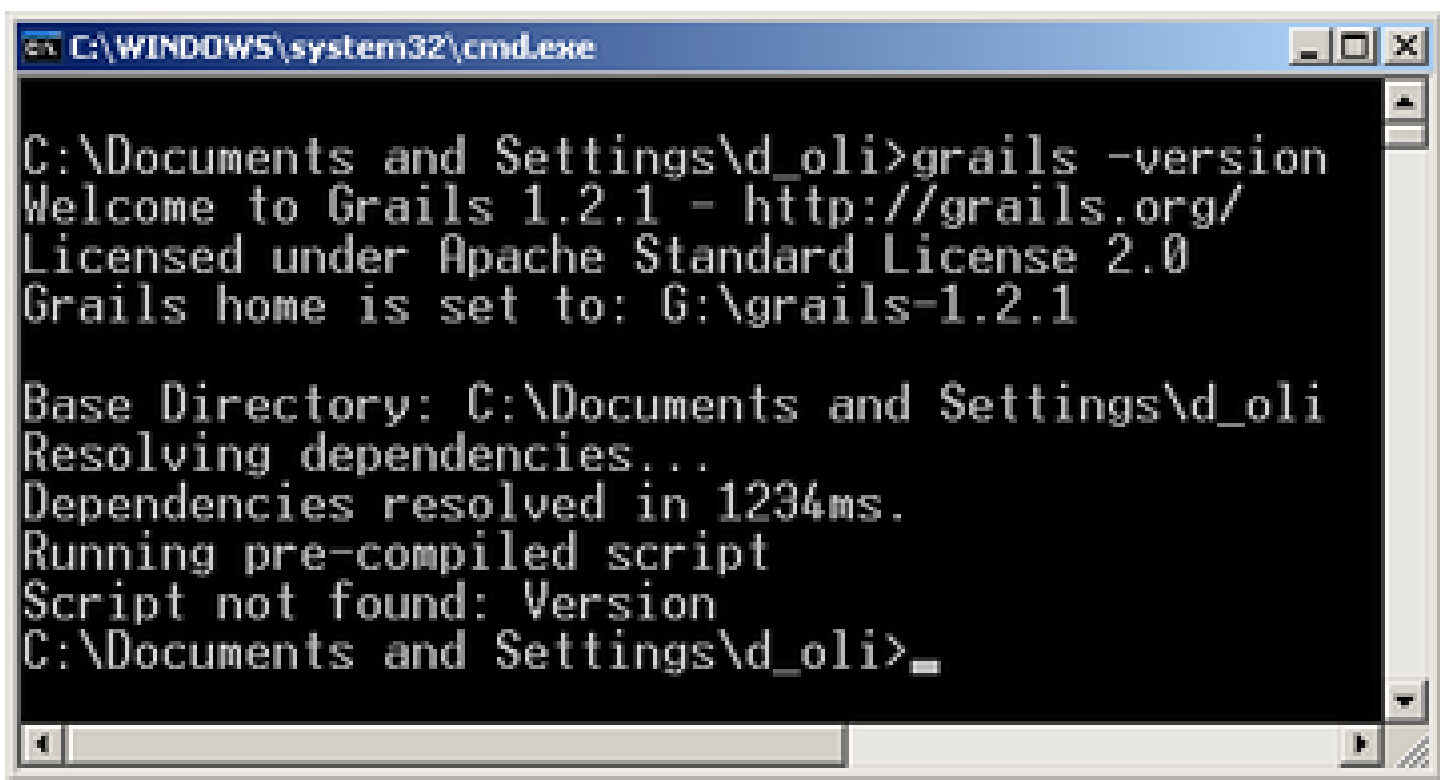

## NetBeans Configuration

1. Download NetBeans 6.8 [here](#) and install it (this version already ships with Grails support, but if you're in the labs chances are that the version 6.7.1 is installed instead, all you'll have to do is go to Tools > Plugins > Available Plugins, select Groovy and Grails from the list and install it).
2. Now let's check-out our project into NetBeans. Click the menu Team, Subversion, and then Checkout... If this is the first time you're using Subversion in NetBeans, it will ask you to download a Subversion client.

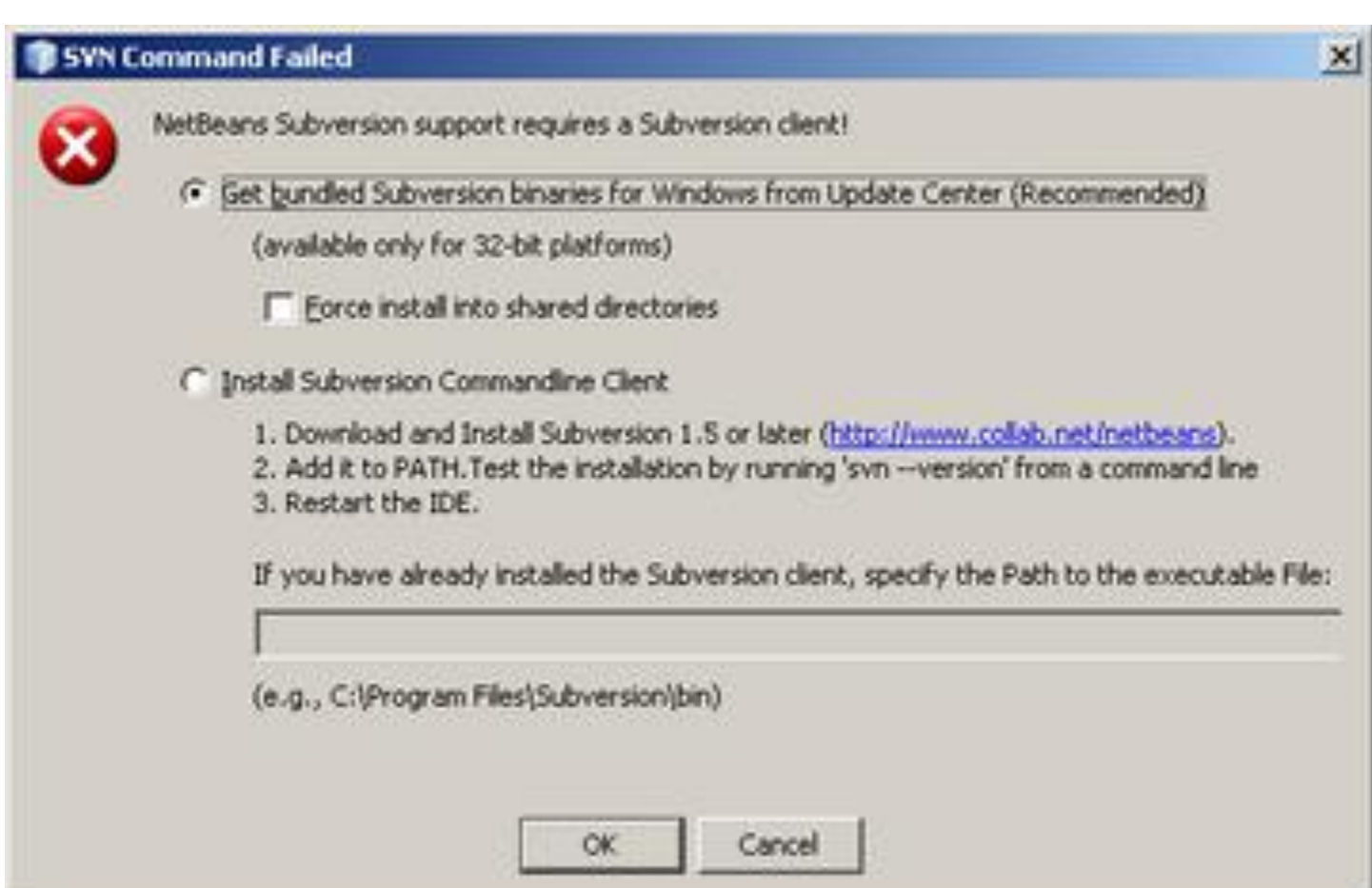

Just click OK that NetBeans take care of it for you.

3. After you restarted NetBeans go back to the menu Team, Subversion, and then Checkout...
4. In Repository URL, provide our repository which is [https://comp5541-team4.svn.sourceforge.net/svnroot/comp5541-team4](https://comp5541-team4.svn.sourceforge.net/svnroot/comp5541-team4)
5. Fill in your SourceForge user and password, click Save Username and Password if you want, then click Next.

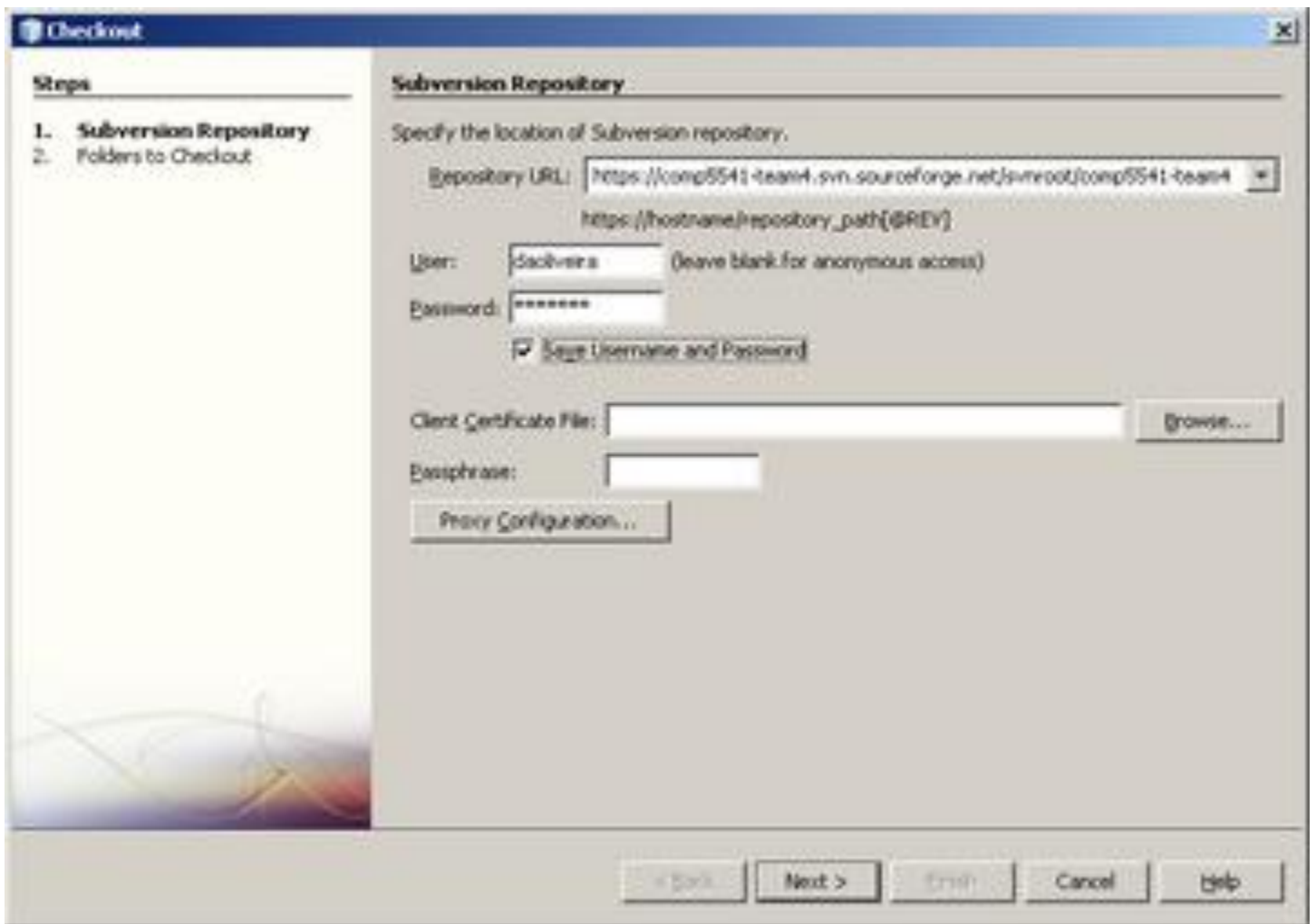

6. In Repository Folder(s) provide app/uuis
7. You can specify a destination (Local Folder), mine is C:\TEMP
8. Click Finish

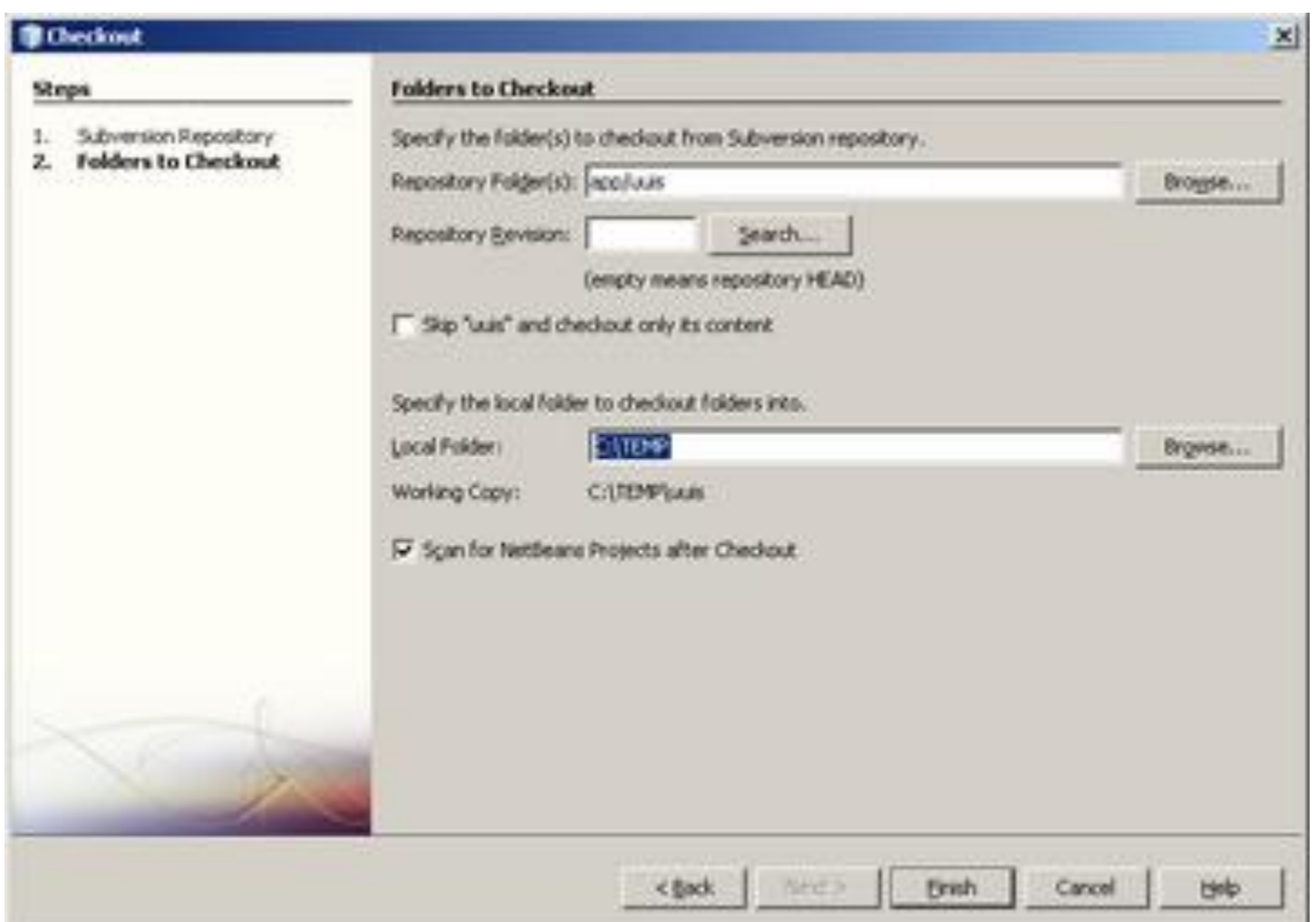

9. When the Checkout finishes, a confirmation dialog is show. Just hit Open Project and we're ready to go!

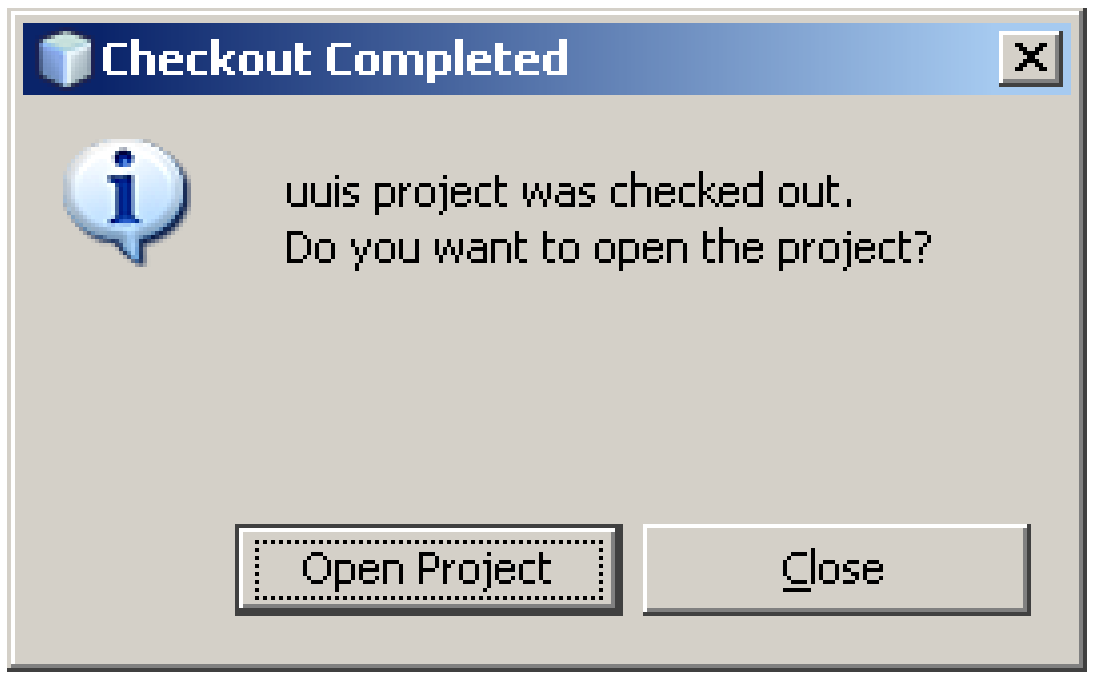

10. To test the project, right click on it then select Run

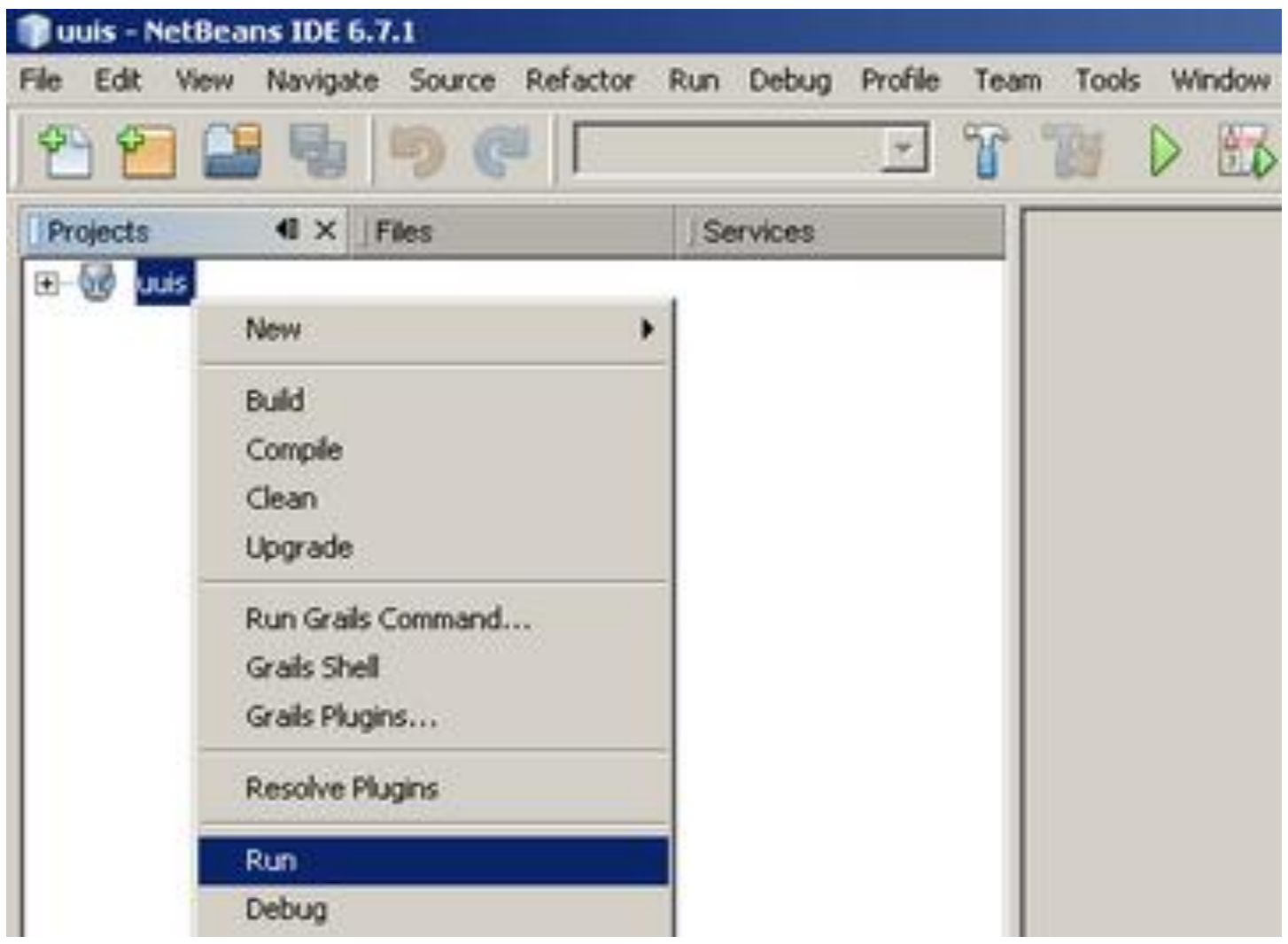

11. After running some build script, a web browser pops up with our application in it

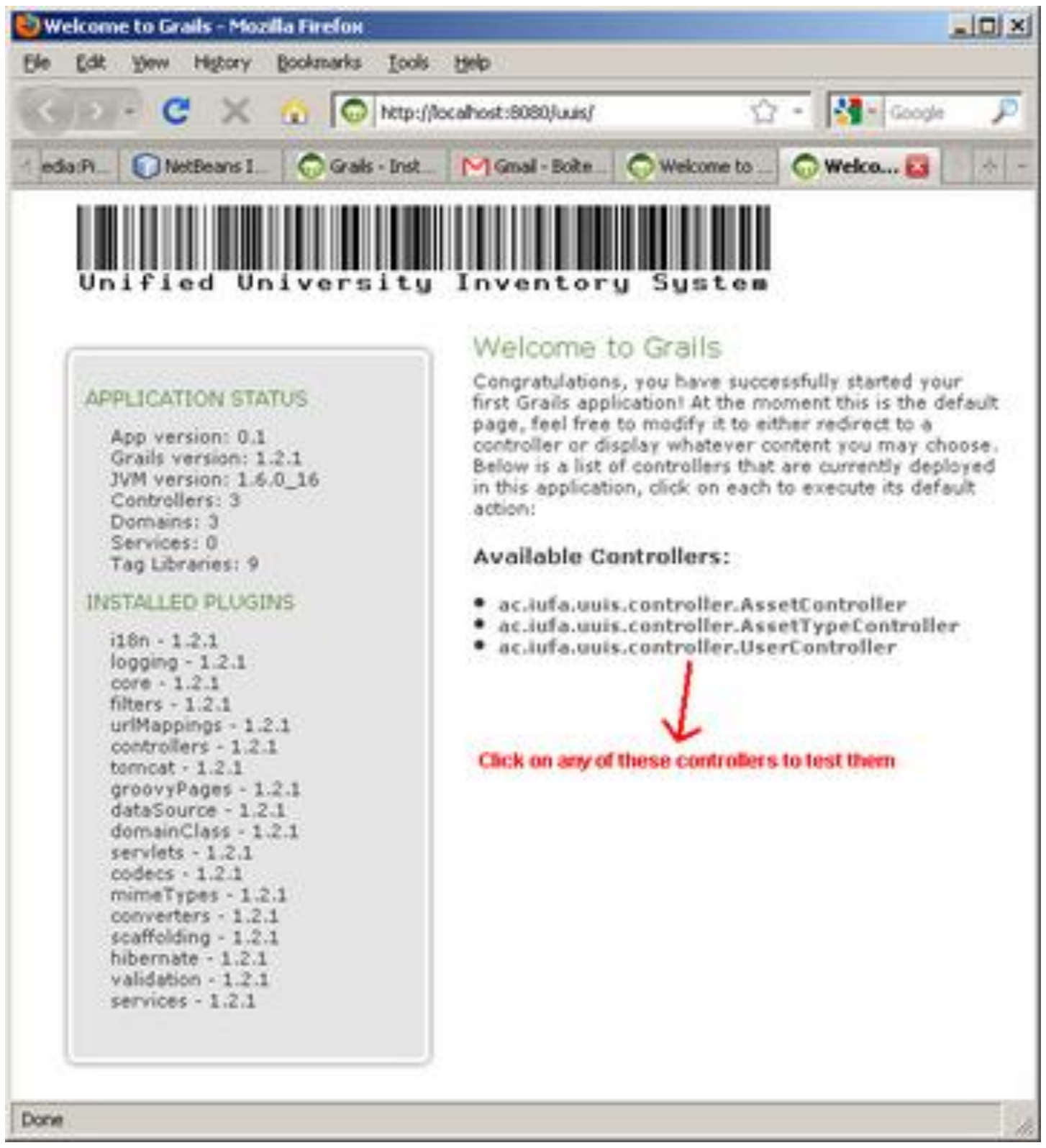

## Application Deployment

To build the application for deployment, database connection information is required in <source path root>/app/uuis/grails-app/conf/DataSource.groovy. The default database type is MySQL, although PostgreSQL could be used with minimal changes. Refer to the Grails documentation for details on specifying database connection drivers.

The DataSource.groovy configuration file requires a database username and password on lines 3 and 4. Alternate data sources for development and production systems can be specified in the DataSource configuration file as well. Refer to the Grails documentation for more details.

To deploy the application type “grails war” from the command line to create a Web Application Archive (WAR) file. With Apache and Tomcat running, browse to the Tomcat server management page and upload the WAR file to deploy it.

# Appendix III: Test cases

**Testing Goal**

The goal of Unified University Inventory System Testing is to ensure that the system performs as per the functional requirements specified by client.

**Testing Tools**

Tested by Rational Quantify Tool

Most cases tested by manual per item.

**Functional Requirements Testing (Black box testing)**

## Code Inspection

| Test case | Input | Test Description | Output / Result |
|---|---|---|---|
| code | gets(str) | Description of the procedures and modify Notes | Return=0 |
| process, function command | lenth = Len(Text1.Text) | Variable, procedure, function command line with the rules | length = 0 |
| Variable, | time1 = Val(Label1.Caption) | Variable, procedure, function command line with the rules | Label1.Caption = 0 |
| Modifying Notes | IUFAID0000000483 | Modify the Notes meets the requirements | Table |
| Class library | jack | Meet the requirements to use the class library | Jack Daniel |
| Request | Requests waiting for approval | Procedures does not meet the format requirements named | No requests available |
| home | My Requests | Screen and report format of the required demand | No requests available |

## Search Testing

| Test Area | Input | Test Description | Output / Result |
|---|---|---|---|
| Search for numbers | Search='10' | Display all the information. | tested |
| Search for a decimal | Search="10.08" | Display no results match | tested |
| Search for an expression | Search="software engineering" | Display all the information or results list | tested |
| Search for an expression of 1024 characters | Search="computer ... science" | Display results list for 1023 characters . | tested |
| Search field contains only spaces | Search="    " | Display all and any information | tested |
| Empty field | Search=" " | Display No results match your criteria | tested |
| Search for a key word | Search="computer " | Display all the information | tested |

## Advanced Search Testing

| Test Area | Input | Test Description | Output / Result |
|---|---|---|---|
| All Search field contains only spaces | Search=" " | Display any all information | tested |
| Search for key word | SearchBoxes1,2,3,or 4="computers" | Display all the information and results | tested |
| Search for sentence or phrase | Searchboxes1,2,3,or4= "computer science" | Display all the information and results | tested |
| If all fields are empty | Search table1to4="" | Display all the information | Tested |
| Using different logical combination of 2 or more fields | Searchboxes1="xyz" and Searchboxes2"wxyz" or Searchboxes3="uvxyz " | May display all the information and results list | tested |
| One or more Search for an expression of 1024 characters | Searchboxes1,2,3,or4= "computer ... programs" | Display all the information and results for 1023 characters | tested |
| One or more Search for a numbers | Searchboxes1,2,3,or4= "10 " | Display all the information and results list | tested |
| One or more Search for a decimal | Searchboxes1,2,3,or4= " 10.08 " | Display no results match | tested |
| One or more Search for an expression of 1024 characters | Searchboxes1,2,3,or4= "software ... engineering" | Display all the information and results for 1023 characters | tested |

**Non- Functional Requirements Testing**

Special methods exist to test non-functional aspects of software.

## Stability testing

Stability testing checks to see if the software can continuously function well in or above an acceptable period. This activity of non-functional software testing is oftentimes referred to as load (or endurance) testing.

## Usability

Usability testing is needed to check if the user interface is easy to use and understand.

This test is used to verify if a user that never use the application is able to search and read result list within a reasonable time.

## Security testing

Security testing is essential for software which processes confidential data and to prevent system intrusion by hackers.

# Appendix IV: Bug List

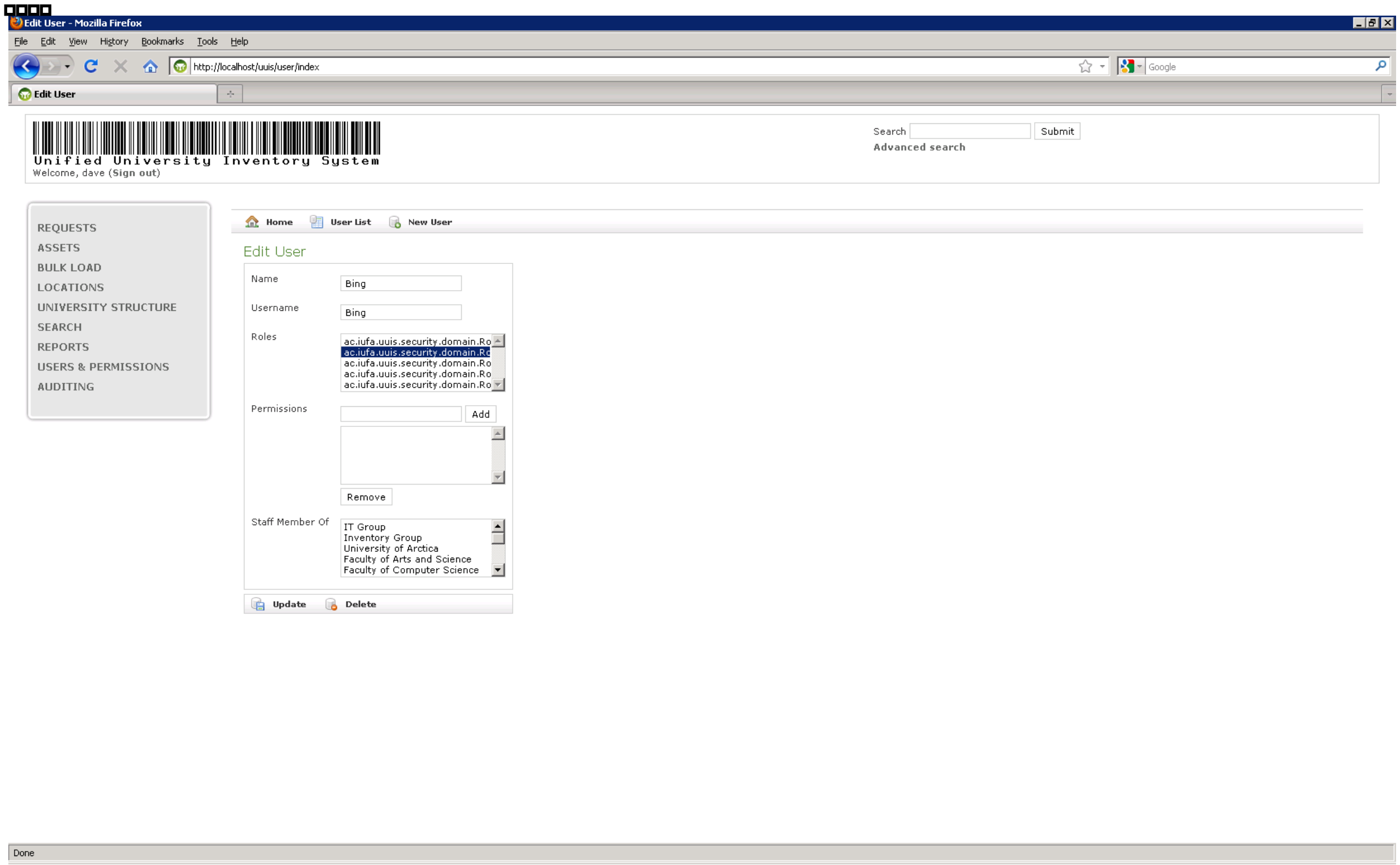

Edit user from the 'users & permissions' module, the roles list is not user friendly.

**FIXED**

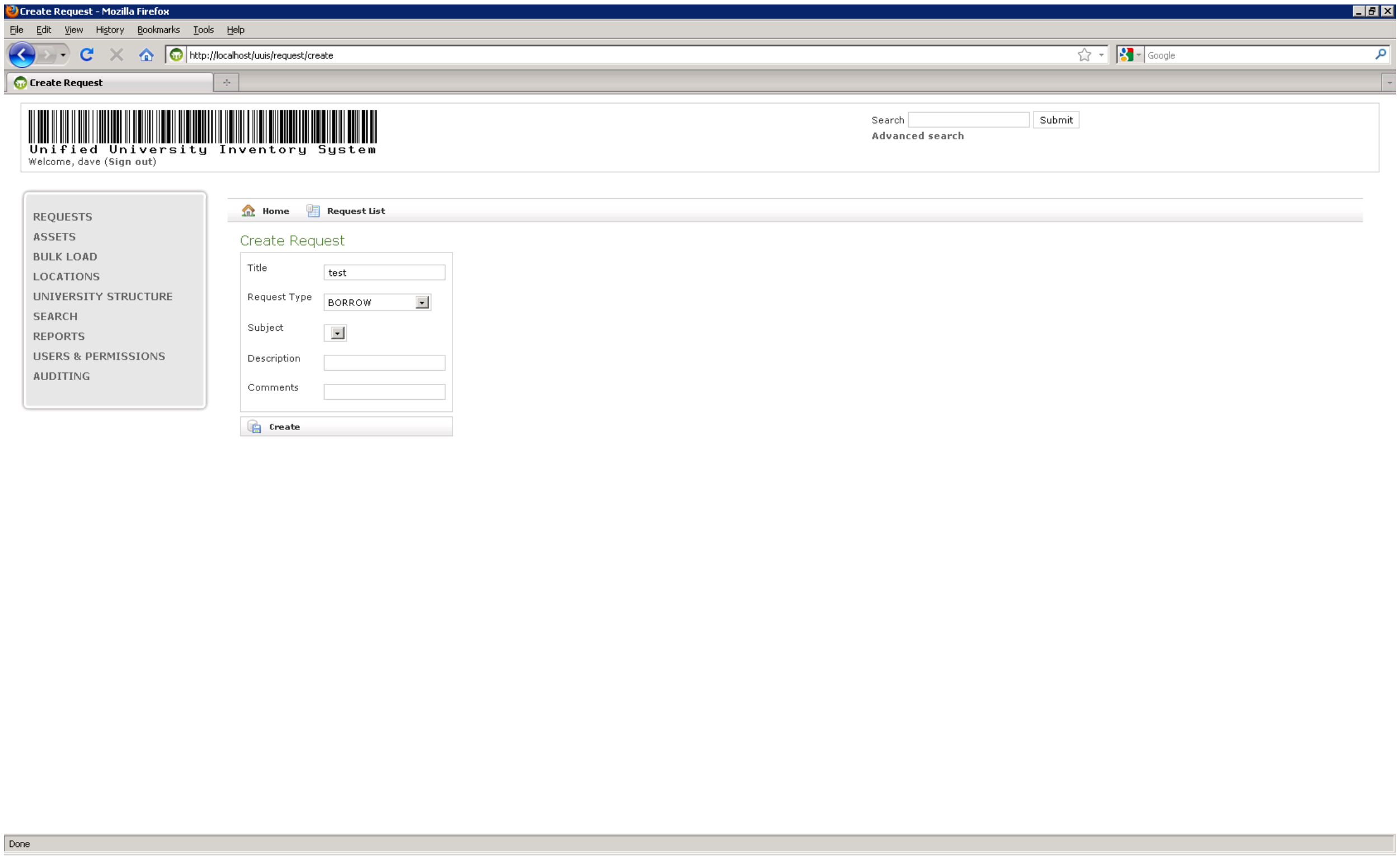

In make new request, the subject drop down list is empty. And when I submit the request, it give me exception.

**FIXED**

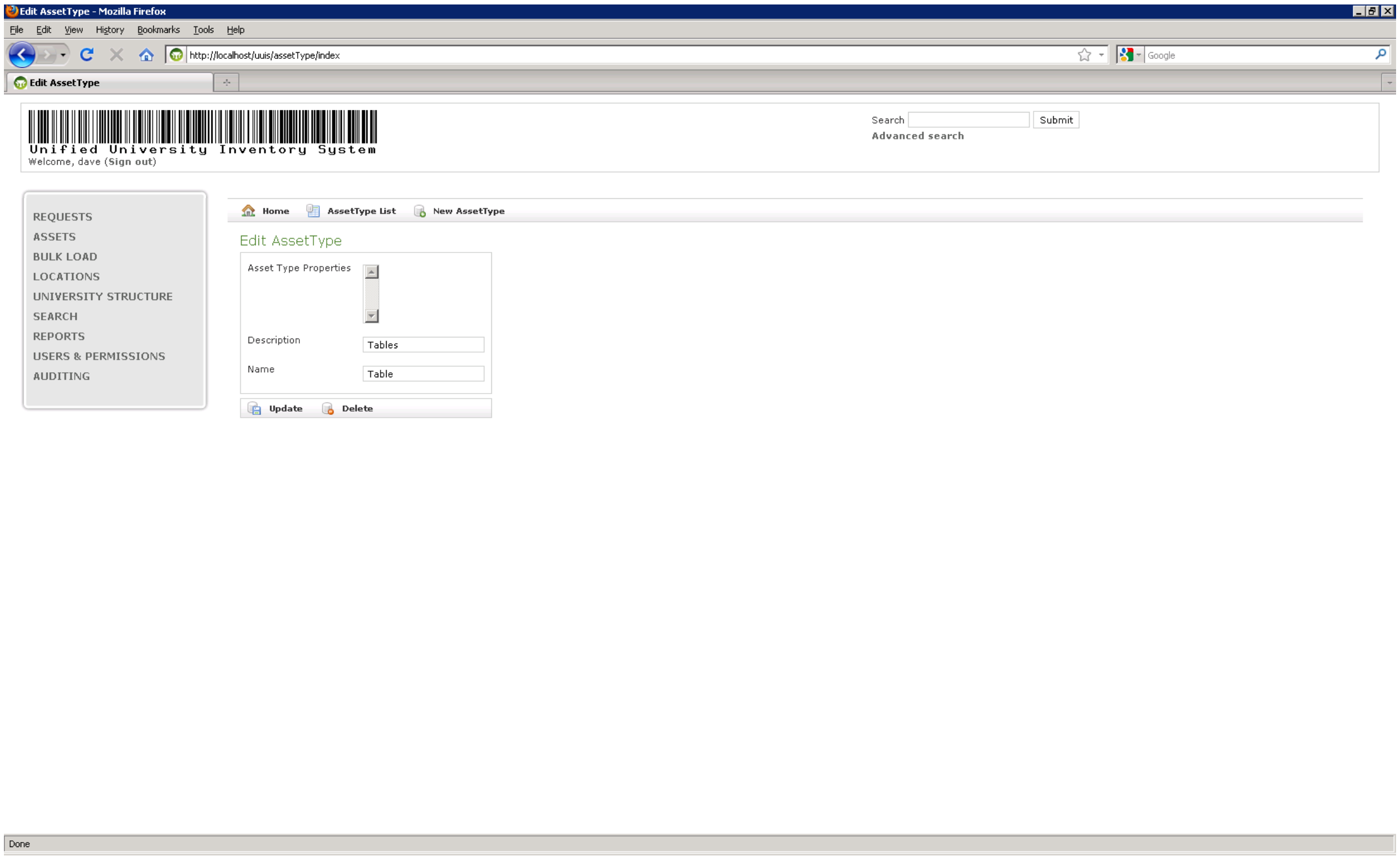

When edit asset type , the properties list is empty

**FIXED**

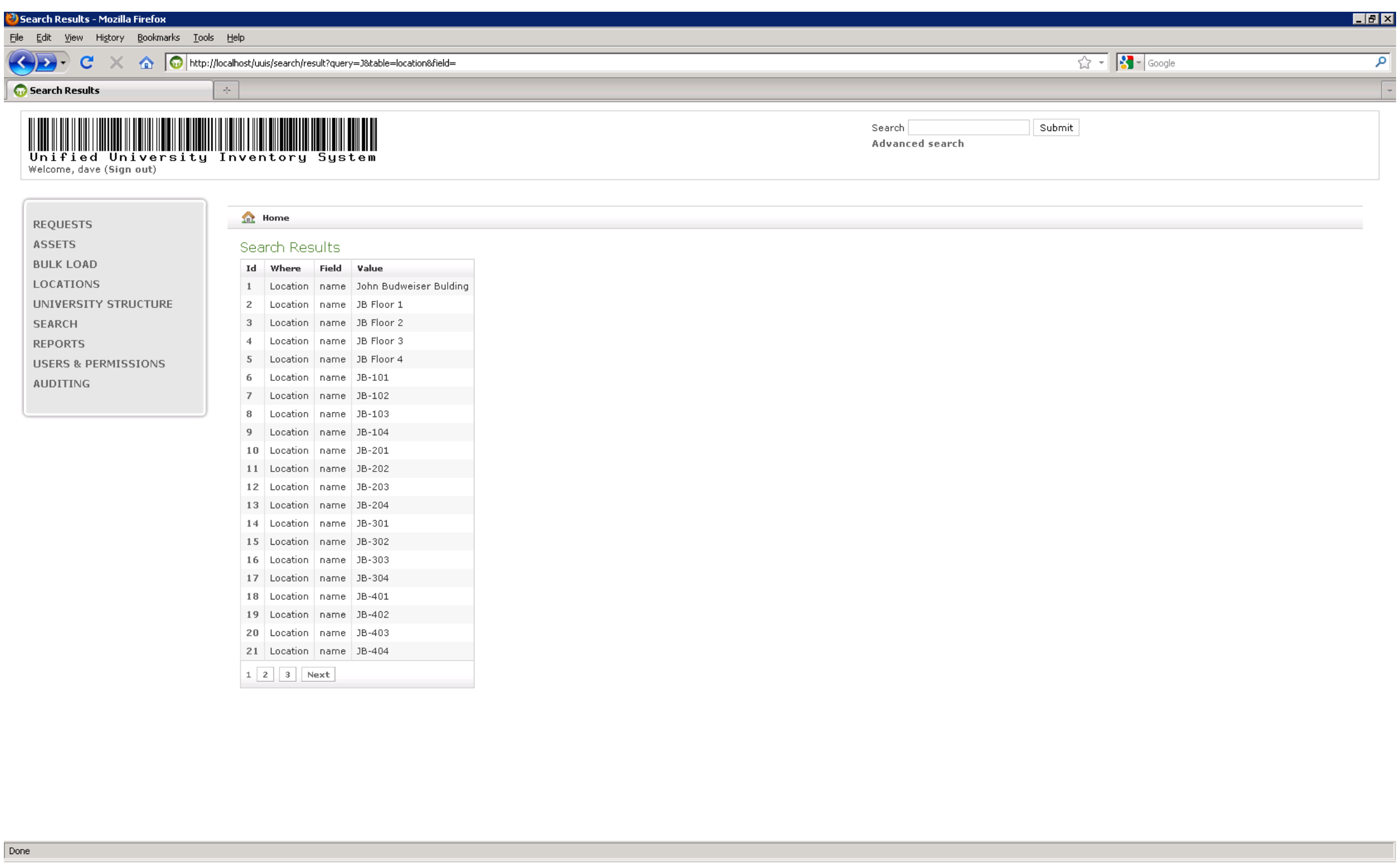

In the search module, search J in the location, I've got this list, and when go to the second page, nothing mached.

**FIXED**

1- when login to the application if type a wrong passwprd for many time it generate an exception and after that you can entre with any password to the system  I attach a print screen of the exception

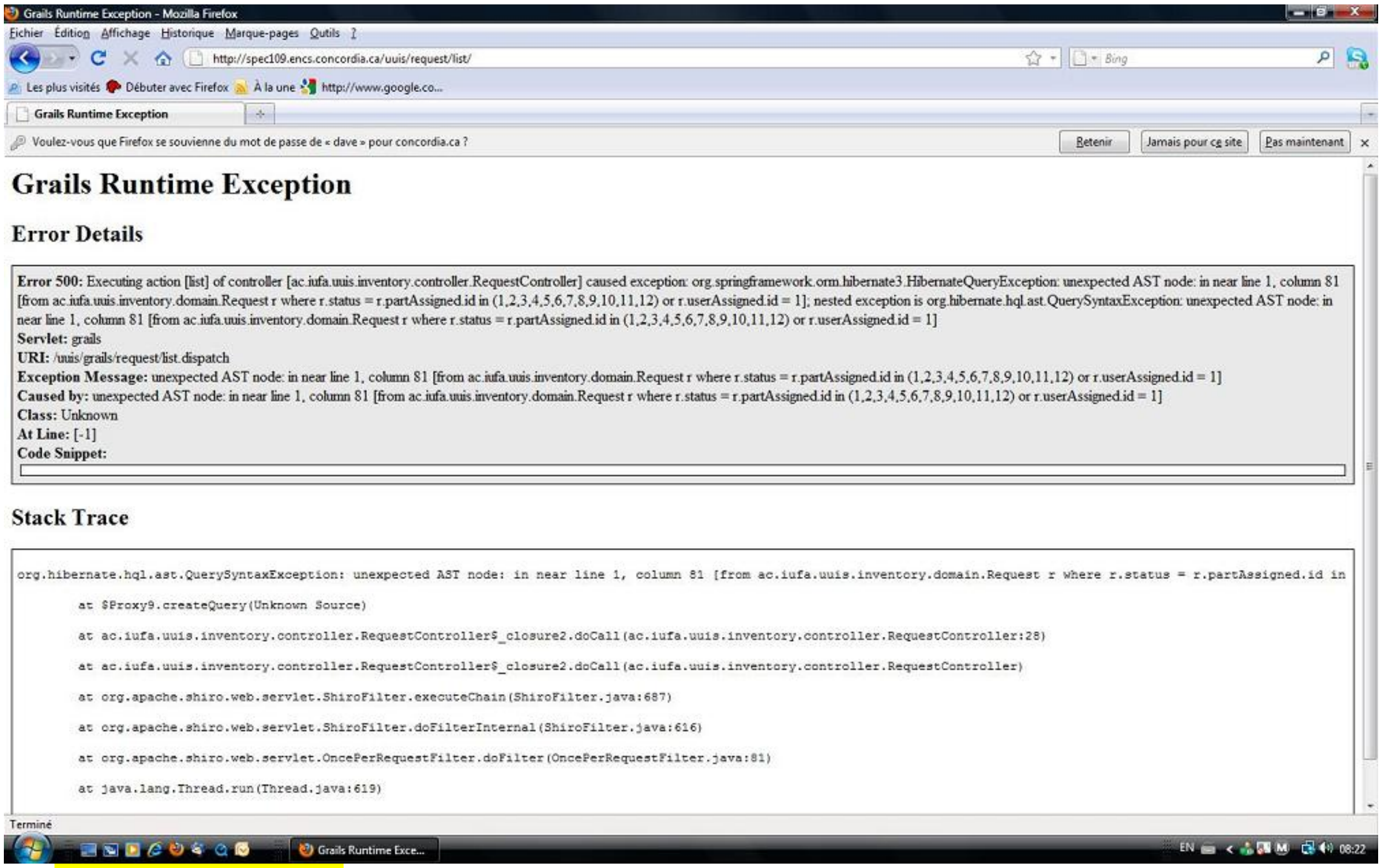

**CANNOT REPRODUCE**

2- on the university structure list if you click on any id an eception is generated (attachment ex2)

**FIXED**

3- if you click home on the user permission list or anu other screen it generates an exception (attachment ex3)

**CANNOT REPRODUCE**

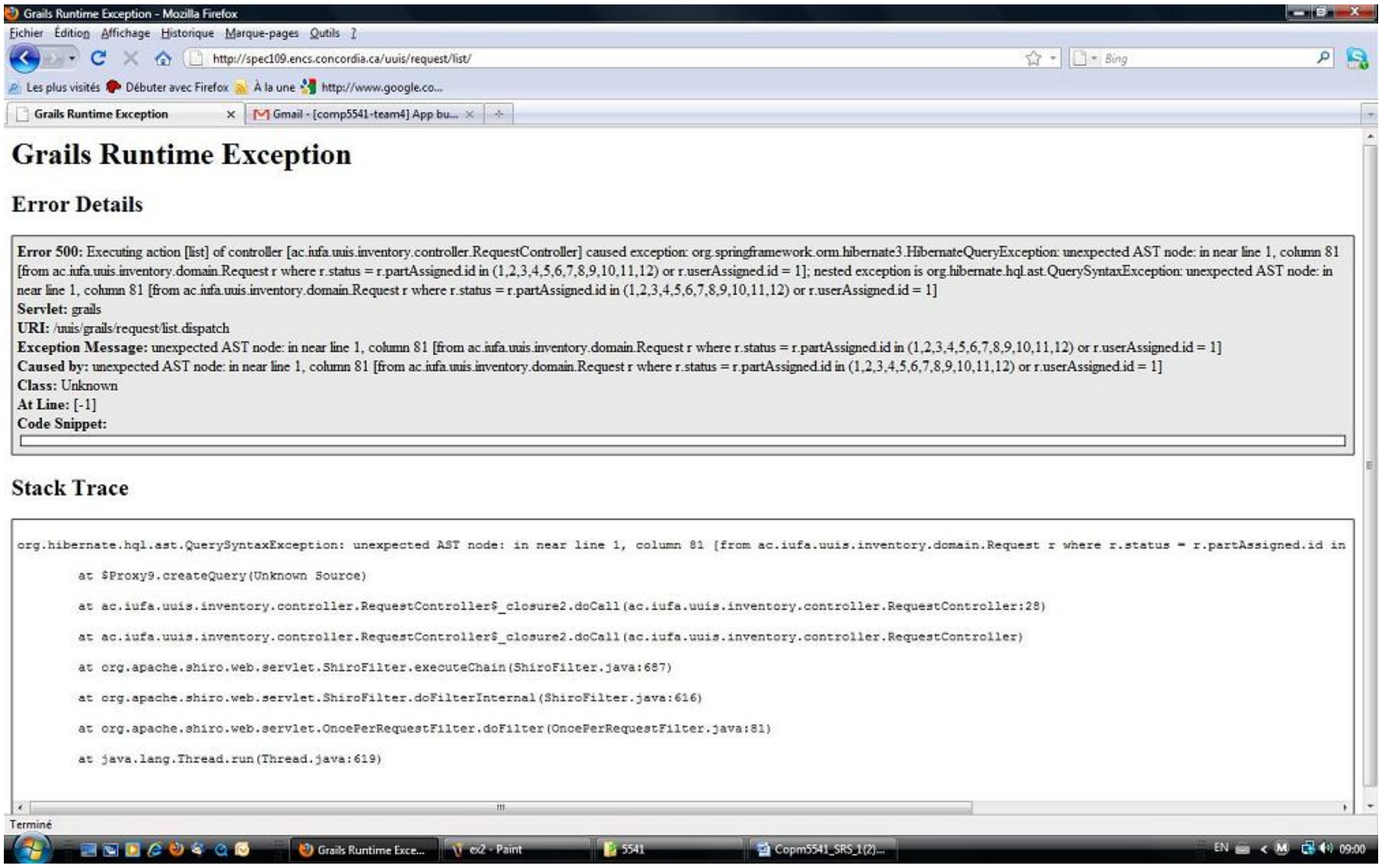

4- how user can change his passord the we cannot edit this field

**INVALID DEFECT:** The application doesn't need to manage users, any functionality related to this (ex: create user) was included with the sole purpose to test the application, we're considering we import these users from an external system.

5- when you use advenced search if no results are found you have to click on advenced seach again to restart searching

**WON'T FIX:** This is a minnor issue that will be addressed in later releases. ;-)

6- when you cret a user you alays edit the user it is not considered like level0 user directly

**INVALID DEFECT:** The level of the user is defined by which University Structure he's head of, if he's head of the University he's level 3, if he's head of a Department he's level 1. When you create a user he's not assigned as head of any University Structure so he's considered level 0 directly. To change this, edit the University Structure to include the user as head.

1- the edit location page, the property list also not user friendly. Please take care of that :)

**FIXED**

2 - In the create new asset page, the parent drop down list should be location name not id

**INVALID DEFECT:** The “Parent” field in the create new asset page is used to “group” many assets, ex: I can have a computer and set it as the parent of assets like keyboard, mouse. What is shown in this field is the IUFAID of the parent asset. Use the field “Location” if you want to set the location of the asset.